# FRET nanoscopy enables seamless imaging of molecular assemblies with sub-nanometer resolution

Jan-Hendrik Budde[1,†], Nicolaas van der Voort[1,†], Suren Felekyan[1], Julian Folz[1], Ralf Kühnemuth[1], Paul Lauterjung[1,2], Markus Köhler[3], Andreas Schönle[3], Julian Sindram[4], Marius Otten[4], Matthias Karg[4], Christian Herrmann[2], Anders Barth[1,*], Claus A. M. Seidel[1,*]

[1] Chair for Molecular Physical Chemistry, Heinrich-Heine-University Düsseldorf, Germany

[2] Physical Chemistry I, Ruhr-Universität Bochum, Germany

[3] Abberior Instruments GmbH, Hans-Adolf-Krebs-Weg 1, 37077 Göttingen, Germany

[4] Lehrstuhl für Kolloide und Nanooptik, Heinrich-Heine-Universität Düsseldorf, 40225 Düsseldorf, Germany

[†] contributed equally

[*] Correspondence to: cseidel@hhu.de, anders.barth@hhu.de

## Abstract

By circumventing the optical diffraction limit, super-resolved fluorescence microscopies enable the study of larger cellular structures and molecular assemblies. However, fluorescence nanoscopy currently lacks the spatiotemporal resolution to resolve distances on the size of individual molecules and reveal the conformational fine structure and dynamics of molecular complexes. Here we establish FRET nanoscopy by combining colocalization STED microscopy with multiparameter FRET spectroscopy. We simultaneously localize donor and acceptor dyes of single FRET pairs with nanometer resolution and quantitatively measure intramolecular distances with sub-nanometer precision over a large dynamic range. While FRET provides isotropic 3D distance information, colocalization measures the projected distance onto the image plane. The combined information allows us to directly determine its 3D orientation using Pythagoras's theorem. Studying two DNA model systems and the human guanylate binding protein hGBP1, we demonstrate that FRET nanoscopy unravels the interplay between their spatial organization and local molecular conformation in a complex environment.

## Introduction

Fluorescence microscopy has made key contributions towards our understanding of biomolecular structures and dynamics at ambient conditions due to being minimally invasive and highly selective to the molecule of interest. An ideal microscope would provide seamless resolution from the dimensions of cells and cellular substructures to the molecular architecture of biomolecular assemblies. By circumventing the diffraction limit of optical microscopy, super-resolved fluorescence microscopy (nanoscopy) techniques have achieved a major step towards such a fluorescence "nanoscope"[1,2]. Established techniques for super-resolved imaging such as stimulated emission depletion (STED)[3], the various single-molecule localization microscopies (SMLM)[4-9] or cryogenic optical localization (COLD)[10] achieve theoretical resolutions of several nanometers. Practically achievable lateral resolutions at ambient conditions are often limited to about 20 nm, although the MINFLUX technique has recently pushed this limit to the single digit nanometer range[11-13]. While nanoscopy is hence well-suited to elucidate the spatial organization of larger cellular structures and molecular assemblies[14], it currently lacks the resolution to determine distances on the size of individual molecules that are required to address the conformation and fine structure of molecular complexes. Recently, colocalization microscopy has been shown to provide nanometer accuracy down to a distance of 8 nm[15], however the resolution of this approach is limited by the occurrence of Förster resonance energy transfer (FRET) between the dyes at distances below ~12 nm. In addition, extensive corrections have to be applied to account for optical aberrations and sample drift during the required long acquisition times.

Here, we combine multiparameter FRET spectroscopy with STED nanoscopy to simultaneously localize single FRET pairs with nanometer resolution and quantitatively measure intramolecular distances with sub-nanometer precision (Fig. 1a-b). Different to previous approaches[16,17], we take full advantage of the potential of FRET to quantitatively resolve distances on the molecular scale with single-molecule resolution[18]. Importantly, the FRET information provides isotropic 3D distance information, while colocalization measures the projected distance onto the xy-plane. Combining these two observables, it becomes possible to determine the 3D orientation of the interdye vector using Pythagoras's theorem (Fig. 1b), without the need for more complex approaches such as tomographic reconstruction[10]. Multiparameter FRET nanoscopy combines three concepts (Fig. 1a-b). First, we apply STED nanoscopy to break the diffraction limit up to a resolution of ~40 nm to distinguish individual molecules. Second, we combine STED with single molecule colocalization (cSTED) to take advantage of the high localization precision for single fluorescent molecules, providing nanometer accuracy in the range of 5-40 nm. The cSTED approach is fast (<10 s) as it does not rely on the accumulation of many localizations per molecule so that no drift correction is needed. Moreover, no corrections for chromatic and spherical aberrations are required because the point spread function (PSF) of the two spectral channels is dominated by the profile of the shared STED beam. Despite the lower photon budget compared to other SMLM techniques, cSTED reaches nanometer localization precisions due to the smaller size of the STED PSF and high scanning speed.

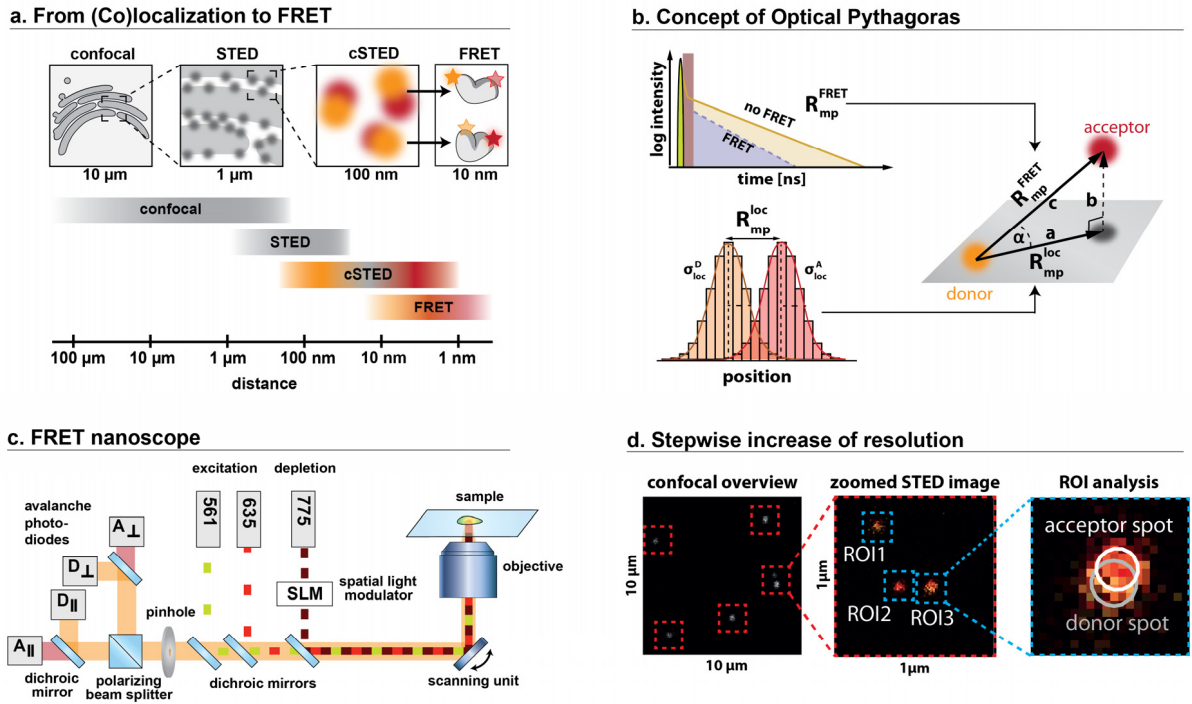

**Figure 1: The concept of FRET nanoscopy. a)** The combination of confocal and STED microscopy with colocalization-STED (cSTED) and FRET spectroscopy of a donor and acceptor dye enables seamless sub-nanometer resolution. **b)** Quantitative FRET-derived distances are obtained from the analysis of the cumulative donor fluorescence decay, providing the average donor acceptor distance ⟨$R_{DA}$⟩ that is converted to the FRET-based physical distance between the mean (center) dye positions $R_{mp}^{FRET}$ (see Suppl. Tab. 8 and Suppl. Fig. 5)[19,20]. In addition, the positions of the donor and acceptor fluorophores are localized from the detected signal in each channel, providing the localization-based distance between the mean dye positions $R_{mp}^{loc}$. By comparing $R_{mp}^{FRET}$ and $R_{mp}^{loc}$, information about the three-dimensional orientation of the molecule is obtained using Pythagoras's theorem. **c)** The experimental setup for FRET nanoscopy is based on a conventional STED setup with two excitation lasers and spectrally resolved detection. A single STED laser (775 nm) depletes both acceptor and donor fluorophores. Polarized excitation and detection provide the anisotropy information and allows to reconstruct the polarization-free fluorescence decay for accurate lifetime analysis (see main text and materials and methods). **d)** The workflow for FRET nanoscopy. Small areas of 1x1 µm² are identified in a confocal overview and imaged under STED conditions. Regions of interest (ROIs) are identified and analyzed separately.

In the third aspect, we take advantage of the occurrence of FRET between a donor and an acceptor dipole at shorter distances as a natural extension of the cSTED approach, allowing us to address molecular distances in the range of 4-12 nm[20,21]. This approach is similar to the recently introduced metal-induced energy transfer (MIET) that provides sub-nanometer axial localization of fluorescence emitters using the distance-dependent quenching by metal[22] or graphene[23-25] over a range of up to 20 nm. However, by measuring the relative distance between two spectrally different fluorophores on the same molecule, FRET provides an internal reference that allows to directly measure the molecular conformation[19,26,27]. We implement FRET nanoscopy with multiparameter fluorescence imaging spectroscopy[28] to obtain spectral, time-resolved and polarization information, allowing us to classify localizations of single molecules using the intensity-based FRET efficiency as well as the fluorescence lifetimes and anisotropies of the donor and acceptor fluorophores. FRET nanoscopy thus combines spatial resolution on the molecular scale with ultimate sensitivity for molecular features by taking advantage of all dimensions of the fluorescence signal[29].

In this study, we benchmarked FRET nanoscopy by two model systems. First, we use DNA origami nanotechnology to place two FRET pairs at a distance of ~75 nm. We show that FRET nanoscopy resolves the geometry of the four dyes on the origami with sub-nanometer precision, while simultaneously providing quantitative FRET information for the two FRET pairs. Second, we assess the accuracy of FRET-derived distances from the intensity and lifetime information using a set of DNA rulers that are immobilized on one or on both ends. The 3D information provided by FRET nanoscopy allows us to address the orientation of the DNA and resolve the roughness of the surface on the molecular scale. Finally, we use FRET nanoscopy to map the human guanylate binding protein 1 (hGBP1) which forms long fiber structures *in vitro* and live cells by oligomerization. Different to its closed conformation in the monomeric state, we show that hGBP1 opens up like a pocketknife and adapts an extended conformation in the assembly, highlighting the potential of FRET nanoscopy to resolve molecular conformation in a complex context.

## Concepts

**Combining STED microscopy and multiparameter FRET spectroscopy.** The multiparameter fluorescence detection STED microscope is based on a modified confocal microscope with pulsed excitation and depletion beams and polarization-resolved detection in two spectral channels (Fig. 1c and Supplementary Fig. 1), so that we register altogether four polarization resolved fluorescence decays, two for each spectral detection window. To probe both donor and acceptor fluorophores in rapid succession, we perform alternating line scans with direct donor and acceptor excitation. We deplete both fluorophores with the same STED pulse that is overlaid with the excitation and has 2 ns duration (Fig. 1c and Supplementary Fig. 1). Efficient depletion of both fluorophores by a single STED laser requires that both fluorophores show sufficient emission at the STED wavelength of 775 nm. Consequently, spectrally close fluorophores are needed. Here we use Atto594 or Alexa594 as FRET donors and Atto647N as FRET acceptor. We additionally apply time-gated detection to increase the spatial resolution for cSTED by discarding any photons that arrive during the depletion pulse[30-32], resulting in typical resolutions of 65 nm and 50 nm for the donor and acceptor channels, respectively (Supplementary Tab. 5). While higher resolutions are achievable with higher depletion power, these numbers represent a compromise to ensure that sufficient photons are detected for the FRET analysis.

**FRET nanoscopy workflow with stepwise increased resolution.** We first record a confocal overview image to determine the location of single molecules or molecular assemblies (Fig. 1d). For each location, a zoomed STED image (1x1 µm) is recorded, wherein regions of interest (ROI) are identified that contain a single dye or a closely spaced donor-acceptor dye pair. Each dye is identified as a 'spot' and the fluorescence signal in a spot pair is used to determine the spectroscopic parameters, i.e., the intensity-based FRET efficiency and the donor fluorescence lifetime and anisotropy. Additionally, the position of the emitters is localized by fitting with a 2D Gaussian function with a typical localization precision of ~3 nm. The localization precision $\sigma_{\mathrm{loc}}$ of the donor or acceptor fluorophore provides an additional criterion to distinguish single molecules and remove aggregates or multi-molecule events. The localization algorithm also identifies the most likely number of emitters based on a maximum likelihood criterion while penalizing for overfitting, allowing us to characterize each ROI by the number of active donor and acceptor fluorophores (spot stoichiometry, Supplementary Fig. 2). These parameters are used for filtering and to identify sub-populations within the ensemble of measured ROIs (Supplementary Fig. 3). For each FRET pair, the length of the projected inter-dye distance vector, $d_{\mathrm{loc}}$, is determined. The distribution of $d_{\mathrm{loc}}$ depends on the distance between the mean positions of the dyes, $R_{\mathrm{mp}}^{\mathrm{loc}}$, and the localization precision and is described by a non-central $\chi$-distribution with two degrees of freedom (see methods eq. 15 and Supplementary Section 'Colocalization analysis')[15,33]. By calibrating the width parameter of the distribution, the mean-position distance $R_{\mathrm{mp}}^{\mathrm{loc}}$ can thus be determined with high precision despite the broad distribution of $d_{\mathrm{loc}}$.

The time-resolved FRET nanoscopy experiment offers two approaches to determine the FRET efficiency and thus interdye distance: either from the detected intensities or the time-resolved fluorescence decays of the donor fluorophore. Due to the flexible dye linkers, an equilibrium distribution of $R_{DA}$ values is measured, so that the FRET efficiency is related to the FRET-averaged apparent donor-acceptor distance, $\langle R_{DA} \rangle_E$, by:

$$E = \frac{1}{1 + \left( \frac{\langle R_{DA} \rangle_E}{R_0} \right)^6},$$

(1)

where $R_0$ is the Förster radius. For each spot pair, the FRET efficiency can be calculated from the corrected photon counts of the donor and acceptor fluorophores after donor excitation, $F_{D|D}$ and $F_{A|D}$, or the fluorescence lifetimes of the donor in the presence and absence of the acceptor, $\tau_{D(A)}$ and $\tau_{D(0)}$. In the absence of dynamics, the intensity-based and lifetime-based estimators of the FRET efficiency are equivalent, given by:

$$E = \frac{F_{A|D}}{F_{D|D} + F_{A|D}} = 1 - \frac{\tau_{D(A)}}{\tau_{D(0)}}.$$

(2)

Slight deviations from this relation occur due to fast dynamics of the dye linkers (see eq. 10 in the methods)[34], in which case the spot-integrated donor fluorescence lifetime corresponds to the fluorescence-weighted average which we denote as $\langle \tau_{D(A)} \rangle_F$ in the following. To estimate accurate fluorescence lifetimes under STED conditions for each spot, we discard the initial part of the donor fluorescence decay during the depletion pulse and fit the tail to a single-exponential model function using a maximum likelihood estimator[35] (see Methods, Supplementary Section 'Determination of spot-integrated fluorescence lifetimes' and Supplementary Fig. 4). The determination of accurate intensity-based FRET efficiencies under STED conditions requires further corrections. A detailed description of the correction procedure for the intensities for crosstalk, direct excitation, differences in the quantum yield and detection efficiencies[20,36], as well as residual signal from partially depleted fluorophores is given in Supplementary Section 'Accurate intensity-based FRET efficiencies under STED conditions'. A more precise estimate of the interdye distance distribution is obtained from the analysis of the sub-ensemble fluorescence decay. Different to the intensity-based approach, lifetime measurements report directly on the mean interdye distance $\langle R_{DA} \rangle$ and the width of the distance distribution (eq. S17). To utilize the complete fluorescence decay in the analysis, we account for the depletion part of the decay as a fast decay component with a lifetime of ~200 ps (Supplementary Tab. 9). Fluorescence decays of the donor in the absence and presence of the acceptor were analyzed globally to increase the robustness of the fit[37] (see Methods and Supplementary Section 'Sub-ensemble fluorescence decay analysis').

## Results

**FRET-nanoscopy resolves DNA origami nanostructures with single base pair resolution.**
We benchmarked the capabilities of FRET nanoscopy by imaging a single-layer rectangular DNA origami platform, O(HF+NF)[38,39], where we placed two FRET pairs (Atto594/Atto647N) with a high and zero FRET efficiency (high-FRET, HF and no-FRET, NF species), respectively (Supplementary Fig. 7). The distance between them amounts to ~75 nm, well below the diffraction limit (Fig. 2a-b). Using negative-stain transmission electron microscopy, we determined the mean dimensions of the origami as (88 ± 3) x (59 ± 4) nm (Supplementary Fig. 8). The DNA origami platforms were immobilized on a PEG-coated glass surface with neutravidin using eight biotin anchors on the lower side to ensure that the origamis are oriented in the xy-plane. We acquired zoomed STED images of 1824 origami platforms and localized individual emitters in the donor and acceptor channels (Fig. 2a). Using the spot stoichiometry of two donors and two acceptors as a filter criterion, we remove incompletely labelled origami platforms and retain only those containing two FRET dye pairs (N=137, Supplementary Fig. 2). To correct for the different orientations of the origamis in the xy-plane, we aligned the structures by rotation and translation of the dye coordinates using the Kabsch algorithm[40] (Fig. 2b, eq. 14). We selected an unbiased reference structure for the alignment by choosing the structure from the dataset with the lowest overall root mean squared deviation (RMSD). This approach requires no prior knowledge about the sample. Defect structures that showed a high RMSD after the alignment step were removed (see gray points in Fig. 2b and Supplementary Fig. 9). The overlay of all aligned structures allowed us to reconstruct the geometric arrangement of the dyes and estimate the mean dye positions and displacement vectors (Fig. 2b). From the standard error of the mean of the dye positions, we obtain a localization precision of < 4.5 Å. This high precision indicated to us that we could utilize our measurements to estimate the structural parameters of the origami platform. Due to the defined attachment points on the platform (Supplementary Fig. 7), the interdye distances can be expressed as the number of base pairs in the horizontal direction and the number of helices in the vertical direction (see eq. S35).

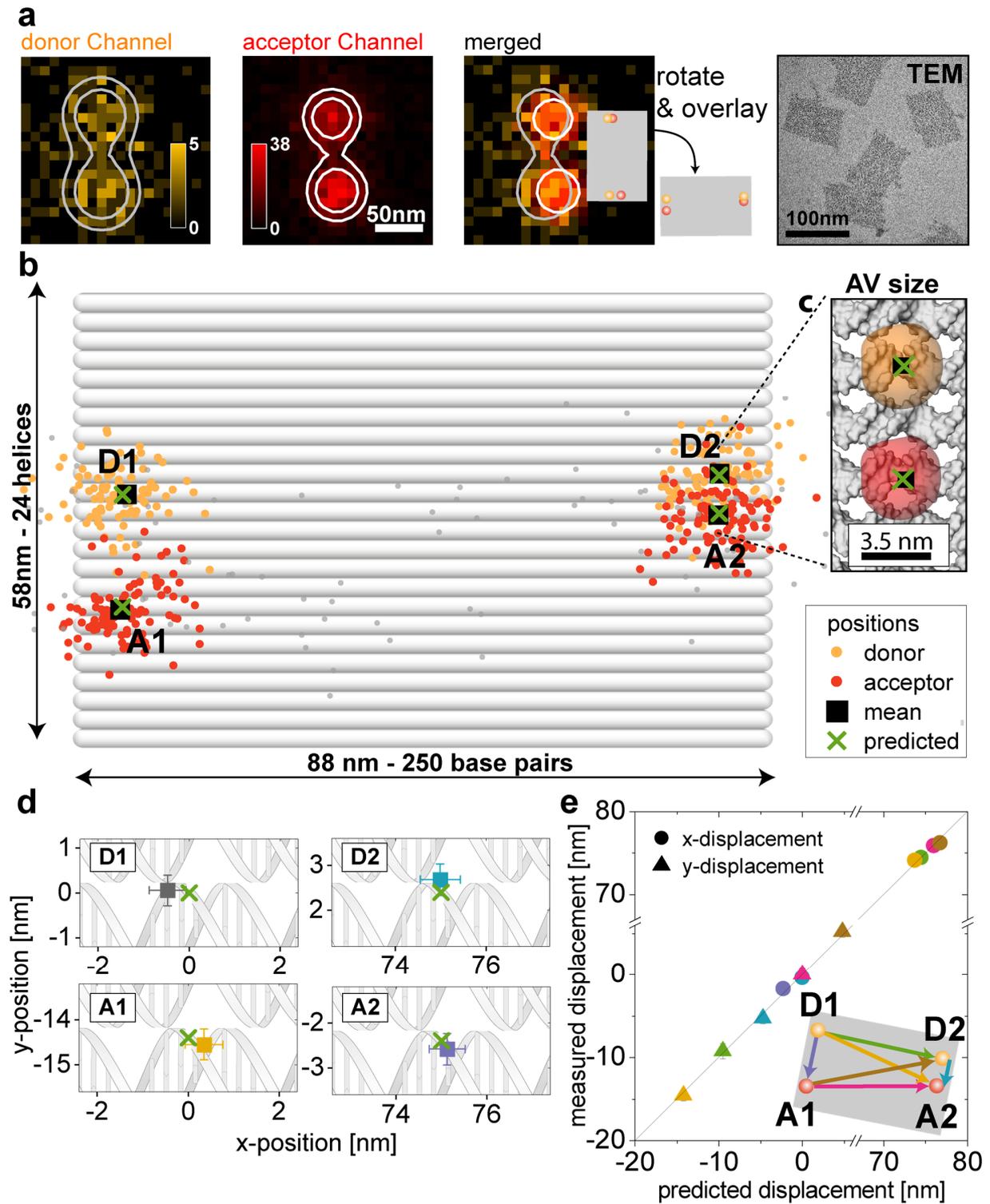

**Figure 2: Colocalization STED (cSTED) analysis on rectangular DNA-origami nanostructures O(HF+NF), labelled with two donor-acceptor dye pairs (Atto594 and Atto647N). (a)** Single origamis are recorded under gated STED conditions (see Supplementary Section 'Data acquisition' and Supplementary Tab. 14), yielding a median of 245 photons for the

donor and 537 for the acceptor fluorophore. The position of the two dyes in each channel is determined by fitting two Gaussian distributions with background. The set of four positions is then rotated and aligned to a common reference structure and each emitter is classified D1, D2, A1 or A2 based on its location and channel. Contour lines indicate one- and two sigma distances from the Gaussian centers. TEM images show intact origami platforms. **(b)** The estimated dye positions for a total of 137 origamis carrying all four dyes are overlaid on the DNA origami platform drawn to scale (gray rods represent DNA double strands). 36 structures had an overall RMSD > 10 nm and are disregarded for further analysis (grey points). The predicted structure is based on an interhelical distance of 2.4 nm and base pair extension of 3.16 Å (see text). The mean structure is overlaid with the predicted structure such that the center-of-mass and A1-A2 unit vector coincide. **(c)** Zoom-in on the D2-A2 dye pair. The clouds indicate the accessible volumes of the flexibly coupled fluorophore due to linker movement. The average dye-dye displacement is 5.2 nm, close to the structural prediction of 4.8 nm. **(d)** Zoom-in of the measured mean positions compared to the predicted values. Cartoon-DNAs are drawn to scale, placing the predicted position at the dye attachment point on the double helix (see main text and Supplementary Fig. 7). **(e)** The six interdye distances are compared to their predicted values. The cartoon structure is rotated such that the A-A vector lies on the x-axis. The values match within a precision of 7 Å over the range of -5 to 75 nm and are listed in Supplementary Tab. 19. Inset coloring matches the data markers.

As an example, the distance between the two acceptor fluorophores is 236 base pairs in the x-direction and 5 helices along the y-direction, for which we measured a distance of 75.8 ± 0.6 nm (Supplementary Fig. 10). We find the best agreement between the structural model and the six measured interdye distances for a rise per base pair of 3.16 ± 0.03 Å and an inter-helical distance of 2.4 ± 0.1 nm (see Supplementary Note 2 and Supplementary Fig. 11). The predicted dye positions based on these model parameters are overlaid in Fig. 2b-d, showing the excellent agreement for all four dye positions. In Fig. 2d, we additionally overlay the estimated dye positions on a structural model of the DNA helices to highlight that our precision of 4.5 Å reaches the dimensions of a single base pair. A potential concern for the localization precision is given by the long linkage used to attach the dyes to the origami platform, allowing them to explore a large accessible volume (AV) on the origami surface as indicated in Fig. 2c (see Supplementary Section 'Accessible volume simulations'). Here, the linkage consists of two units: an unpaired nucleotide at the 3' end of a short staple strand and the usual flexible dye tether, resulting in a slightly increased linker length of ~29 Å (Supplementary Fig. 7 and Supplementary Tab. 6). However, as the movement of the dyes is fast compared to the acquisition time of the experiment, the localization approach measures the average position of the dye within its sterically accessible volume (AV). Interestingly, the spread of the localizations of ~4.5 nm exceeds the theoretical limit based on photon statistics of 2 nm, which is unlikely to originate from registration error and suggests that we are sensitive to structural heterogeneities of the DNA origami constructs (see Supplementary Note 3). In summary, the cSTED approach resolves distances between individual fluorophores on DNA origami nanostructures with a localization precision of < 5 Å over a wide range of displacements along the x and y direction from 5-80 nm (Fig. 2e).

The estimated average interhelical distance of 2.4 nm is well in the range of the experimental and theoretical observations. Due to the 'chicken-wire' structure of the DNA origami, the interhelical distance fluctuates between 1.85 nm at the junction and a maximum of 3.6 nm[41,42]. In our design, the dyes are placed 7 bp away from the nearest junction and thus half-way between the points of maximum and minimum interhelical distance. Our value agrees well with theoretical estimates from MD simulations[42,43] but is smaller than the values reported from cryoEM or small-angle X-ray scattering of 2.6-2.7 nm[41,44]. Interestingly, our estimate for the rise per base pair of 3.16 Å is slightly lower compared to values for individual double-stranded DNA (of ~3.32 ± 0.19 Å from crystallographic data[45] and 3.29 ± 0.07 Å from scattering interference measurements[46]). Different to these approaches, we measured the average rise per base pair over long distances of 237 bp across the origami nanostructure. Hence, the lower value indicates that the origami is slightly compressed along the long axis, potentially due to a breathing of the chicken-wire structure. A comparable deviation towards shorter distances has also been previously reported for interdye distances on DNA origami helix bundles measured by DNA-PAINT microscopy[47].

**Resolving distinct FRET species within the diffraction limit.** We then defined regions of interest of 70 x 70 nm containing single donor-acceptor FRET pairs for each DNA origami and computed two FRET indicators, the corrected intensity-based FRET efficiency $E$ and the fluorescence-weighted average donor fluorescence lifetime $\langle \tau_{D(A)} \rangle_F$ (see Methods, spectroscopic analysis). Their correlation is shown in a two-dimensional frequency histogram of analyzed spots (Fig. 3a). Different to the colocalization analysis, we also included all constructs which contained at least one donor-acceptor dye pair in the FRET analysis to obtain higher statistics (N=1391). The $E$ - $\langle \tau_{D(A)} \rangle_F$ diagram reveals two species with zero and high FRET efficiency that follow the expected relation given by the static FRET-line (solid line in Fig. 3a, see eq. 10 in the methods and Supplementary Table 20). Due to the low photon number in the donor channel for the high-FRET (HF) dye pair, we detect a larger fraction of no-FRET (NF) dye pairs. The tailing from the HF population towards the NF population is a result of photobleaching of the acceptor fluorophore during the acquisition time that is also evident in control measurements of the HF FRET-pair in the reference construct (Supplementary Fig. 13a). Consequentially, the FRET efficiency of the HF species (E=0.45) is slightly underestimated compared to solution-based single-molecule FRET control measurements of origamis carrying only one of the two dye pairs (E=0.53, Supplementary Fig. 12a). However, the intensity-based FRET efficiency is mainly used here for separating the two species, and we show below that accurate FRET-derived distances are obtained from the sub-ensemble analysis of the fluorescence decays.

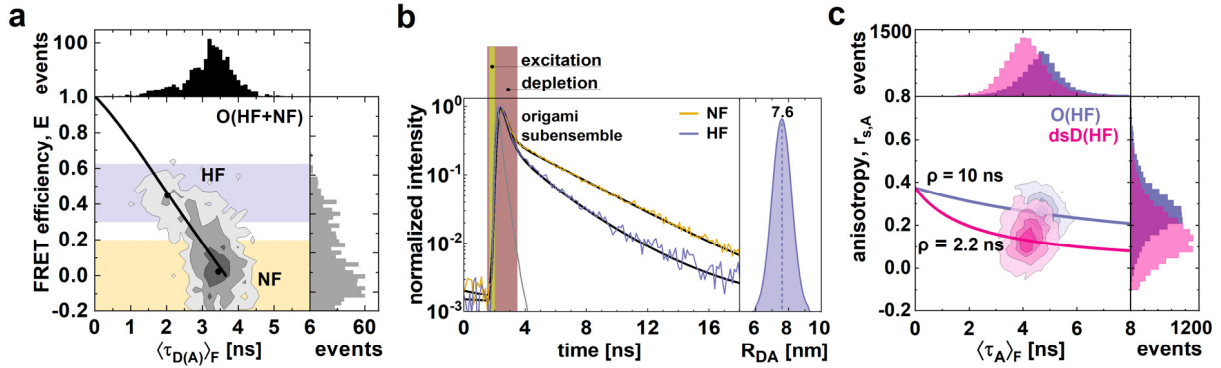

**Figure 3: FRET nanoscopy distinguishes two FRET pairs for the O(HF+NF) sample within the diffraction limit. (a)** Two-dimensional frequency histogram of the ROI-integrated intensity-based FRET efficiency, E, and the fluorescence-weighted average donor fluorescence lifetime, $\langle \tau_{D(A)} \rangle_F$. Dashed lines indicate the expected FRET efficiencies from single-molecule measurements in solution. The intensity-based FRET efficiency, E, was corrected for STED conditions as described in the methods. The static FRET-line (solid line, eq. 10 in the methods and Supplementary Table 20) shows the theoretic relation between lifetime and FRET efficiency. Only complete FRET pairs from all emitter stoichiometries were considered ($N = 1391$), while emitter localizations that did not belong to a FRET pair were removed. The HF population is found to have an $E = 0.45$, corresponding to a FRET-averaged distance $\langle R_{DA} \rangle_E = 7.9$ nm that matches the confocal single-molecule FRET measurements with $E = 0.53$ and $\langle R_{DA} \rangle_E = 7.5$ nm (Supplementary Fig. 12a). An intermediate population is visible, indicating mixing of the two FRET species. This population is absent in origami samples carrying one of the two FRET pairs (see Supplementary Fig. 13a). **(b)** Sub-ensemble fluorescence decays of the donor for the high and low FRET-efficiency species in **a**, selected by $E > 0.3$ and $E < 0.2$ for the HF and NF species, respectively. The decays are analyzed globally by eq. 6-7, yielding a donor only lifetime of 3.7 ns and a mean interdye distance $\langle R_{DA} \rangle = 7.3$ nm, matching the result from solution-based single-molecule FRET experiments of 7.8 nm. **(c)** Comparison of the two-dimensional frequency histograms for the ROI-integrated scatter corrected acceptor anisotropy, $r_{s,A}$, and acceptor fluorescence lifetime, $\langle \tau_A \rangle_F$, after direct excitation for the origami sample O(HF) and the DNA ruler dsD(HF) (see Supplementary Fig. 14). The distinct Perrin lines are given by $r_{s,A} = r_{0,A} \left( 1 + \frac{\langle \tau_A \rangle_F}{\rho} \right)^{-1}$, where $r_{0,A}$ is the fundamental anisotropy of the acceptor fluorophore taken to be 0.374 and $\rho$ is the rotational correlation time. The Perrin lines demonstrate that the mobilities of the acceptor dye, characterized by the mean rotational correlation time, $\rho$, is significantly increased for O(HF+NF).

We then selected dye pairs with a high or low FRET efficiency (HF: E>0.3, NF: E<0.2) and generated sub-ensemble decays of the donor fluorescence (Fig. 3b). For the fluorescence decay of the NF species, we obtained a donor fluorescence lifetime of 3.7 ± 0.1 ns. While the intensity-based FRET efficiency of the HF species was underestimated due to acceptor photobleaching, we can account for these artifacts by including a donor-only fraction in the lifetime analysis. Using a distance distribution model function (see Supplementary Section 'Sub-ensemble fluorescence decay analysis'), we obtain a distance between the average donor-acceptor positions $R_{mp}^{FRET}$ of 7.3 ± 0.3 nm, in good agreement with single-molecule measurements of freely diffusing molecules in solution ($R_{mp}^{FRET}$ = 7.3 ± 0.1 nm). As an additional control, we measured origamis that only contained the high-FRET or no-FRET dye pair and determined FRET-based distances using the lifetime and intensity information under single-molecule and STED conditions, which yielded consistent mean-position distances in the range of 7.2 -7.6 nm (Supplementary Fig. 13d and Supplementary Tabs. 9 and 10).

While we obtained consistent FRET-based distances under STED conditions, the FRET-derived mean position distance of 7.3 nm deviates significantly from the localization-based estimate of 5.2 nm and exceeds the expected distance for dyes that are placed two helices apart of ~4.8 nm. FRET measurements are additionally sensitive to the orientation of the transition dipole moments of the donor and acceptor fluorophores, expressed in the orientation factor $\kappa^2$ that enters the calculation of the Förster radius[20,48]. In the case of fast and free rotation of the fluorophores, an average value of $\langle \kappa^2 \rangle$=2/3 is usually assumed[48]. Hübner et al. showed that the dye Atto647N, when linked to DNA origami structure, tends to position itself between two DNA helices, which restricts the diffusion and rotation of fluorophore[49,50]. Therefore, we use the available fluorescence anisotropy information to check the validity of this assumption for our sample. Indeed, confocal single-molecule control measurements show that the fluorescence lifetime and anisotropy of the positively charged Atto647N on the origami are significantly shifted to higher values as compared to the dsDNA, while the properties of the donor dye are unchanged (Fig. 3c and Supplementary Fig. 14, Supplementary Tab. 11). The restriction of the rotation induces an uncertainty of ~0.5 nm on the Förster radius (Supplementary Note 4), which is insufficient to explain the discrepancy. In addition, the interaction with the origami surface potentially displaces the acceptor dye with respect to the mean position within the AV (Fig. 2c). As our localization-based distance estimate of the interhelical distance is however in good agreement with previous reports, the displacement must occur predominantly in the axial (z) direction. Indeed, the rotational correlation times ρ determined from the fluorescence anisotropy (Supplementary Fig. 14a) indicate that the donor dye is mobile (ρ = 0.7 ns) while the acceptor dye is trapped (ρ = 10 ns). Hence, a likely explanation for the larger FRET-derived distance is that the acceptor dye is trapped between two helices, moving it away from the donor dye, while the donor dye is pointing upwards away from the origami surface and remains mobile. Based on this model (Supplementary Note 4), we estimate a z-displacement between the dyes of 5.0 ± 1.2 nm, which is consistent with the combined length of the dye linkers, the additional unpaired nucleotide, and the thickness of the origami platform (Supplementary Figs. 7e and 15, Supplementary Note 4). This indicates that FRET-nanoscopy is not only capable of resolving different FRET species within a diffraction-limited spot, but also provides accurate FRET distances. Notably, the combined information of FRET and the localization distances together with Pythagoras's theorem (Fig. 1b) allows us to assess the 3D orientation of the interdye distance vector.

**Accurate FRET measurements of dsDNA rulers.** To test the accuracy of the FRET analysis under STED conditions, we performed measurements on short double-stranded DNA rulers (dsD) labeled with the dyes Alexa594 and Atto647N at distances from 7 to 15 nm. We verified that both dyes are freely moving in all dsD constructs (Supplementary Fig. 14b) so that we can exclude a significant influence of specific dye orientations and positions on the FRET-based interdye distances. The orientation of the interdye distance vector can be controlled by immobilizing the DNA rulers using biotin-neutravidin binding either with a single biotin (sb) at one end or doubly (db) with a biotin at each end (Fig. 4a). As a control, we performed the experiments using two different surface passivation methods based on bovine serum albumin (BSA) or polyethylene glycol (PEG, see methods). A typical confocal overview image and zoomed STED images are shown in Figure 1c for an interdye distance of 15 nm (dsD(NF)) and double-biotin immobilization using BSA. Measuring the dsD constructs with increasing interdye distances on BSA (Fig. 4b) and PEG (Supplementary Fig. 19) surfaces, the recovered values of the two FRET indicators $E$ and $\langle \tau_{D(A)} \rangle_F$ in the two-dimensional histograms follow the static FRET-line for decreasing FRET efficiency and increasing donor fluorescence lifetime. The experimental average FRET efficiencies of the populations match well with the predicted values based on AV simulations (black and red dashed lines, respectively, in Fig. 4b, Supplementary Tab. 13) To resolve the FRET rate directly and consider also photobleached dye species, we performed sub-ensemble TCSPC analysis using a distance distribution model function (Fig. 4c and Supplementary Fig. 21).

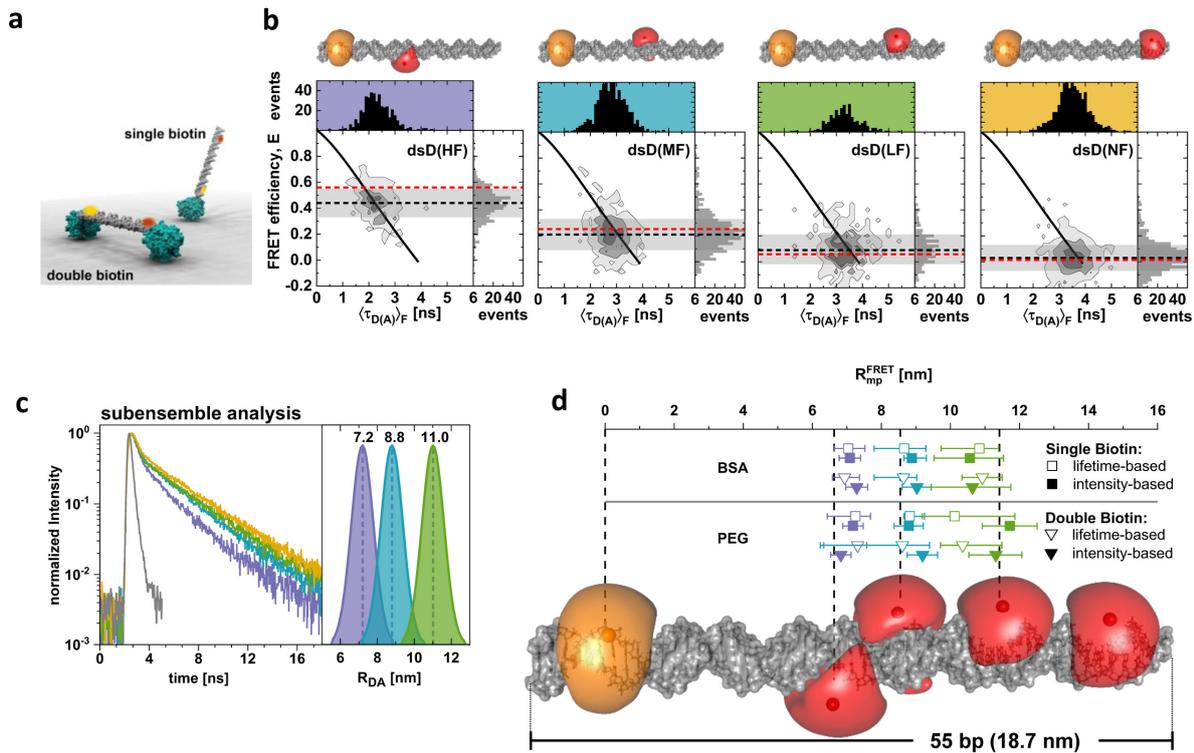

**Figure 4: Accurate FRET analysis under STED conditions using DNA rulers. a)** Cartoons of the double-stranded DNA rulers (dsD, 55 bp) labeled with Alexa594 and Atto647N on the functionalized glass surface. The DNA rulers are immobilized on one side (single biotin) or both sides (double biotin) by binding to immobilized avidin (turquoise, PDB-ID: 2AVI). The fluorophore positions are highlighted in orange and red. (details see Supplementary Tab. 1) **b)** Two-dimensional histograms of the FRET-efficiency, $E$, and fluorescence-weighted average donor fluorescence lifetime $\langle\tau_{D(A)}\rangle_F$ for the dsD(HF), dsD(MF), dsD(LF) and dsD(NF) DNA rulers using biotinylated BSA for immobilization. Measured mean FRET efficiencies are indicated by dashed black lines and predictions based on AV calculations by dashed red lines. The intensity-based FRET efficiency, $E$, was corrected for STED conditions (see Supplementary Section 'Accurate intensity-based FRET efficiencies under STED conditions'), placing it on the static FRET-line (black line, see eq. 10 in the methods and Supplementary Table 20). The selection of ROIs for sub-ensemble fluorescence decays of the donor in **c** is displayed as grey shaded area. Structural models for the different DNA rulers, indicating the accessible volumes (AV) of the fluorophores, are shown above. **c)** Sub-ensemble fluorescence decays of selected ROIs are shown in the colors corresponding to **b**. Decays were fitted to Gaussian distributed distances model with fixed width of $\sigma_{DA}$ = 0.6 nm (eq. 6-7). The depletion part of the decay was described by a short lifetime component, as described in the Supplementary Section 'Sub-ensemble fluorescence decay analysis'. **d)** Comparison of intensity-based (filled symbol) and lifetime-based (open symbol) experimentally FRET-based distances between the mean dye positions, $R_{mp}^{FRET}$ for immobilization under distinct conditions: (i) a single (square) and two (triangle) biotins, and (ii) BSA-layer (top), PEG-layer (bottom). Measured values are very close to the structural mean distances expected from AV simulations (vertical dashed lines). For **d**, the measured FRET-averaged distance, $\langle R_{DA}\rangle_E$ (from the intensities), and mean interdye distance, $\langle R_{DA}\rangle$ (from the lifetime), are converted into the mean-position distance $R_{\mathrm{mp}}^{\mathrm{FRET}}$ as described in Kalinin et al.[18] (see Methods, Supplementary Tab. 8 and Supplementary Fig. 5).

As the intensity-based and lifetime-based estimates of the interdye distance correspond to different averages over the accessible volume of the fluorophores, we convert both distance measures into a physical distance between mean dye positions $R_{\mathrm{mp}}^{\mathrm{FRET}}$ (Fig. 4d, Supplementary Fig. 5 and Supplementary Tab. 8)[19]. Within error, both methods agree well with the structural predictions based on a B-DNA structure. Notably, we obtained consistent results between single and double biotin samples and for the different surfaces, illustrating that FRET indeed measures the interdye distance independent of the orientation of the molecule.

**Colocalization analysis resolves surface heterogeneities.** Next, we assessed whether distinct immobilization of the single- and double-biotin dsD samples result in characteristic features for localization distances $d_{\mathrm{loc}}$ between donor and acceptor positions that could be resolved by cSTED. The obtained probability densities of $d_{\mathrm{loc}}$ for the BSA surface in Fig. 5a (see Supplementary Fig. 18 for the PEG surface) clearly show that larger distances are observed for the double-biotin immobilization. Theoretically, the distribution of colocalization distances between two fixed emitters follows a $\chi$-distribution whose width depends on the localization precision (see Supplementary Section 'Model-based analysis of localization-based distance distributions', eq. S30)[15,33]. However, all measured distance distributions showed excess broadening as they could not be described by a single component, which suggests a heterogeneous distribution of inclination angles (Supplementary Note 7 and Supplementary Tab. 12). To exclude that this heterogeneity is dynamic (e.g., caused by temporary sticking to the surface), we performed repeated localizations of the same molecule throughout the measurement. No large jumps were observed, and the standard deviation of the localization agreed with the localization precision (Supplementary Note 5 and Supplementary Fig. 22). This indicates a static heterogeneity wherein the DNA molecules experience distinct environments on the surface.

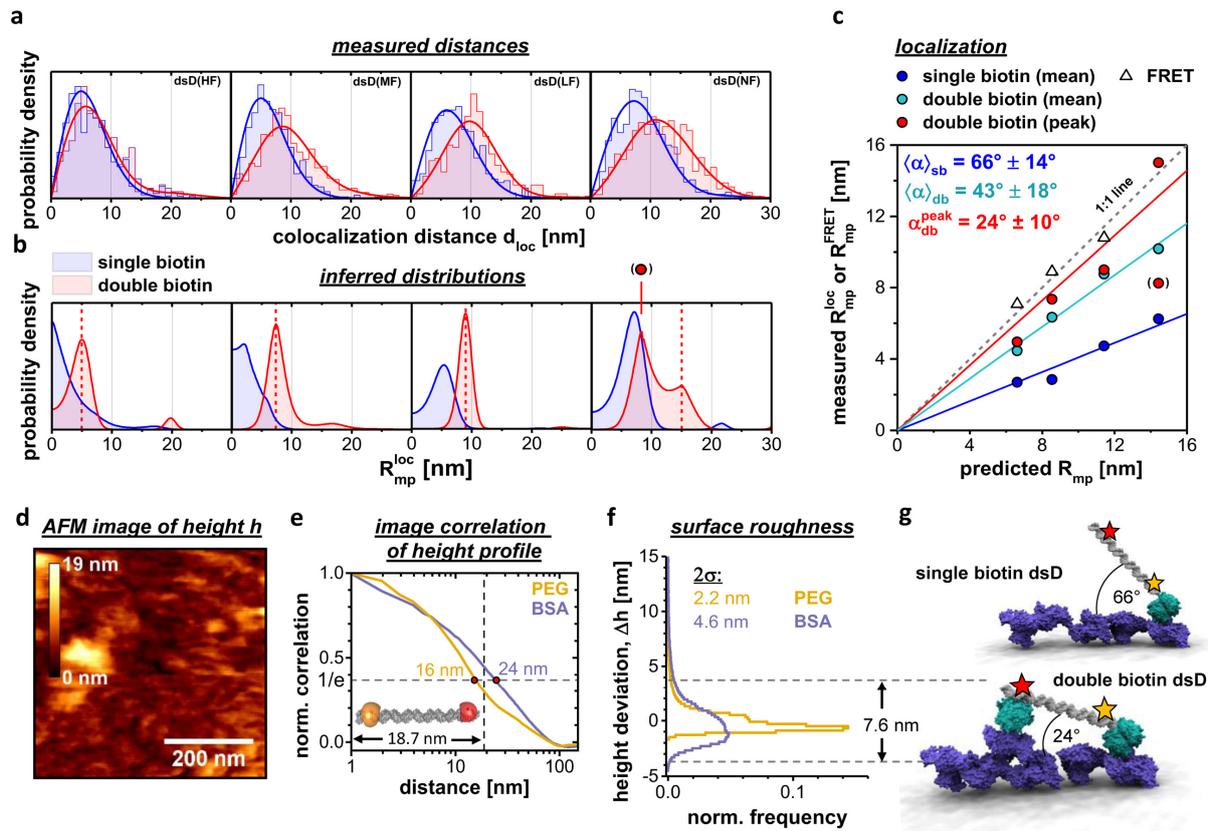

**Figure 5: Colocalization analysis of DNA rulers. a)** Distributions of the measured donor-acceptor distances from localization analysis for the single-biotin (blue) and double-biotin (red) samples (from left to right: dsD(HF), dsD(MF), dsD(LF), dsD(NF) on the BSA surface (see Supplementary Fig. 18 for the PEG surface). Fitted distributions based on maximum entropy analysis in **b** are shown as solid lines. **b)** We used MEM to infer the distribution of the center distance $R_{mp}^{loc}$ using a superposition of $\chi$-distributions with fixed width $\sigma_\chi = 4.4$ nm to describe the measured distance distributions in **a**. The expected distances between the mean positions of the fluorophores based on the AV model are shown as red dashed lines. **c)** Distance-distance plots of the measured mean or peak (maximum) values (red dashed lines in b) of the inferred distance distributions against the predicted distances. Solid lines are linear fits to the data according to eq. S51 (Supplementary Note 9). The slopes of the fits define the inclination angle $\alpha$ which is calculated based on the mean, $\langle\alpha\rangle$, and width of the distance distribution, $\sigma$, or the peak value of the double-biotin population, $\alpha_{db}^{peak}$. For the FRET distances, the four estimates obtained for each sample (intensity- and lifetime-based, for single biotin and double biotin) were converted to $R_{mp}$ and the average is reported (all four individual values were in overall agreement, see Supplementary Tab. 9 and 10). For the mean angles, the error is estimated from the width of the distance distribution. The red circle (in brackets) indicates the second maximum of the dsD(NF) sample originating from DNA bound only with one biotin. **d)** AFM height images of BSA-neutravidin coated surfaces reveal an increased roughness of the molecular surface. The data was recorded using the QI mode and denoised using a 10-pixel median filter. The images for all conditions are shown in Supplementary Fig. 23. **e)** Normalized correlation curves of the height images were computed to provide estimates for the size of the elevated objects (see Supplementary Note 8).

The observed characteristic decay constants are close to the size of our AFM tip of ~15 nm and the size of the DNA ruler. **f)** Normalized histograms of the roughness in the AFM height images for BSA (violet) and PEG (yellow) functionalized cover slides after addition of neutravidin. The roughness was computed by subtracting the average of the measured heights from the individual values. We characterized the histograms by the standard deviation $2\sigma$ (root-mean-square roughness). For comparison we depict the expected height difference between the ends of the double-immobilized DNA based on the inclination angle of $\alpha \approx 24°$ determined in **c**. **g)** Potential molecular cartoons for the orientation of DNA rulers on the molecular surface, where BSA and neutravidin are shown in purple and turquoise, respectively. DNA immobilized by a single anchor resulted a preferential orientation at an angle of ~60-70°. Immobilizing the dsDNA on both ends, a horizontal DNA orientation is expected, but height variations due to surface heterogeneities result in angles of up to ~24° between the attachment points.

To describe the experimental $d_{loc}$ distributions, we employed the maximum entropy method (MEM) that allows us to infer the distribution of the distance between the mean positions of the dyes based on the $\chi$-distribution model function (Supplementary Note 6)[51]. The recovered distributions confirm the shift to larger distances for the double biotin samples (Fig. 5b). Interestingly, the analysis also reveals a visible peak for the single biotin low (dsD(LF)) and no FRET samples (dsD(NF)) at approximately half of the maximum $R_{mp}^{loc}$ distance, implying that DNA is not randomly oriented on the surface. Similarly, the double-biotin sample does not lie flat on the surface as the peak distances are shorter than expected from AV simulations (dashed lines in Fig. 5b). The presence of a second peak at shorter distance for the double-biotin dsD(NF) sample is assigned to a residual population of single-bound molecules, which is also observed for the data measured on the PEG surface (Supplementary Fig. 18). The found fractions of doubly bound dsD correspond well to an estimate of 40 ± 10% based on the density of neutravidin molecules on the surface (Supplementary Note 8 and Supplementary Fig. 24). The consistent results for BSA and PEG surfaces (Supplementary Fig. 18), illustrate the reproducibility of the cSTED approach. Moreover, they suggest that the observed features are due to surface roughness associated with the biotin-neutravidin immobilization rather than specific interactions.

Thus, we applied our Optical Pythagoras procedure by studying the correlation between the mean distances of the inferred distribution for the single-biotin and double-biotin samples and the expected distance (Fig. 5c). While $R_{mp}^{FRET}$ follows the expected 1:1 line (open triangles in Fig. 5c), colocalization-based $R_{mp}^{loc}$ have slopes < 1, as expected for similar distributions of the inclination angle $\alpha$ for the different DNA rulers. From the slopes of the regression lines, we estimate an average inclination angle of 66° ± 14° for the single-biotin sample and 43° ± 18° for the double-biotin sample on the BSA surface, and 58° ± 19° and 44° ± 24° for the PEG surface (Supplementary Note 9). To describe the contribution of completely immobilized molecules in the double-biotin sample, we additionally determined the peak distances belonging to the double-bound fraction, yielding a small inclination angle of 24° ± 10° for the BSA surface and 20° ± 4° for the PEG surface. The corresponding height difference between the ends of the DNA of ~5-7 nm is similar to the molecular dimensions of BSA and neutravidin, indicating that the double-biotin DNA does not assume a perfectly horizontal orientation due to the roughness of the surface on the molecular scale. To test this hypothesis, we performed atomic force microscopy (AFM) of the functionalized surfaces (Fig. 5d and Supplementary Fig. 23) that revealed a heterogeneous height profile on a spatial scale of ~20 nm (Fig. 5e) with a root-mean-square roughness of 2-5 nm (Fig. 5f). It should be noted that most features on the molecular scale are smoothed over because they are smaller than the width of the AFM tip (~15 nm, see Supplementary Fig. 23e). Remarkably, the addition of neutravidin to the slide increased roughness and hardness of the surface in a correlated manner (Supplementary Fig. 23).

Overall, the data obtained by AFM and FRET nanoscopy give a consistent view that the functionalized surfaces are rough on the scale of the DNA rulers. Moreover, FRET nanoscopy provides further detailed insights into the potential interactions of DNA with functionalized surfaces (Fig. 5g). For double-biotin DNA, we did not observe a fully flat configuration. The small angle of ~24° could be explained by the roughness on a molecular scale causing a height difference between the two attachment points (Fig. 5g). For the single-biotin DNA, the absence of a double-biotin like population indicates that there is no sticking of the unbound end. However, instead of assuming a standing-up conformation, the single-biotin DNA showed a preferred orientation of ~60-70° that remained stable over the acquisition time (Fig. 5g). A similar angle of 43 ± 1° was found for double-stranded DNA rulers anchored to a lipid membrane using cholesterol, which was attributed to steric hindrance at the attachment point preventing the ruler from lying down[25]. Similarly, we propose that the preferred orientation of the single-biotin DNA might originate from a preferential orientation of neutravidin on the surface that is propagated to the orientation of the single-biotin DNA via steric constraint at the biotin binding site.

**FRET nanoscopy resolves the conformation hGBP1 in complex assemblies.** Lastly, we applied FRET nanoscopy also to proteins, in particular to the challenging research area of protein oligomers and assemblies, to study their conformational transitions during oligomerization. A highly relevant system is the human guanylate-binding protein 1 (hGBP1) that plays a major role in innate immunity[52,53] and belongs to the dynamin superfamily of large GTPases. hGBP1 has a molecular weight of ~67 kDa. In the monomer state of hGBP1, its three domains assume a compact formation of ~12 nm in length (Fig. 6a). Upon addition of the GTP analogue GDP-AlF$_x$, farnesylated hGBP1 polymerizes into fiber rings with disc-like assemblies of hGBP1 for which Shydlovskyi *et al.*[54] proposed that hGBP1 adopts an extended conformation. To image these assemblies, we randomly labeled hGBP1-wt on the native lysines and cysteines with reactive Alexa594 and Atto647N dyes, respectively (see Supplementary Section 'hGBP1 expression and labeling'). After mixing the randomly labeled hGBP1-wt with unlabeled protein in a ratio of 1:10 and triggering oligomerization, we observed symmetric ring-like structures (Fig. 6b). To further improve the STED resolution, we applied a deconvolution algorithm (Supplementary Fig. 28 and Supplementary Section 'Assessment of hGBP1 fiber diameter'). From the deconvolved images, we estimated a fiber diameter of 73 ± 4 nm, in agreement with previous reports (Fig. 6c)[54].

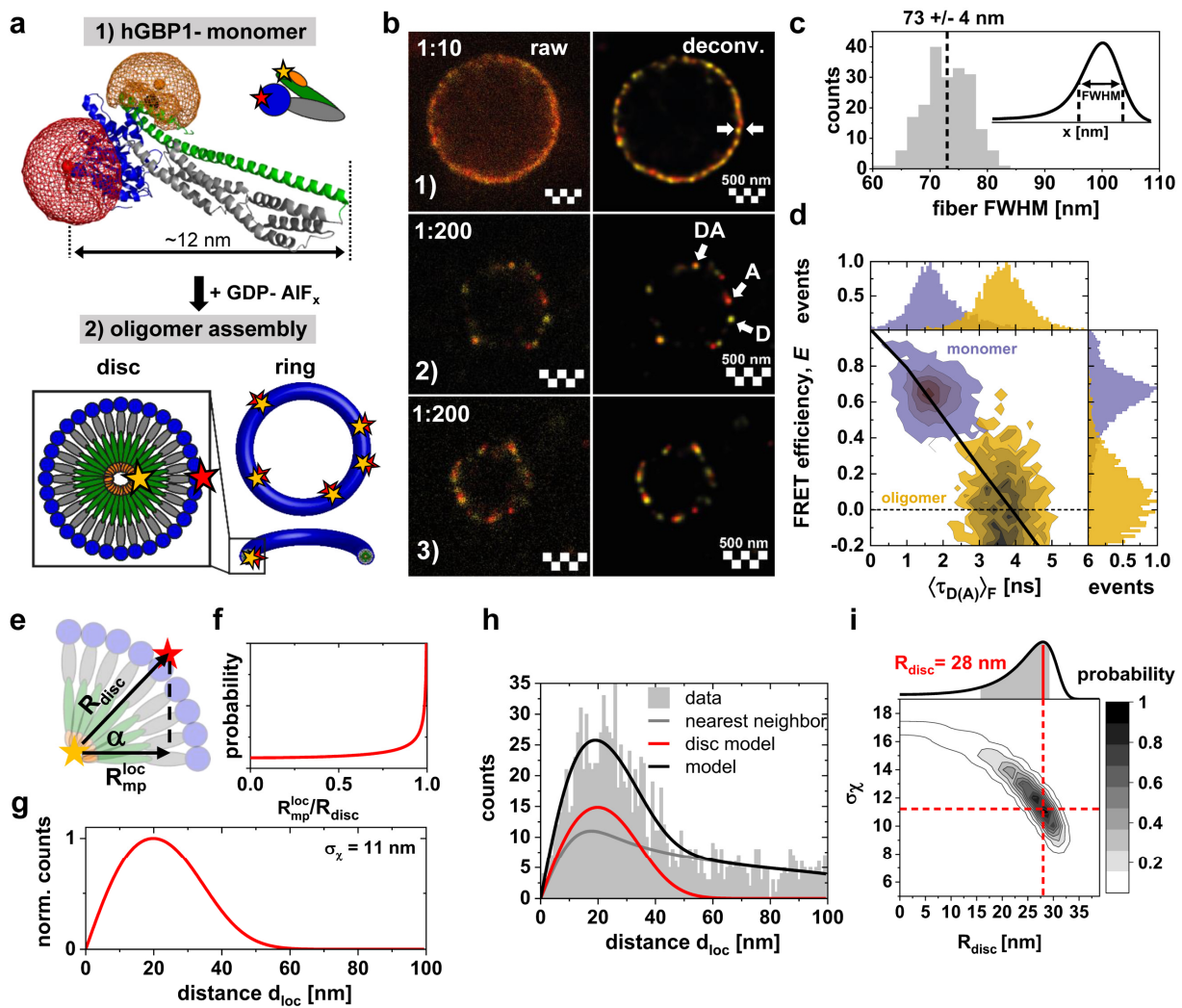

**Figure 6: FRET nanoscopy reveals the conformation of hGBP1 in ring-like fibers. a)** 1) The closed conformation of farnesylated hGBP1 in the Apo state with fluorophores attached at residue 18 (Atto647N, red) and 577 (Alexa594, yellow) is displayed as cartoon of the crystal structure PDB-ID: 6K1Z[53]. The protein is divided into the large GTPase (LG) domain (blue), middle domain (MD, grey), helical GTPase effector domain (GED, green) and the farnesyl moiety (orange). 2) Upon activation with GDP-AlF$_x$, the protein polymerizes into ring-like fibers composed of discs of

hGBP1 in an extended conformation[52,54]. **b)** STED measurements of hGBP1 polymers. Raw STED images are shown to the left and corresponding deconvolved STED images to the right. Single dyes are shown as yellow (Donor D, Alexa594) and red (Acceptor, A, Atto647N) spots. When both dyes colocalize they are shown as an orange spot (DA). 1) Measurement of 1:10 labeled to unlabeled hGBP1-wt. Wild type hGBP1 is randomly labelled with Atto647N on cysteines and Alexa594 on lysines to obtain a high labelling degree without mutation. 2-3) Measurements with 1:200 specifically labeled hGBP1 18-577 to unlabeled hGBP1-wt. Arrows indicate single donor-only (D), acceptor-only (A) and double-labeled (DA) molecules. **c)** Quantification of the fiber width of labeled hGBP1-wt. The full width at half maximum (FWHM) of the fibers was determined from line profile of the deconvolved images (see pair of white arrows in B). **d)** Two-dimensional frequency histograms of the FRET-efficiency, E, and fluorescence-weighted average donor fluorescence lifetime $\langle \tau_{D(A)} \rangle_F$ are shown for the hGBP1 18-577 monomer in purple (obtained from free-diffusion measurements) and for GDP-AlF$_x$ induced oligomers in yellow (obtained from FRET nanoscopy). The static FRET-line is shown in black (see eq. 10 in the methods and Supplementary Table 20). **e)** Imaging in the xy-plane observes the mean-position distance $R_{mp}$ of the projected disc radius $R_{disc}$ for all angles $\alpha$. **f)** The resulting probability distribution function of $R_{mp}$ of the projected distance $R_{disc}$ for the disc-like arrangement within the fibers. **g)** The distribution of the interdye distance $d_{loc}$ experimentally observed by cSTED is related to $R_{mp}$ (Supplementary Note 10, eq. S57), which is broadened by the localization uncertainty and sample heterogeneities (Supplementary Section 'Colocalization analysis', eq. S30). We display the disc model distribution of $d_{loc}$ for a disc radius $R_{disc}$ of 28 nm and a standard deviation of $\sigma_\chi$ of 11 nm. **h)** Experimental nearest-neighbor distances $d_{loc}$ between donor and acceptor fluorophores in the fiber structures at 1:200 dilution (gray histogram). The overall model function (black) consists of the disc model shown in g and a baseline due to false-positive pairing in the nearest-neighbor algorithm due to randomly distributed donor- and acceptor-only molecules throughout the ring (gray line) (Supplementary Note 10, eq. S57). **i)** Parameter scan of the model $R_{disc}$ and $\sigma_\chi$ fit parameters. The marginal probability density of the disc radius $R_{disc}$ is shown at the top. The 68% confidence interval is indicated as grey shaded area.

To gain insight into the structure of the assemblies and the conformation of hGBP1 in its oligomeric state, we site-specifically labeled hGBP1 on opposite ends of the protein at residues 18 and 577 (Fig. 6a). We reduced the fraction of labeled hGBP1 to 1:200 to localize single molecules in the images (Fig. 6b) and apply our FRET nanoscopy workflow. In contrast to the high FRET efficiency observed for the monomer from free-diffusion single-molecule FRET experiments, we observe a FRET efficiency of zero for hGBP1 oligomers that clearly reveals an extended conformation of hGBP1 (Fig. 6d). As the interdye distance in the extended state is outside the FRET range, we can specify a lower boundary of ~12 nm and an upper boundary set by the dimensions of the extended protein of ~30 nm. To resolve the interdye distance in this range more accurately, we applied our established cSTED approach. As the colocalization measures the projected distance $d_{loc}$, it is necessary to consider the 3D orientation of the protein in the disc-like assemblies (Fig. 6e). By assuming a uniform distribution of the inclination angle $\alpha$, the distribution of the projected mean-position distance $R_{\mathrm{mp}}^{\mathrm{loc}}$ is described by $P\left(R_{\mathrm{mp}}^{\mathrm{loc}}\right) = \frac{2}{\pi}\left(1 - \left(\frac{R_{\mathrm{mp}}^{\mathrm{loc}}}{R_{\mathrm{disc}}}\right)^2\right)^{-\frac{1}{2}}$, where $R_{\mathrm{disc}}$ is the disc radius (Fig. 6f and Supplementary Note 10). The experimentally observed localization distance $d_{loc}$ is given by a non-central $\chi$-distribution distribution of $R_{\mathrm{mp}}^{\mathrm{loc}}$ that is broadened due to the localization uncertainty defined by $\sigma_\chi$ (Supplementary eq. 30, Fig. 6g). The experimental distance distributions are well described by this model (Fig. 6h), yielding $R_{\mathrm{disc}}$ = 28 nm (68% confidence interval: 16–29 nm, Fig. 6i) with $\sigma_\chi$ ~11 nm. These results clearly support the model that oligomerized hGBP1 adopts an extended conformation as suggested by the absence of FRET.

**Summary/Outlook**

In summary, we demonstrated that FRET nanoscopy provides seamless resolution from micrometers to the sub-nanometer range, with an accuracy of < 5 Å. Moreover, the combination of the localization and FRET information provides 3D information by the Optical Pythagoras and the multiparametric fluorescence analysis, as demonstrated for the origami and dsDNA samples. The resolution is also sufficient to resolve 3D orientation of single molecules in heterogeneous environments, such as functionalized surfaces or protein fibers.

While we obtained sufficient signal for single molecules to define the spectroscopic parameters and achieve a high localization precision of ~3 nm, the observation time for individual molecules is fundamentally limited by photobleaching. A promising approach that would allow the prolonged and repeated probing of a single molecule is the combination of DNA-PAINT with STED microscopy, utilizing the repeated hybridization of short labeled oligos to complementary sequences on the molecule of interest[55]. Different to other SMLM techniques, our approach is fast (< 10 s imaging time) and instead relies on the consecutive imaging of many ROIs. Hence, it does not necessitate the fixation of the specimen and is potentially applicable to the study of transient assemblies in living cells. The high time resolution of the confocal time-resolved detection[26,27] can further be utilized to resolve also fast molecular dynamics in a wide time range from nanoseconds to seconds by the workflow of multiparameter fluorescence imaging spectroscopy[28,56].

FRET nanoscopy is readily applicable to any biological system wherein molecules have a defined orientation. Further promising applications could be the study of the structural dynamics of membrane proteins, such as transmembrane receptors or transporters. We further envision that the combination of colocalization STED microscopy with particle alignment and averaging will provide novel insights into the nanoscale organization of higher-order, symmetric biological assemblies, such as the nuclear pore complex, centrosome, or chromatin fibers. FRET nanoscopy is ideally suited to determine the precise location of individual components within such systems, while simultaneously probing their conformation using the FRET information. Finally, we note that FRET nanoscopy can be implemented on any STED microscope with two color and time-resolved detection, which are available commercially and in microscope facilities around the world.

**Methods**

*STED-FRET microscope*

FRET-nanoscopy imaging was performed on a custom Abberior Instruments Expert Line system with polarization-sensitive readout and single-photon counting abilities. Briefly, linear polarized excitation lasers (561 nm and 635 nm, pulse width < 100 ps) and a circular polarized depletion laser (775 nm, pulse width 1.2 ns) are synchronized and overlaid using notch filters and focused by a 100x oil objective (Supplementary Figure 1). The fluorescence signal is separated by polarization and color and detected using avalanche photodiodes and single photon counting electronics. Alternating line excitation was used to sequentially excite the donor and acceptor fluorophores. More details are given in the Supplementary Section 'Data acquisition – STED-FRET microscope'.

*Sample preparation*

Single-layer DNA origami nanostructures are based on published designs and were assembled as described in Schnitzbauer et al.[38,39]. Donor and acceptor fluorophores (Atto594, Atto647N) were attached to the 3'-end of the respective oligos with an additional thymine base pair as a spacer. All DNA strands are listed in Supplementary Table 1 and the origami design is illustrated in Supplementary Figure 7. Sample was immobilized on a PEG-biotin coated surface and measured in folding buffer containing 10 mM Tris, 1 mM EDTA, 12.5 mM $MgCl_2$ and 1 mM Trolox at pH 8.0. More details are given in the Supplementary Section 'Sample preparation – DNA origami'.

For dsDNA rulers, the position of the donor dye (Alexa594) was kept fixed while the acceptor (Atto647N) position on the complementary strands was varied to achieve several interdye distances. The donor strands were labelled with a single biotin (sb) or a double Biotin (db) at the 3' and 5' ends for surface immobilization. Samples were immobilized on either BSA-biotin or PEG-biotin surfaces and measured in PBS buffer containing 1 mM Trolox at pH 7.6. The DNA sequences are listed in Supplementary Table 1. More details are given in the Supplementary Section 'Sample preparation – dsDNA'.

Wild-type hGBP1 was labeled with Alexa594 succinimidyl ester and Atto647N maleimide at the native lysines and cysteines, respectively. For site-specific labeling of hGBP1, first all native cysteine residues were mutated (i.e., C12A, C82A, C225S, C235A, C270A, C311S, C396A, C407S, C589S), as used before[57]. Reversal of the C-terminal C589S mutation allowed for site-specific farnesylation, and the residues N18 and Q577 were mutated to cysteines for fluorophore labeling. After farnesylation, hGBP1 was labeled with Alexa594 maleimide, purified by anion exchange chromatography and then labeled with Atto647N maleimide. Oligomerization was induced by the addition of 200 µM GDP in the presence of AlF$_x$ in buffer containing 50 mM Tris-HCl, 150 mM NaCl, 5 mM MgCl$_2$, 300 µM AlCl$_3$ and 10 mM NaF at pH 7.9. Polymer structures were visible after 10-40 min after which they would remain stable for hours. The ratio of labelled Cys8 hGBP1-18-577 (or labelled hGBP1-wt) to unlabeled wild-type hGBP1 was adjusted depending on the desired labelling density in hGBP1 polymers. More information is given in the Supplementary Section 'Sample preparation – hGBP1 expression and labeling'.

*Image Spectroscopy.*

**Intensity-based spectroscopic parameters.** Correction factors accounting for the spectral crosstalk of the donor fluorescence into the acceptor detection channel, the direct excitation of the acceptor by the donor excitation laser and the different detection efficiencies for the donor and acceptor were determined following the approach outlined in Lee et al.[36] and described in the Supplementary Section 'Determination of intensity-based correction factors'. Quantum yields $\Phi_F$ of donor and acceptor were estimated from the fluorescence lifetime by comparison to a known reference. Correction of the detected signals was performed as described in the Supplementary Section 'Intensity-based spectroscopic parameters'.

To estimate accurate intensity-based FRET efficiencies under STED conditions, we used the total intensity detected in the donor and acceptor channels without any time gating. The ungated signal contains contributions from both (partially) depleted and undepleted molecules, which exhibit different quantum yields. Donor and acceptor fluorophores are also generally depleted to a different extent. We assume the FRET efficiency of the depleted molecules to be zero due to the increased donor de-excitation rate through the depletion pulse. Under these assumptions, the accurate, distance-related FRET efficiency $E$ is given by:

$$E = E' \cdot \left(1 + \frac{x_D^d \cdot \Phi_{F,D}^d}{x_D^0 \cdot \Phi_{F,D}^0}\right), \tag{3}$$

where $x_D^d$ and $x_D^0$ are the fractions of depleted and undepleted donor fluorophores and $\Phi_{F,D}^d$ and $\Phi_{F,D}^0$ are the respective quantum yields. The modified FRET efficiency $E'$ is related to the measured fluorescence signals as:

$$E' = \frac{F_{A|D}}{\gamma' \cdot F_{D|D} + F_{A|D}}, \tag{4}$$

where $F_{D|D}$ and $F_{A|D}$ are the signals detected in the donor and acceptor channels after donor excitation and $\gamma'$ is a factor that corrects for the relative detection yield if the donor and acceptor fluorophores and taking the fraction of undepleted molecules into account. See Supplementary Section 'Accurate intensity-based FRET efficiencies under STED conditions' for details and a general expression for non-zero FRET efficiency of the depleted species. The quantum yields and fractions under STED conditions are obtained from bi-exponential lifetime fits by assuming that the short lifetime component of ~200-400 ps originates from depleted fluorophores (see Supplementary Table 9). Error estimation was performed by propagating the uncertainty of all correction factors as described in Hellenkamp et al.[20]. The correction parameters of all measurements are compiled in Supplementary Table 3.

**Sub-ensemble fluorescence decay analysis.** To minimize polarization effects, the total fluorescence decay was approximated from the detected parallel and perpendicular signals (see Supplementary Section 'Sub-ensemble fluorescence decay analysis' eq. S15). We fit the fluorescence decay of the donor-only sample, $f_{D|D}^{(D0)}(t)$, using a bi-exponential model to account for non-depleted and partially depleted molecules given by:

$$f_{D|D}^{(D0)}(t) = x_{STED}^{(D0)} \cdot e^{-t/\tau_{D0,STED}} + \left(1 - x_{STED}^{(D0)}\right) \cdot e^{-t/\tau_{D0}}, \tag{5}$$

where $t$ is the TCSPC delay time, $x_{STED}^{(D0)}$ is the apparent fraction of undepleted donor fluorophores and $\tau_{D0}$ and $\tau_{D0,STED}$ are the donor-only lifetimes of the undepleted and depleted donor fluorophore.

The FRET-induced donor fluorescence decay, $f_{D|D}^{(DA)}(t)$, is described by a normal distribution of the interdye distance, $R_{DA}$, arising due to the flexible dye linkers[34], with the addition of a donor-only fraction to account for bleaching:

$$f_{\text{D|D}}^{(\text{DA})}(t) = x_{\text{STED}}^{(\text{DA})} \cdot e^{-t/\tau_{\text{D,STED}}} + \left(1 - x_{\text{STED}}^{(\text{DA})}\right) \cdot$$

$$\left( (1 - x_{\text{D0}}) \cdot \int_0^\infty x_{\text{FRET}}(R_{\text{DA}}) \cdot e^{-\left(\tau_{\text{D0}}^{-1} + k_{\text{FRET}}(R_{\text{DA}})\right)t} dR_{\text{DA}} + x_{\text{D0}} \cdot e^{-t/\tau_{\text{D0}}} \right), \qquad (6)$$

where $x_{\text{STED}}^{(\text{DA})}$ and $\tau_{\text{D,STED}}$ are the apparent fraction and lifetime of the depleted donor fluorophore in the presence of the acceptor, respectively. $x_{\text{D0}}$ is the donor-only fraction and $x_{\text{FRET}}(R_{\text{DA}})$ is the fraction of molecules with interdye distance $R_{\text{DA}}$, corresponding to a FRET rate constant of $k_{\text{FRET}}(R_{\text{DA}}) = \frac{1}{\tau_{\text{D(0)}}} \cdot \left(\frac{R_0}{R_{\text{DA}}}\right)^6$. The distance distribution $x_{\text{FRET}}(R_{\text{DA}})$ is given by a normal distribution centered at the mean interdye distance $\langle R_{\text{DA}} \rangle$ with width $\sigma_{\text{DA}}$:

$$x_{\text{FRET}}(R_{\text{DA}}) = \frac{1}{\sqrt{2\pi}\sigma_{\text{DA}}} \cdot \exp\left(-\frac{(R_{\text{DA}} - \langle R_{\text{DA}} \rangle)^2}{2\sigma_{\text{DA}}^2}\right) \qquad (7)$$

Donor-only and FRET-sensitized decays are analyzed globally to improve the robustness of the fit. Error estimates are obtained from an analysis of the reduced $\chi^2$ surface. For details, see Supplementary Sections 'Sub-ensemble fluorescence decay analysis' and 'Error estimation of FRET-derived distances'.

**Spot-integrated fluorescence lifetimes.** To estimate accurate spot-integrated fluorescence lifetimes of the undepleted fluorophores, we fitted the tail of the spot-integrated fluorescence decay $f(t)$ to a single-exponential model according to:

$$f(t) = f_0 e^{-t/\tau} + bg \qquad (8)$$

where $f_0$ is the initial amplitude of the decay and $bg$ is a constant background term. To accurately account for the polarized detection, a global fit of the parallel and perpendicular decays is performed that incorporates the depolarization of the fluorescence signal due to molecular rotation, polarization mixing by the objective and the polarization of the scattered background signal (Supplementary Section 'Determination of spot-integrated fluorescence lifetimes'). The optimization is performed using a maximum likelihood estimator that correctly accounts for the counting statistics by minimizing the $2I^*$ parameter defined by:[35]

$$2I^* = 2\sum_i n_i \ln\frac{n_i}{f_i}, \qquad (9)$$

where $n_i$ is the number of photons detected in TCSPC bin $i$ and $f_i$ is the average value of the model function in TCSPC bin $i$.

**Linker-corrected relationship between the FRET efficiency and donor fluorescence lifetime.**
The static FRET-line as given in eq. 2 describes the ideal relation between the fluorescence-weighted average FRET efficiency $E$ and the fluorescence-weighted fluorescence donor lifetime $\langle \tau_{D(A)} \rangle_F$ in the absence of dynamics. Deviations from eq. 2 occur due to the flexibility of the dye linker, causing an apparent distance distribution width of 6 Å as determined from reference measurements of freely-diffusing DNA rulers[34]. Linker-corrected static FRET lines were approximated using fourth-order polynomials as described previously[34]:

$$E = 1 - \frac{\sum_{i=0}^{4} a_i \langle \tau_{D(A)} \rangle_F^i}{\tau_{D(0)}}, \tag{10}$$

where $\tau_{D(0)}$ is the fluorescence lifetime of the donor in the absence of the acceptor and $a_i$ are the polynomial coefficients. The polynomial coefficients used in this work are compiled in Supplementary Table 20.

**Accessible volume (AV) simulations and distance conversion.** The sterically accessible volumes (AVs) of the fluorophores were determined using the *FRET positioning and screening (FPS)* software[19]. From this the distance between the mean positions $R_{mp}$, the average distance $\langle R_{DA} \rangle$, the average FRET efficiency $\langle E \rangle$ and the FRET-averaged distance $R_{\langle E \rangle}$ are obtained. See Supplementary Section 'Accessible volume simulations' for details and Supplementary Table 6 for the used parameters. Conversion of the measured distances $\langle R_{DA} \rangle$ (from the lifetime information) and $R_{\langle E \rangle}$ (from the fluorescence intensities) into the mean-position distance $R_{mp}$ was performed as described previously[19]. Briefly, AVs for the donor and acceptor fluorophores were calculated for dsDNA. By randomly orienting the AVs and calculating the parameters $R_{mp}$, $\langle R_{DA} \rangle$ and $R_{\langle E \rangle}$, conversion functions based on a polynomial approximation are obtained (see Supplementary Table 8 and Supplementary Figure 5).

*Image processing in FRET nanoscopy*

**Spot localization.** The point spread function (PSF) of the STED microscope is modelled as a 2D Gaussian function. To account for multiple spots in origami and hGBP1 data, up to three 2D Gaussian functions were fitted. The model function where $G(x, y)$ at pixel $(x, y)$ is given by:

$$G(x, y) = \sum_{i=1}^{3} A_i \cdot \exp\left[\frac{(x - x_{0,i})^2 + (y - y_{0,i})^2}{2\sigma_{PSF,i}^2}\right] + bg, \tag{11}$$

where $A_i$ is the amplitude, $x_{0,i}$ and $y_{0,i}$ are the center coordinates and $\sigma_{\mathrm{PSF},i}$ is the width of spot $i$, and $bg$ is a constant background term. To correctly account for the Poisson statistics of photon detection, spot localization is performed using a maximum likelihood estimator by maximizing the *2I\** parameter given by:

$$2I^* = -2\sum_i n_i - G_i + n_i \ln \frac{G_i}{n_i}, \tag{12}$$

where $n_i$ and $G_i$ are the number of photons and the value of the model function in pixel $i$, respectively (see Supplementary Section 'Colocalization analysis - Spot localization' for details). To estimate the number of spots in a ROI, the fit with the lowest *2I\** was chosen while adding a constant penalty for additional free parameters (Supplementary Section 'Colocalization analysis - Spot stoichiometry'). The theoretical localization precision was predicted based on the photon counting statistics as described in Mortensen et al.[58] (see Supplementary Section 'Colocalization analysis - Predicted localization precision').

**Data filtering.** Data filtering was performed based on imaging parameters (e.g., the spot width and the number of spots), as well as spectroscopic parameters (e.g., the donor fluorescence lifetime, FRET efficiency and stoichiometry). Origami data was filtered based on the number of spots detected in the donor and acceptor channels (spot stoichiometry) and a minimal number of photons per spot of 20 (see Supplementary Table 14 and Supplementary Section 'Spectroscopic analysis - Filtering procedure for origamis'). dsDNA data was filtered by selecting spots with a width of less than 35 nm and am apparent stoichiometry, $S_{\mathrm{app}}$, of 0.5 ± 0.1, defined by:

$$S_{\mathrm{app}} = \frac{I_{\mathrm{D|D}} + I_{\mathrm{A|D}}}{I_{\mathrm{D|D}} + I_{\mathrm{A|D}} + I_{\mathrm{A|A}}}, \tag{13}$$

where $I_{\mathrm{D|D}}$ and $I_{\mathrm{A|D}}$ are the detected signal in the donor and acceptor detection channels after donor excitation and $I_{\mathrm{A|A}}$ is the detected signal in the acceptor detection channel after acceptor excitation. For more details, see Supplementary Table 15 and Supplementary Section 'Filtering procedure for DNA rulers'.

**Alignment of DNA origami nanostructures.** Particle averaging on origami sample was done in three steps. Firstly, only fully labelled structures were selected. Second, each dye was characterized as being either D1, D2, A1 or A2 as defined in Figure 2b. This is possible because all structures are bound to the surface with the same side due to the placement of the biotin anchors. Third, the Kabsch algorithm was used to align the structures with respect to a reference structure by translation and rotation by minimizing the root-mean-square-displacement (RMSD):[40]

$$RMSD = \sqrt{\frac{1}{4} \sum_{j \in \{D1,D2,A1,A2\}} \left(x_j - x_{\text{ref},j}\right)^2 + \left(y_j - y_{\text{ref},j}\right)^2} , \tag{14}$$

where $x_j$ and $y_j$ are the x- and y-coordinates of fluorophore $j$ and the subscript 'ref' refers to the reference structure. A model-free reference structure was obtained by selecting the structure from the experimental dataset, which provided the lowest RMSD over all structures. Outliers in the RMSD score were removed as shown in Supplementary Figure S9. A complete description of the alignment procedure is given in the Supplementary Section 'Colocalization analysis - Alignment and particle averaging for origami measurements'.

**Colocalization analysis.** The uncertainty of the localization estimation is described by a normal distribution whose width is determined by the localization precision. The resulting distribution of the localization-based interdye distances $d_{\text{loc}}$, however, is not normally distributed and assumes an asymmetric form in the case that the interdye distance $R_{\text{mp}}^{\text{loc}}$ is comparable to the localization precision, which is the case in the cSTED analysis. It is given by a noncentral $\chi$-distribution $P_\chi$ with two degrees of freedom, as has been described previously:[15,33]

$$P_\chi\left(d_{\text{loc}}|R_{\text{mp}}^{\text{loc}}, \sigma_\chi\right) = \left(\frac{d_{\text{loc}}}{\sigma_\chi^2}\right) \cdot \exp\left(-\frac{d_{\text{loc}}^2 + R_{\text{mp}}^{\text{loc}}{}^2}{2\sigma_\chi^2}\right) \cdot I_0\left(\frac{d_{\text{loc}}R_{\text{mp}}^{\text{loc}}}{\sigma_\chi^2}\right) , \tag{15}$$

where $d_{\text{loc}}$ is the measured colocalization distance, $R_{\text{mp}}^{\text{loc}}$ is the mean-position distance, $\sigma_\chi$ is a width parameter and $I_0(x)$ is the modified Bessel function of zero-th order (for details, see Supplementary Note 7). The width parameter of the distribution $\sigma_\chi$ depends on the localization precisions for the donor and acceptor fluorophores ($\sigma_{\text{loc,D}}$ and $\sigma_{\text{loc,A}}$) and additional registration error ($\sigma_{\text{reg}}$):

$$\sigma_\chi = \sqrt{\sigma_{\text{loc,D}}^2 + \sigma_{\text{loc,A}}^2 + \sigma_{\text{reg}}^2} \tag{16}$$

Here, we estimate a typical width of $\sigma_\chi = 4.4$ nm based on the defined distribution of the acceptor-acceptor distance obtained for the DNA origami nanostructures (Supplementary Figure 10), which was fixed for the analysis of the distance distributions for the dsDNA rulers.

The maximum entropy method (MEM) is an approach to extract the most unbiased distribution of a given parameter that provides a satisfactory fit to the experimental data[59-61]. Instead of minimizing the reduced chi-square, $\chi_r^2$, the following functional is maximized:

$$\Theta = vS - \chi_r^2 \tag{17}$$

where $v$ is a constant scaling factor and $S$ is the entropy functional of the parameter distribution. Without prior knowledge, the entropy $S$ of a discrete probability distribution $p_i$ is defined by:

$$S = -\sum_i p_i \log p_i,\qquad(18)$$

We subject the MEM analysis to the mean-position distance $R_{\mathrm{mp}}^{\mathrm{loc}}$ to extract the distribution $p\left(R_{mp}^{\mathrm{loc},(i)}\right)$ The experimental distribution of colocalization distances $H(d_{\mathrm{loc}})$ is described as superposition of non-central -distributions $P_\chi$ as defined in eq. 15 by:

$$H(d_{\mathrm{loc}}) = \sum_i p\left(R_{\mathrm{mp}}^{\mathrm{loc},(i)}\right) P_\chi\left(d_{\mathrm{loc}}\middle|R_{\mathrm{mp}}^{\mathrm{loc},(i)}, \sigma_\chi\right)\qquad(19)$$

where the set of kernel functions $\left\{P_\chi\left(d_{\mathrm{loc}}\middle|R_{\mathrm{mp}}^{\mathrm{loc},(i)}, \sigma_\chi\right),\ i = 1 \dots N\right\}$ is defined over a range of mean-position distance $R_{\mathrm{mp}}^{\mathrm{loc}}$ from 0 to 30 nm. The reduced chi-squared $\chi_r^2$ is defined as:

$$\chi_r^2 = \frac{1}{K}\sum_k \frac{1}{w_k^2}\left(H\left(d_{\mathrm{loc}}^{(k)}\right) - M\left(d_{\mathrm{loc}}^{(k)}\right)\right)^2\qquad(20)$$

where $M$ is the measured histogram, $K$ is the number of bins on the histogram and $w_k$ are the weights of data points given by $w_k = \sqrt{M\left(d_{\mathrm{loc}}^{(k)}\right)}$ for Poissonian counting statistics. Maximization of Θ is performed as described in Vinogradov and Wilson[51]. See Supplementary Note 6 for details.

*Processing of STED images*

hGBP1 images were deconvolved and analyzed in the Huygens software (Scientific Volume Imaging) using the CMLE algorithm with a signal to noise ratio of 10 and 40 iterations (see Supplementary Section 'Assessment of hGBP1 fiber diameter').

*TEM imaging of DNA origami nanostructures*

Origami samples for transmission electron microscopy were negatively staining using uranyl acetate and imaged using a JEM-2100PLUS instrument (JEOL) at 80 kV acceleration voltage (see Supplementary Section 'Transmission electron microscopy').

*AFM imaging of functionalized surfaces*

AFM images were recorded using a soft cantilever (k = 0.04 N/m) in the QI mode on a NanoWizard 4 instrument (JPK Instruments). See Supplementary Section 'Atomic force microscopy' for details.

*Confocal single-molecule FRET measurements of freely diffusion molecules*

Confocal single-molecule FRET measurements were performed on a two-color multiparameter fluorescence detection setup with pulsed-interleaved excitation as described previously[62]. See Supplementary Section 'Confocal single-molecule measurements' for details.

## Acknowledgements

This work was funded in part by the German Research Foundation (DFG) within the Collaborative Research Center SFB 1208 "Identity and Dynamics of Membrane Systems - From Molecules to Cellular Functions" (T.P. A08 to C.S.) and within the Priority Program SPP2191 "Molecular mechanisms of functional phase separation" (project SE 1195/21-1 to C.S.) as well as by the European Research Council through the Advanced Grant 2014 hybridFRET (number 671208) to C.S. The studies of hGBP1 were supported by DFG via projects HE 2679/6-1 to C.H. and SE 1195/17-1 to C.S. We acknowledge the DFG and the state NRW for establishing two microscopes within program Major Research Instrumentation as per Art. 91b GG that were essential for this study: (i) the FRET nanoscope with the funding ID (INST 208/741-1 FUGG) to C.S., and (ii) the cryo-TEM with the f unding ID (INST 208/749-1 FUGG) to M.Ka.. J.B. and J.F. acknowledge the support of the International Helmholtz Research School of Biophysics and Soft Matter (IHRS BioSoft). We thank Oleg Opanasyuk for help with theoretical modeling of distance distributions, and Laura Vogel and Michelle Rademacher for performing the sample preparation for AFM imaging. We thank Costanza Girardi (Dipartimento di Fisica "G.Occhialini", Università di Milano Bicocca, Italy) for helping to test and establish measurement conditions and analysis procedures for dsDNA rulers with our STED nanoscope during her research stay for her master thesis funded within the Erasmus program.

## Author contributions

C.S., A.B. and C.H. designed research. C.S., R.K., S.F. and A.S. designed the multiparameter STED microscope. M.Kö. and A.S. realized the hardware and software and assisted in operation of multiparameter STED microscope. J.B., N.V. R.K. and S.F established the operation and measurement procedures for the STED nanoscope. N.V. performed and analyzed measurements of DNA origami nanostructures. A.B. designed and assembled the DNA origami nanostructures. J.B. and C.G. performed and analyzed experiments on DNA rulers. P.L. and J.B. performed and analyzed experiments on hGBP1. A.B., N.V. and J.B. contributed analytical tools and oversaw data analysis of experiments on DNA rulers and hGBP1. S.F. and N.V. developed software for molecule identification and localization. J.F. performed single-molecule FRET control experiments of DNA origami nanostructures and hGBP1. M.O. and M.Ka. performed TEM imaging of DNA origami nanostructures and analyzed TEM images. J.S. and M.Ka. performed AFM experiments of functionalized surfaces and analyzed AFM images. A.B., C.S., N.V. and J.B. discussed and prepared figures. A.B. wrote the initial draft of the manuscript with extensive input from N.V. and J.B. and C.S. defined and expanded the initial draft, contributed extensively to the writing and revising of the manuscript. All authors contributed to and approved the final version of the manuscript. C.S. initiated, supervised and coordinated the project.

**Competing interests**

The authors declare no competing interests.

# Supporting Information

## FRET nanoscopy enables seamless imaging of molecular assemblies with sub-nanometer resolution

Jan-Hendrik Budde[1,†], Nicolaas van der Voort[1,†], Suren Felekyan[1], Julian Folz[1], Ralf Kühnemuth[1], Paul Lauterjung[1,2], Markus Köhler[3], Andreas Schönle[3], Julian Sindram[4], Marius Otten[4], Matthias Karg[4], Christian Herrmann[2], Anders Barth[1,*], Claus A. M. Seidel[1,*]

[1] Chair for Molecular Physical Chemistry, Heinrich-Heine-University Düsseldorf, Germany

[2] Physical Chemistry I, Ruhr-Universität Bochum, Germany

[3] Abberior Instruments GmbH, Hans-Adolf-Krebs-Weg 1, 37077 Göttingen, Germany

[4] Lehrstuhl für Kolloide und Nanooptik, Heinrich-Heine-Universität Düsseldorf, 40225 Düsseldorf, Germany

[†] contributed equally

[*] Correspondence to: cseidel@hhu.de, anders.barth@hhu.de

## Table of contents

# Materials and Methods

## Data acquisition

### *STED-FRET microscope*

Super-resolved STED-FRET imaging was performed on a custom-designed Abberior Instruments Expert Line microscope (Abberior Instruments, Göttingen, Germany) using an Olympus IX83 microscope body equipped with an *easy 3D* module based on a spatial light modulator (SLM), that is sketched in Supplementary Fig. 1a. The excitation lasers with wavelengths (561 nm and 640 nm, pulse width < 100 ps) and the STED depletion laser (775nm, pulse width 1.2 ns) were synchronized at 40 MHz and spatially overlaid by notch filters N (Notches: N1 for 561 nm excitation: NF03-594E (Semrock) with tuned angle; N2 for 640 nm excitation: NF03-658E (Semrock) with tuned angle and N3 for 775 nm STED depletion (Abberior), for alternative wavelengths excitation not used here: N1' for 488 nm excitation: NF03-514E (Semrock) with tuned angle; N2' for 518 nm excitation: NF01-532U (Semrock) with tuned angle). Extension of excitation beam- and detection path by quarter wave plates guarantees a linear polarization of excitation laser, while the STED- profile remains circular in the focal plane. The fluorescence signal was split by polarization and color using a broad band polarizing beam splitter (PBS, Abberior) and two dichroic mirrors at 640 nm (DM1 & DM2, Abberior) and detected by four APDs (Excelitas, SPCM-AQRH-13-TR). Further spectral selection in the donor and acceptor channels was achieved by dye-specific band pass filters. For donor fluorescence (BP2 & BP4): (i) dsDNA-Alexa594 and origami-Atto594 (Semrock ET615/20) and hGBP1-Alexa594 (AHF, Germany, ET608/45). For acceptor fluorescence (BP1 & BP3, Abberior ET685/70M) of Atto647N for all samples. The detected signal was split into its parallel and perpendicular components using a polarizing beam splitter. Finally, photon arrival times were recorded with picosecond resolution on an additional external time-correlated single photon counting (TCSPC) unit (Hydra Harp 400, PicoQuant GmbH, Berlin, Germany). All measurements were performed with an oil-immersion objective (NA 1.4, UPLSAPO 100XO, Olympus, Germany). The point-spread function was measured using 150 nm gold nanoparticles (Sigma-Aldrich, St. Louis, US, see Supplementary Fig. 1b,c), while the achievable STED resolution was determined using dye-filled polystyrene beads (crimson beads; 40 nm, Thermo Fisher Scientific, Waltham, US).

To further increase the achievable resolution, time-gated fluorescence detection was performed during data processing using a microtime window from 0.9 -12.8 ns relative to the STED depletion pulse ) Time gating resulted in an increase of the STED resolution by up to 20% (see Supplementary Fig. 1d). While the instrument is capable of achieving a higher resolution, this comes at the cost of fewer photons available for FRET analysis. The resolution shown here represents a good tradeoff. The spectral crosstalk α of the donor fluorescence into the acceptor channels ranges between 0.49 and 0.56 (see Supplementary Tab. 3 and Supplementary Fig. 1e). To avoid crosstalk in the acceptor channel and to achieve alternating excitation of the donor and acceptor fluorophores, we performed line-interleaved donor and acceptor excitation. Because the movement in the y-direction of the scanner is constant over time, the acceptor emission upon acceptor excitation channel is shifted by half a pixel, or 5 nm in the y direction. All acceptor localizations are corrected for this shift. No other corrections to the localization data are applied. To minimize the effects of spherical aberrations and drift, data was recorded close to the optical axis (< 20 µm) in a small ROI (1 µm x 1 µm) for fast image acquisition. The instrument response function (IRF), approximating the timing response of the system to an infinitely short lived fluorophore, was determined under following conditions: (i) excitation at 561 nm: aqueous solution of erythrosine in 5 M potassium iodide as quencher[1], and (ii) excitation at 640 nm: aqueous solution of malachite green[2](see Supplementary Fig. 1f).

## *Data recording*

To automate the acquisition of zoomed STED images of single molecules as depicted in Fig. 1d, we implemented a custom-written spot finding algorithm for the Imspector software (http://www.imspector.de), written in the Python programming language. First, we acquire an overview image (20 µm x 20 µm) by direct excitation of the acceptor (640nm, 13.6 µW, dwell time: 100µs), allowing us to identify molecules with active acceptors. Spot detection was performed using an intensity threshold and by requiring a minimal distance between spots. Additional statistical tools were employed to avoid measuring aggregation/crowding effects of dsDNA on surface. Second, the total number of detected acceptor-labeled spots per overview image was used to estimate the spot surface density and finally calculate the probability of single- and multi-spot events per image under assumption of randomly distributed spots on the surface. If the probability of multi events goes below 5%, the recorded overview image was taken for the acquisition of STED images of single spots. Further, the detected spots were filtered by applying thresholds on the photon number and minimal distance. After filtering, the positions of spots containing single molecules were stored. Consecutively, for each position a 1 µm x 1 µm image was then recorded for 61 frames and saved in the PicoQuant PTU file format.

To optimize the use of the available signal before photobleaching, different strategies were used for the extraction of the FRET parameters and the localization of the fluorophores. For the FRET signal, photobleaching of either donor or acceptor is a limiting factor. On the other hand, localization does not require both dyes to be active. In the case of high FRET, acceptor photobleaching even aides the donor localization precision as the donor signal is increased. Thus, only the first 100 µs or 20 frames are used for the calculation of the FRET-related spectroscopic parameters, while the total collected signal is used to estimate the fluorophore position.

## *DNA origami and DNA ruler measurements*

All data was recorded using a 1 µm x 1 µm image (100 x 100 pixel) with a 5 µs dwell time, amounting to a frame time of 300 ms including the delay caused by the resonant scanner. For the origami measurements, the excitation powers were 6.4 µW for the donor excitation (561 nm), 3.3 µW for the acceptor excitation (640 nm) and 35 mW for the STED depletion (775 nm) lasers, respectively. For the DNA ruler measurements, the excitation powers were 4.2 µW for the donor excitation (561 nm), 2.7 µW for the acceptor excitation (640 nm) and 42 mW for the STED depletion (775 nm) lasers, correspondingly. The analysis settings for the measurements of the different DNA origami constructs containing only the noFRET dye pair O(NF), only the highFRET dye pair O(HF), or both dye pairs O(HF+NF), are listed in Supplementary Tab. 14.

### *hGBP1 measurements*

Super-resolved STED FRET measurements of Cys8 hGBP1-18-577 were performed with line-alternating excitation of donor (Alexa594) and acceptor (Atto647N) dyes, excited with 561 nm laser (5 µW) and 640 nm laser (2 µW), respectively, at a pulse rate of 40 MHz. Depletion was done by 775 nm STED laser at a power of 42 mW. Images of hGBP1 rings were taken at a distance of 300-500 nm above the surface to ensure minimal diffusion of the structures while reducing background signal from the surface. The analysis settings are summarized in Supplementary Fig. 29.

## Sample preparation

### *dsDNA*

To screen a broad range of FRET efficiencies, we designed highFRET (dsD(HF)), midFRET (dsD(MF)), lowFRET (dsD(LF)) and noFRET (dsD(NF)) samples of donor-acceptor pairs on dsDNA (Supplementary Tab. 1). All single DNA strands were synthesized and labeled by IBA GmbH (Göttingen, Germany), followed by HPLC purification. The position of the donor dye (Alexa594) was kept fixed while the acceptor (Atto647N) position on the complementary strands was varied. The dye derivatives with an activated NHS-ester were coupled to the amino group of C6-amino-linker connected a thymine (T) or cytosine (C). Additionally, a single or two biotin anchors were attached to the 3' and 5' ends of the donor strands (single biotin was attached to the 5' end only) for surface immobilization.

Hybridization of complementary single stranded DNA was performed inside a thermocycler (primus 96 advanced, pegLab, Erlangen). Donor only strands were mixed with reversed acceptor only strands in excess 1:3 in buffer (20mM MgCl$_2$, 100mM KCl, 20 mM KH$_2$PO$_4$/K$_2$HPO$_4$, pH 6.5) and quickly heated up to 85° C and fast cooled up to 52° C with 0.1° C/s. The solution was kept at this temperature for 2 hours, subsequently cooled down to 4° C and stored at this temperature. Detailed information about sample properties is listed in Supplementary Tab. 1.

### *DNA origami*

We used single-layer DNA origami nanostructures are based on published designs and were assembled as described in Schnitzbauer et al[3]. The structural design is depicted in Supplementary Fig. 7. Folding of structures was performed in buffer containing 10 mM Tris, 12.5 mM MgCl$_2$ and 1 mM EDTA at pH 8.0 in a one-pot reaction containing 10 nM p7249 M13 single-stranded DNA scaffold (tilibit nanosystems, Munich, Germany); 100 nM core staple strands, 100 nM biotinylated staple strands and 1 µM fluorescently labeled staple strands in a total volume of 40 µl (the used sequences are listed in Supplementary Tab. 2). Unlabeled and biotinylated staple strands were purchased from Biomers (Ulm, Germany). Fluorescently labeled staple strands were purchased from IBA (Göttingen, Germany). Annealing was performed by heating the mixture to 80 °C and cooling it using a temperature gradient from 60 to 4 °C in steps of 3 min 12 s per °C. The correct assembly of the structures was checked by gel electrophoresis (2% agarose) and TEM imaging. Purification was performed by precipitating the origami nanostructures by adding one volume of buffer containing 15% PEG 8000 (w/v), 5 mM Tris, 1mM EDTA and 500 mM NaCl[4]. The sample was thoroughly mixed and centrifuged at 16.000 g for 25 min in a microcentrifuge at room temperature. After the supernatant was removed with a pipette, the pellet was dissolved in imaging buffer (10 mM Tris, 12.5 mM $MgCl_2$, 1 mM EDTA, pH 8.0) and the precipitation step was repeated one more time, after which the pellet was incubated in imaging buffer over night at room temperature to fully recover the structures.

Fluorophores were attached to the 3'-end of the respective oligos with an additional thymine base pair as a spacer. Fluorophore positions were determined with the Picasso software[3]. All strands are listed in Supplementary Tab. 2. The origami design as obtained from the caDNAno software is shown in Supplementary Fig. 7[5].

### Surface preparation

DsDNA or DNA origami nanostructures were immobilized on the surface using biotinylated NHS-PEG or biotinylated BSA (see Supplementary Tab. 13). Both immobilization protocols start with a cleaning procedure of a single cover slide (Marienfeld, Precision cover glasses thickness No. 1.5H). Cover slides were sonicated in 5% Hellmanex for 20 min, washed 10-times with water to remove the Hellmanex and dried under nitrogen flow. Cleaned cover slides were activated for 10 minutes in an oxygen plasma (FEMTO Plasma Cleaner, Diener electronic).

_NHS-PEG-Biotin._ Cleaned cover slides were incubated with ethanolamine hydrochloride (3M in dimethyl sulfoxide, DMSO, prepared by dissolving 10.24 g ethanolamine hydrochloride in 35ml DMSO) for least 12 h at room temperature. After removal of remaining ethanolamine hydrochloride from the surface by washing with water and drying with nitrogen, 200µl NHS-PEG-Biotin (IRIS Biotech GmbH) in chloroform was added and sandwiched between a second cover slide. After 1 h, the sandwiched cover slides were separated carefully, cleaned with acetonitrile, and again dried with nitrogen. Finally, the surface was incubated for 10 min with neutravidin (Invitrogen, 20µg/ml in water) and washed with PBS to remove remaining/ unbound neutravidin molecules.

_BSA-Biotin._ After the cleaning procedure, the surface was incubated for 10 min with biotinylated BSA (Sigma-Aldrich, 3 mg/ml in PBS), washed up to 5 times with PBS, incubated for 10 min with neutravidin (Invitrogen, 20µg/ml in water), and washed again with buffer to remove remaining/ unbound neutravidin molecules.

The overnight measurements of immobilized samples on the cover slide took up to 24 h. In order to reduce evaporation of the sample solution, the coated slides were glued to IBIDI sticky slides VI 0.4 (ibidi, Gräfelfing, Germany). Here 6 defined chambers are created with a volume of 60 µl. After sample injection, the chamber was washed 5 times with PBS before being filled with imaging buffer solution.

Additional chemicals and hardware used for functionalizing and handling of surfaces are compiled in Supplementary Tab. 16.

### hGBP1 expression and labeling

*Plasmids.* The wild-type of hGBP1 (hGBP1-wt) harbors 9 cysteine residues with one being blocked by the farnesyl moiety after modification. To site-specifically attach fluorescence labels for FRET measurements, all wild-type cysteines were mutated in our previous work leading to the mutant termed Cys9 hGBP1 (i.e., C12A, C82A, C225S, C235A, C270A, C311S, C396A, C407S, C589S)[6]. The mutation C589S was reversed using the QuikChange site-directed mutagenesis kit with KOD Hot Start DNA polymerase (Merck, Millipore) resulting in so termed Cys8 hGBP1. This allows for the farnesylation of the C-terminus of the protein. To introduce two cysteine residues for FRET studies, residues N18 and Q577 were replaced by a cysteine. This mutant termed Cys8 hGBP1-18-577 is used throughout this study. The success of mutagenesis was verified by sequencing (3130xl sequencer, Applied Biosystems).

*Protein expression and purification.* Expression and purification of wild type and mutant hGBP1 was performed as described previously[7,8]. In brief, the DNA was expressed in E. coli strain Rosetta™ (DE3)pLysS using a pQE-80L vector (Qiagen). For affinity chromatography, Cobalt-NTA-Superflow was used followed by size exclusion chromatography with a Superdex 200 column. To ensure the stability of the protein, the buffer for size exclusion chromatography and storage included 2 mM DTT. This was removed prior to labelling reactions by applying the solutions to spin concentrators (Vivaspin™) in three repetitive cycles. Both non-farnesylated hGBP1-wt and the non-farnesylated mutant Cys8 hGBP1-18-577 were farnesylated using the protocol for enzymatic modification[7] leading to hGBP1-wt and Cys8 hGBP1-18-577, respectively. Protein absorption was measured at 280 nm using a NanoDrop™ 2000 spectrophotometer (ThermoFisher Scientific, USA) and the concentration was calculated with an extinction coefficient of 45000 $M^{-1}cm^{-1}$. Purity of the farnesylated protein was checked by SDS-PAGE and the activity of the protein was verified by the turbidity assay using standard conditions[7] which demonstrates both enzymatic activity and the formation of polymers.

*Protein labelling.* Human GBP1-wt was unspecifically labelled by mixing 100 µM protein with 150 µM Alexa594 succinimidyl ester (Life Technologies GmbH) on lysines and 150 µM Atto647N maleimide (ATTO TEC GmbH) on cysteines by incubation on ice for 30 minutes. Unbound dye was removed with spin concentrators (Vivaspin™). Cys8 hGBP1-18-577 was labelled by mixing 100 µM protein with 130 µM Alexa 594 C5 maleimide (Life Technologies GmbH) in buffer $C_{Label}$ (50 mM Tris/HCl, 150 mM NaCl, 5 mM $MgCl_2$, pH 7.4) and by incubation for 60 min. The unbound dye was removed, and the buffer was exchanged to low salt buffer (50 mM Tris/HCl, 5 mM $MgCl_2$, pH 7.4) with spin concentrators (Vivaspin™). Different labelled species were isolated by anion exchange chromatography using a ResourceQ column (GE Healthcare) and running a gradient from 0-200 mM NaCl at pH 7.4. The labelled species with approximately 100 % labelling efficiency were mixed with 3 eq of Atto647N maleimide (ATTO TEC GmbH) and incubated for 90 min. The unbound dye was removed with a HiPrep 26/20 S25 desalting column. The labelling efficiencies for both dyes for labelled hGBP1-wt (Donor: 42%; Acceptor: 78%) and labelled Cys8 hGBP1-18-577 (Donor: 118%; Acceptor: 92%) were determined by using their respective extinction coefficients (Atto647N $\varepsilon(646 \text{ nm}) = 150000 \text{ M}^{-1}cm^{-1}$; Alexa594 $\varepsilon(590 \text{ nm}) = 92000 \text{ M}^{-1}cm^{-1}$), accounting for the spectral overlap and comparing the resulting dye concentrations to the protein concentration. Afterwards, the protein was stored at concentrations of 10-30 µM in buffer C (50 mM Tris-HCl, 150 mM NaCl, 5 mM MgCl$_2$, pH 7.9) after addition of 2 mM DTT at -80°C.

*Triggering hGBP1 oligomerization*. A 10 µl sample containing a total of 10 µM protein was prepared in buffer C with AlF$_x$ (50 mM Tris-HCl, 150 mM NaCl, 5 mM MgCl$_2$, 300 µM AlCl$_3$, 10 mM NaF, pH 7.9). The mixture of labelled Cys8 hGBP1-18-577 (or labelled hGBP1-wt) and hGBP1-wt varied depending on the desired labelling density in hGBP1 polymers. Most measurements where single fluorescence spots could be identified were taken at a Cys8 hGBP1-18-577 concentration of 0.05 µM and a hGBP1-wt concentration of 9.95 µM. To induce the polymerization reaction, 200 µM GDP was added and the reaction solution was incubated for 15 min. The reaction solution was mixed and 1 µl was diluted in 1 ml buffer C with AlF$_x$ and 200 µM GDP. After mixing the diluted sample, 300 µl were transferred into a Nunc™ Lab-Tek™ II Chamber Slide™ (ThermoFisher Scientific), which was previously passivated with a 1 g/l BSA solution for 20 minutes. Polymer structures were visible after 10-40 min after which they would remain stable for hours.

## Spectroscopy and image analysis

### Filtering procedures

*Filtering procedure for DNA rulers.* The processed data file contains a list of parameters for each spot (see SI Supplementary Tab. 18). We distinguish between image parameters (such as the localization precision, spot symmetry and spot stoichiometry) and spectroscopic parameters (e.g., the intensity-based stoichiometry, the FRET efficiency and the fluorescence lifetimes of the donor and acceptor dyes). The filtering procedure is schematically illustrated in Supplementary Fig. 3. First, spots with a spot width larger than a given threshold were discarded (see SI Supplementary Tab. 15). Large spot widths occur for insufficient photon numbers (low localization precision) or if multiple donor or acceptor fluorophores are present in the ROI (multi-molecule events, aggregation). In the second step, double-labeled spots containing one donor and one acceptor were identified based on the intensity-based stoichiometry by selecting spots with $S \approx 0.5$.

*Filtering procedure for origamis.* The origami datasets were fitted with multiple circular 2D Gaussians and filtering was performed based on the detected number of spots for each color (spot stoichiometry). Dim spots with less than 20 photons were discarded. Only completely labelled structures were used for particle averaging, whereas for spectroscopic analysis also ROIs with two or three spots for donor and acceptor were used (Supplementary Fig. 2) for better statistics. An additional cut in the acceptor brightness and spot stoichiometry was applied to the only data O(HF) to remove structures affected by acceptor photobleaching. For the particle averaging, we additionally used the RMSD with respect to the reference structure for filtering (Supplementary Fig. 9). For the analysis of sub-ensemble TCSPC decays, an additional cut of the FRET efficiency was performed. All filtering parameters are listed in Supplementary Tab. 17 and the settings are given in Supplementary Tab. 14.

*Filtering procedure for hGBP1.* For hGBP1, a manual pre-selection of data was done by visual inspection of the ring structures with respect to the labeling density and the definition of the spots (blurring).

Analogous to the origami analysis, the hGBP1 data was fitted with multiple circular Gaussians. Dim spots with less than 20 photons were discarded. Only spots with at least one detected donor and acceptor spot (spot stoichiometry >0) and an uncorrected FRET efficiency (proximity ratio, PR) > 0.35 were selected for further analysis (Supplementary Fig. 29a). All filtering thresholds are given in Supplementary Tab. 14.

### *Determination of intensity-based correction factors*

Spectral crosstalk of the donor fluorescence into the acceptor detection channel, $\alpha$, the relative excitation flux, $\beta$, and the direct excitation of the acceptor by the donor excitation laser, $\delta$, where calculated following the approach outlined in[9]. Briefly, $\alpha$ was determined from the ratio of signal in the donor and acceptor channel in the absence of an acceptor dye using confocal measurements of the noFRET origami sample. $\delta$ was determined under STED conditions by placing the acceptor-only population at zero stoichiometry and $\beta$ was determined from each dataset directly by placing the double-labeled population at a stoichiometry of 0.5. Donor-only and acceptor-only constructs were selected based on the number of detected spots for each color (spot stoichiometry) (Supplementary Fig. 2). Spectra were obtained from the manufacturer[10] and combined with the transmission spectra of the dichroic mirrors, notch filters and emission filters to obtain estimated detection efficiencies using the homebuilt program *Detection Efficiencies*[11]. Quantum yields were estimated based on the fluorescence lifetimes of donor-only and acceptor-only samples. To estimate the lifetime of the undepleted fluorophores under STED conditions, the fluorescence decays were fitted with two lifetimes. In the model, the short lifetime accounts for the fast depletion by the STED pulsed, while the long lifetime is assigned to the fluorescence of the undepleted fluorophores. The quantum yield is determined from the measured lifetime $\tau$ by:

$$\Phi_{\mathrm{F}} = \Phi_{F,\mathrm{ref}} \frac{\tau}{\tau_{\mathrm{ref}}} \, , \qquad (S1)$$

where $\Phi_{\mathrm{F}}$ is the quantum yield and the subscript *ref* refers to a known reference. Here, the reference values for both Atto594 and Atto647N are taken from reference[10]. The species fractions $x_{\mathrm{D}}^0$, $x_{\mathrm{D}}^{\mathrm{d}}$, $x_{\mathrm{A}}^0$, $x_{\mathrm{A}}^{\mathrm{d}}$ are obtained from the sub-ensemble lifetime fits for each dataset separately. See SI Supplementary Tab. 3 for an overview of the correction factors.

### *Intensity-based spectroscopic parameters*

In order to estimate qualitatively FRET inside your data it is not always necessary to account for all correction parameter, immediately. Thus, the proximity ratio $E_{PR}$:

$$E_{\mathrm{PR}} = \frac{{}^{i}I_{\mathrm{Aem|Dex}}}{{}^{i}I_{\mathrm{Dem|Dex}} + {}^{i}I_{\mathrm{Aem|Dex}}} \qquad (S2)$$

is calculated from the uncorrected (raw) intensities ${}^{i}I$ in the donor channel after donor excitation (${}^{i}I_{\mathrm{Dem|Dex}}$) and the acceptor channel after donor excitation (${}^{i}I_{\mathrm{Aem|Dex}}$). To determine the accurate FRET efficiency, the uncorrected intensities need to be corrected as follow. As a first step, the signals are corrected for constant background $I^{(\text{BG})}$:

$$
\begin{aligned}
{}^{ii}I_{\text{Dem|Dex}} &= {}^{i}I_{\text{Dem|Dex}} - I^{(\text{BG})}_{\text{Dem|Dex}} \\
{}^{ii}I_{\text{Aem|Dex}} &= {}^{i}I_{\text{Aem|Dex}} - I^{(\text{BG})}_{\text{Aem|Dex}} \\
{}^{ii}I_{\text{Aem|Aex}} &= {}^{i}I_{\text{Aem|Aex}} - I^{(\text{BG})}_{\text{Aem|Aex}}
\end{aligned}
\tag{S3}
$$

Here, ${}^{i}I_{\text{Aem|Aex}}$ describes the uncorrected measured intensity in the acceptor channel after acceptor excitation. It is assumed that the spectral properties of the fluorophores (extinction coefficient, absorption spectrum, emission spectrum) do not change under STED conditions. Corrections for spectral crosstalk of the donor into the acceptor detection channel and direct excitation of the acceptor by the donor excitation laser can thus be performed as for conventional FRET experiments[9,12], yielding the corrected acceptor fluorescence after donor excitation $F_{\text{A|D}}$:

$$
F_{\text{A|D}} = {}^{ii}I_{\text{Aem|Dex}} - \alpha\,{}^{ii}I_{\text{Dem|Dex}} - \delta\,{}^{ii}I_{\text{Aem|Aex}}
\tag{S4}
$$

where $\alpha$ is the correction factor for crosstalk and $\delta$ for direct excitation. No further corrections have to be applied to the other two channels. For convenience, we change the notation to indicate that the intensities ${}^{ii}I$ are corrected fluorescence signals:

$$
\begin{aligned}
F_{\text{D|D}} &= {}^{ii}I_{\text{Dem|Dex}} \\
F_{\text{A|A}} &= {}^{ii}I_{\text{Aem|Aex}}
\end{aligned}
\tag{S5}
$$

The uncorrected (apparent) ROI – stoichiometry, ${}^{i}S_{\text{app}}$, is given by:

$$
{}^{i}S_{\text{app}} = \frac{I_{\text{Aem|Dex}} + I_{\text{Dem|Dex}}}{I_{\text{Aem|Dex}} + I_{\text{Dem|Dex}} + I_{\text{Aem|Aex}}}
\tag{S6}
$$

Additionally, the observed intensities need to be corrected for different excitation intensities and absorption cross sections ($\beta$ correction factor), donor and acceptor quantum yields ($\Phi^{0}_{\text{F,D}}, \Phi^{0}_{\text{F,A}}$) and detection efficiencies ($\gamma$ correction factor). The $\beta$ factor is determined by the excitation spectra of the donor and acceptor and the excitation intensity ratio as:

$$
\beta = \frac{\sigma_{\text{A|R}}}{\sigma_{\text{D|G}}} \frac{I_{\text{Aex}}}{I_{\text{Dex}}}
\tag{S7}
$$

The γ factor is given by:

$$
\gamma = \frac{g_{\text{A|A}}}{g_{\text{D|D}}} \frac{\Phi^{0}_{\text{F,A}}}{\Phi^{0}_{\text{F,D}}}
\tag{S8}
$$

Here, the detection efficiency ratio $\frac{g_{\text{A|A}}}{g_{\text{D|D}}}$ is calculated based on the emission spectra of the donor and acceptor and the transmission of the optical elements. The corresponding quantum yields of donor- and acceptor-only sample are determined under experimental conditions by sub-ensemble lifetime fitting. The fully corrected ROI stoichiometry is then explicitly given by:

$$
S = \frac{\gamma F_{\text{D|D}} + F_{\text{A|D}}}{\gamma F_{\text{D|D}} + F_{\text{A|D}} + {}^{1}\!/_{\beta}\, F_{\text{A|A}}}
\tag{S9}
$$

The FRET population should be symmetrically distributed at $S = 0.5$, the donor only population at $S = 1$ and the acceptor only population at $S = 0$. The correction parameters of all measurements are compiled in Supplementary Tab. 3.

For the FRET efficiency histogram, spots were selected based on the selection criteria given in Supplementary Tab. 14 for origami samples and Supplementary Tab. 15 for DNA Rulers.

### Accurate intensity-based FRET efficiencies under STED conditions

To estimate accurate intensity-based FRET efficiencies, we used the total intensity detected in the donor and acceptor channels without any time gating. This is necessary to avoid that the effective detection efficiency of the donor fluorescence (i.e., the fraction of the fluorescence decay that falls within the time gate) depends on the fluorescence lifetime, which in turn depends on the rate of energy transfer. In addition, the available signal is maximized. However, without the application of time gating, the measured fluorescence signals contain contributions from both (partially) depleted and undepleted molecules, which exhibit different quantum yields. Donor and acceptor fluorophores are also generally depleted to a different extent. The measured fluorescence signals after donor excitation can be described by:

$$F_{D|D} = \sigma_{D|D} \, I_{Dex} \, g_{D|D} \left[ x_D^0 \, \Phi_{F,D}^0 \, (1 - E_0) + x_D^d \, \Phi_{F,D}^d \, (1 - E_d) \right]$$
$$F_{A|D} = \sigma_{D|D} \, I_{Dex} \, g_{A|A} \left[ x_A^0 \, \Phi_{F,A}^0 \, E_0 + x_A^d \, \Phi_{F,A}^d \, E_d \right] \tag{S10}$$

where $\sigma_{D|G}$ is the absorption cross-section of the donor at the donor excitation wavelength, $I_{Dex}$ is the excitation intensity of the donor excitation laser, $g_{D|D}$ and $g_{A|A}$ are the detection efficiencies of the donor fluorophore in the donor detection channel and the acceptor fluorophore in the acceptor detection channel, $x_D^0$ and $x_A^0$ are the fractions of undepleted donor and acceptor molecules with respective quantum yields $\Phi_{F,D}^0$ and $\Phi_{F,A}^0$ and FRET efficiency $E_0$, and $x_D^d$ and $x_A^d$ are the fractions of depleted or partially depleted donor and acceptor molecules with respective quantum yields $\Phi_{F,D}^d$ and $\Phi_{F,A}^d$ and FRET efficiency $E_d$. Note that the FRET efficiency of the depleted molecules is expected to be reduced due to the increased donor de-excitation rate through the depletion pulse, resulting in a negligible FRET efficiency for depleted donors. Accordingly, we have not accounted for the possibility that a depleted donor might transfer its energy to a non-depleted acceptor. We also do not consider the rare event that a non-depleted donor might transfer its energy to a depleted acceptor.

In the lifetime analysis of sub-ensemble decays, we describe the contribution of depleted fluorophores with a short lifetime component. From the time-resolved information of donor- and acceptor- only molecules, we can thus obtain the species-fraction of the depleted and undepleted molecules, $x^d$ and $x^0 = 1 - x^d$, and estimate the respective quantum yields from the fitted lifetimes, $\Phi_F^d$ and $\Phi_F^0$ for the donor and acceptor fluorophore. This allows us to solve for the FRET efficiency $E_0$ in the absence of depletion by taking the ratio of the measured fluorescence signals:

$$\frac{F_{A|D}}{F_{D|D}} = \frac{g_{A|A}}{g_{D|D}} \cdot \frac{x_A^0 \cdot \Phi_{F,A}^0 \cdot E_0 + x_A^d \cdot \Phi_{F,A}^d \cdot E_d}{x_D^0 \cdot \Phi_{F,D}^0 \cdot (1 - E_0) + x_D^d \cdot \Phi_{F,D}^d \cdot (1 - E_d)} \tag{S11}$$

from which we obtain

$$E_0 = E_0^{'} \cdot \left(1 + \frac{x_D^d \cdot \Phi_{F,D}^d}{x_D^0 \cdot \Phi_{F,D}^0}\right) - E_d \cdot \left(E_0^{'} \frac{x_D^d \cdot \Phi_{F,D}^d}{x_D^0 \cdot \Phi_{F,D}^0} + \left(1 - E_0^{'}\right) \cdot \frac{x_A^d \cdot \Phi_{F,A}^d}{x_A^0 \cdot \Phi_{F,A}^0}\right) \qquad (S12)$$

The modified FRET efficiency $E_0^{'}$ is related to the measured fluorescence signals as:

$$E_0^{'} = \frac{F_{A|D}}{\frac{x_A^0 \cdot g_{A|A} \cdot \Phi_{F,A}^0}{x_D^0 \cdot g_{D|D} \cdot \Phi_{F,D}^0} F_{D|D} + F_{A|D}} = \frac{F_{A|D}}{\gamma^{'} \cdot F_{D|D} + F_{A|D}} \qquad (S13)$$

where $\gamma^{'} = \frac{x_A^0}{x_D^0} \cdot \overset{\gamma}{\overbrace{\frac{g_{A|A} \cdot \Phi_{F,A}^0}{g_{D|D} \cdot \Phi_{F,D}^0}}} = \frac{x_A^0}{x_D^0} \cdot \gamma$ is a modified $\gamma$-factor that corrects for the relative detection yield of the donor and acceptor fluorophores, taking the fraction of undepleted molecules into account.

We further assume that the FRET efficiency under depletion conditions is zero, as the de-excitation of the excited donor fluorophore is faster than the energy transfer, yielding the simplified equation as used in the main text:

$$E_0 = E_0^{'} \cdot \left(1 + \frac{x_D^d \cdot \Phi_{F,D}^d}{x_D^0 \cdot \Phi_{F,D}^0}\right) \qquad (S14)$$

The correction parameters of all measurements are compiled in Supplementary Tab. 3.

### Determination of average intensity-based FRET efficiencies

To estimate the average intensity-based FRET efficiency from single-molecule histograms, the one-dimensional FRET efficiency distribution was first fitted to a normal distribution. To reduce the influence of outliers, the average FRET efficiency was calculated based only on those spots that fell within one standard deviation of the mean. The results of all sample moelcules are compiled in Supplementary Tab. 10.

### Sub-ensemble fluorescence decay analysis

Selecting molecules with similar ROI-integrated spectroscopic properties (i.e., FRET efficiency or donor fluorescence lifetime) allows us to perform a quantitative analysis of the interdye distance distributions. Sub-ensemble fluorescence decays were generated by binning the microscopic arrival times of the donor photons of all selected spots. As we use polarization-resolved detection, the unpolarized total fluorescence decay $f(t)$ is constructed according to:

$$f(t) = f_{\parallel}(t) + 2 \cdot G \cdot f_{\perp}(t) \qquad (S15)$$

where $f_{\parallel}(t)$ and $f_{\perp}(t)$ are the fluorescence decays measured in the parallel and perpendicular detection channels, respectively[39], and the polarization correction factor $G = \frac{\eta_{\parallel}}{\eta_{\perp}}$ accounts for the instrument's polarization dependent transmission where $\eta_{\parallel}$ and $\eta_{\perp}$ are the detection efficiencies of the parallel and perpendicular detection channels (see Supplementary Tab. 3). Note that this approach (eq. S15) works for random orientations. Considering immobilized and hence partially aligned molecules, $f(t)$ is only approximated.

The energy transfer due to FRET increases the relaxation rate of the excited donor fluorophore and thus reduces the excited state lifetime. The rate of energy transfer due to FRET, $k_{RET}$, depends on the sixth power of the interdye distance $R_{DA}$[13]:

$$k_{\mathrm{RET}}(R_{DA}) = \frac{1}{\tau_{D(0)}} \cdot \left(\frac{R_0}{R_{DA}}\right)^6 \qquad (S16)$$

The interdye distance can thus be measured by the reduction of the fluorescence lifetime of the donor-acceptor labeled sample compared to a sample containing only the donor fluorophore (Donor-only, D0). In the absence of STED, the fluorescence decay of the donor in the presence of the acceptor $f_{D|D}^{(DA)}$ can then be described as:

$$f_{\mathrm{D|D}}^{(\mathrm{DA})}(t) = \int_0^\infty x_{\mathrm{FRET}}(R_{DA}) \cdot e^{-(k_{\mathrm{D0}} + k_{\mathrm{RET}(R_{DA})})t} \cdot \mathrm{d}R_{\mathrm{DA}} \qquad (S17a)$$

where $k_{D0}$ is the deexcitation rate of the donor in the absence of the acceptor and $x_{\mathrm{FRET}}(R_{DA})$ describes the fraction of molecules with interdye distance $R_{DA}$. Here, we assume that the dyes rotate fast ($\kappa^2 = 2/3$) but diffuse slowly compared to the fluorescence lifetime. The distribution of interdye distances arising due to the flexible dye linkers can then be described by a normal distribution with width $\sigma_{DA}$:

$$x_{\mathrm{FRET}}(R_{DA}) = \frac{1}{\sqrt{2\pi}\sigma_{\mathrm{DA}}} \cdot \exp\left(-\frac{(R_{\mathrm{DA}} - \langle R_{\mathrm{DA}}\rangle)^2}{2\sigma_{\mathrm{DA}}^2}\right) \qquad (S17b)$$

where $\langle R_{\mathrm{DA}}\rangle = \frac{1}{NM}\sum_{i=1}^{N}\sum_{j=1}^{M}|R_{\mathrm{D,i}} - R_{\mathrm{A,j}}|$ is the mean interdye distance averaged over all possible donor and acceptor positions.

To utilize the full range of the fluorescence decay, we account for the signal from partially depleted molecules using a fast decay component. We describe the fluorescence decay of the donor-only sample $f_{\mathrm{D|D}}^{(\mathrm{D0})}$ with a bi-exponential model:

$$f_{\mathrm{D|D}}^{(\mathrm{D0})}(t) = \overbrace{x_{\mathrm{STED}}^{(\mathrm{D0})} \cdot e^{-k_{\mathrm{D0,STED}}t}}^{\text{Depletion}} + \overbrace{\left(1 - x_{\mathrm{STED}}^{(\mathrm{D0})}\right) \cdot e^{-k_{\mathrm{D0}}t}}^{\text{Donor only}} \qquad (S18)$$

where $k_{\mathrm{D0}}$ is the deexcitation rate of the non-depleted molecules, and $k_{\mathrm{D0,STED}}$ and $x_{\mathrm{STED}}^{(\mathrm{D0})}$ are the deexcitation rate and fraction of partially depleted molecules in the donor-only sample. Analogously, we include a fast decay component in the description of the donor fluorescence decay in the presence of the acceptor:

$$f_{\mathrm{D|D}}^{(\mathrm{DA})}(t) = \overbrace{x_{\mathrm{STED}}^{(DA)} \cdot e^{-k_{\mathrm{D,STED}}t}}^{\text{Depletion}} + \left(1 - x_{\mathrm{STED}}^{(\mathrm{DA})}\right) \cdot$$

$$\left(\overbrace{(1 - x_{\mathrm{noFRET}}) \cdot \int_0^\infty x_{\mathrm{FRET}}(R_{DA}) \cdot e^{-(k_{\mathrm{D0}} + k_{\mathrm{RET}(R_{DA})})t}\mathrm{d}R_{\mathrm{DA}}}^{\text{Gaussian distribution}} + \overbrace{x_{\mathrm{noFRET}} \cdot e^{-k_{\mathrm{D0}}t}}^{\text{Donor only}}\right) \qquad (S19)$$

where $k_{D,STED}$ and $x_{STED}^{(DA)}$ are the de-excitation rate and fraction of partially depleted molecules in the double-labeled sample, and $x_{noFRET}$ describes the fraction of molecules that lack the acceptor due to bleaching or blinking of ~5-10% (see Supplementary Tab. 9).

For a quantitative analysis, it is crucial to precisely know the de-excitation rate of the donor in the absence of the acceptor $k_{D0}$. In this work, we take the population of noFRET molecules for the origami and the noFRET sample for the DNA rulers as the donor-only reference. This choice of a reference is preferred to a pure donor-only reference as it allows us to select intact molecules by the presence of the acceptor that is placed outside of the FRET range. To further stabilize the fit, we globally analyzed the donor-only and FRET-induced donor decays by linking the donor-only de-excitation rate $k_{D0}$ but optimizing the STED-related parameters individually. The width of the Gaussian distance distribution was fixed to a value of 6 Å that satisfies benchmark experiments on similar systems[14] and was confirmed for the solution-based FRET measurements of the DNA rulers.

Sub-ensemble fluorescence decays were analyzed with the ChiSurf software package (http://www.fret.at/tutorial/chisurf/)[15] using the iterative re-convolution approach.

Model decays are convoluted with the experimental instrument response function (*IRF*). In addition, a constant number of background counts N$_{bg}$ (offset) and contribution of scattered laser counts N$_{sc}$ is considered, according to:

$$F_{exp}(t) = N_0 \cdot IRF(t) \otimes f(t) + N_{sc} \cdot IRF(t) + N_{bg} \qquad (S20)$$

, where $N_0$ is the number of initial photon counts of the decay.

The measured data and sub-ensemble fluorescence fits shown in Supplementary Figs. 12 (origami, confocal), 13 (origami, STED) and 20 (DNA ruler, STED), 21 (DNA ruler, confocal). The fitted results are compiled in Supplementary Tab. 9.

*Determination of spot-integrated fluorescence lifetimes*

The intensity-weighted average fluorescence lifetime for single emitters was obtained from a single-exponential tail-fit of the spot-integrated fluorescence decays (part of the histogram after depletion pulse) using maximum likelihood estimation. In general the measured fluorescence pattern is described by a convolution of the instrument response function (IRF) with an exponential decay model function, $f(t)$. In our particular case with an additional STED pulse, the rising and fast dropping parts of decays are not within fit range and, in order to reduce computation time of the model function, the IRF here is replaced by a $\delta$-function (1 channel convolution).

Assuming a single depolarization process, the ideal model functions for the fluorescence decays detected on the parallel and perpendicular detectors are approximated by:

$$f_\parallel(t,\tau,\rho) = \frac{f_0}{3} e^{-t/\tau} \left[ 1 + r_0(2 - 3l_1)e^{-t/\rho} \right] \qquad (S21a)$$

$$f_\perp(t,\tau,\rho) = \frac{f_0}{3} e^{-t/\tau} \left[ 1 - r_0(1 - 3l_2)e^{-t/\rho} \right] \qquad (S21b)$$

where $\tau$ denotes the fluorescence lifetime, $f_0$ is the normalization factor, $r_0$ the fundamental anisotropy of the fluorophore, $\rho$ is the rotational correlation time and the correction factors $l_1, l_2$ account for the mixing of the polarizations by the objective[40]. The rotational correlation time, $\rho$, is related to the steady state anisotropy $\langle r \rangle$, the fundamental anisotropy $r_0$ and the fluorescence lifetime $\tau$ by the Perrin equation:

$$\rho = \tau \left( \frac{r_0}{\langle r \rangle} - 1 \right)^{-1} \qquad (S21c)$$

The final model functions for the parallel and perpendicular signal, $F_{\parallel}$ and $F_{\perp}$, are constructed from ideal, area normalized patterns ($f_{\parallel,\mathrm{norm}}$, $f_{\perp,\mathrm{norm}}$) by accounting for the scatter fraction, $x_{sc}$, of the area normalized stacked scatter decay patterns ($S_{\parallel,norm}$, $S_{\perp,norm}$), the polarization correction factor $G$ and the constant background fraction, $x_{bg}$ (see eq. S20):

$$F_{\parallel}(t; \tau, \rho, \gamma_{sc}) = F_{\parallel}(0) \cdot \left\{ \left[ (1 - x_{sc}) \cdot f_{\parallel,norm}(t; \tau, \rho) + x_{sc} \cdot S_{\parallel,norm}(t) \right] \cdot (1 - x_{bg}) + x_{bg} \right\} \qquad (S21d)$$

$$F_{\perp}(t; \tau, \rho, \gamma_{sc}) = F_{\perp}(0) \cdot \left\{ G \cdot \left[ (1 - x_{sc}) \cdot f_{\perp,norm}(t; \tau, \rho) + x_{sc} \cdot S_{\perp,norm}(t) \right] \cdot (1 - x_{bg}) + x_{bg} \right\} \qquad (S21e)$$

Here, $F_{\parallel}(0)$ and $F_{\perp}(0)$ are initial amplitudes, respectively. By the polarization correction factor, $G$, the detection efficiencies differences in parallel and perpendicular channels were accounted as the fluorescence ratio of parallel channel to perpendicular one ($G = \frac{\eta_{\parallel}}{\eta_{\perp}}$ accounts for the instrument's polarization dependent transmission where $\eta_{\parallel}$ and $\eta_{\perp}$ are the detection efficiencies of the parallel and perpendicular detection channels) . For a detailed model description, see Schaffer et al[39].

The fit function uses a maximum likelihood estimator for Poisson statistics as described previously[16]. However, the intensity-weighted average fluorescence lifetime still contains information on the shape of the fluorescence decay by means of the second moment of the lifetime distribution, which can be exploited to detect conformational dynamics on timescales faster that the imaging time[14].

Background photons originate from three sources: i) scattered photons of the excitation laser that leak through the emission filters, which mostly coincide with the rise term of our fluorescence decay; ii) background fluorescence, which can be reduced by working with clean surfaces and was negligible here; and iii) uncorrelated background noise, *bg*, from detector dark counts or background photons, which is present in our data but predictable. The uncorrelated background signal expected amount scales linearly with the acquisition time and follows a flat pattern (constant *offset*).

We estimate the uncorrelated background amount based on the background count rate in an empty surface area (see Supplementary Tab. 14). Overall, the uncorrelated background contribution was low (<1%) for the donor channel and slightly higher (<5%) due to higher dark counts for the acceptor.

### *Accessible volume simulations*

The model of the dsDNA is generated by the Nucleic Acid builder version 04/17/2017 of AmberTools[17]. The accessible volume (AV) is defined by modelling of the dye molecule by a geometrical approach that considers sterically allowed dye positions within a defined linker length from the attachment point with equal probability (see Supplementary Tab. 6) for the used dye parameters). The AVs are generated using the *FRET positioning and screening (FPS)* software[18]. From the AVs of the donor and acceptor fluorophores, the distance between the mean positions, $R_{mp}$, the average distance $\langle R_{DA} \rangle$, the average FRET efficiency $\langle E \rangle$ and the FRET-averaged distance $R_{(E)}$ are obtained (Supplementary Tabs. 7 and 8 and Supplementary Fig. 5). We compared these values to experimentally measured values of the dsDNA rulers.

### Error estimation of FRET-derived distances

*Fluorescence decay sub-ensemble analysis.* To estimate the distance uncertainty of the sub-ensemble fluorescence decay analysis, we performed a parameter scan of the mean interdye distance $\langle R_{DA} \rangle$ while optimizing all other parameters (support plane analysis). To determine the confidence intervals, we performed an F-test that compares the best fit (null hypothesis) to the alternate model where $\langle R_{DA} \rangle$ is changed from the optimal value. The confidence interval is obtained from the reduced chi-squared value $\chi_r^2$ where the change compared to the optimal value $\chi_{r,0}^2$ cannot be explained by the loss of a degree of freedom within a certain confidence. The threshold value is determined by the F-test as [19]:

$$\chi_{r,\text{threshold}}^2 = \chi_{r,0}^2 \cdot \left(1 + \frac{p}{N-p} \cdot F(\alpha, p, \nu)\right) \tag{S22}$$

where $p$ is the number of parameters, $N$ is the number of datapoints and $F(\alpha, p, N-p)$ is the inverse cumulative F-distribution with confidence level $\alpha$, i.e., the value at which the cumulative F-distribution reaches the value $\alpha$. If not specified otherwise, we report the 95% confidence intervals.

*Intensity-based FRET efficiency.* Error estimation for the distances obtained from intensity-based FRET efficiencies was performed by propagating the uncertainty of all correction factors as described in Hellenkamp et al.[12]. The assumed uncertainties of the correction factors are summarized in Supplementary Tab. 3.

### Colocalization analysis of FRET pairs

The colocalization-STED analysis encompasses three main steps. Firstly, spots are identified and cropped from the larger taken zoomed STED image for the confocal overview image (Fig. 1d). Secondly, the spot is fitted with a single or multiple 2D Gaussians and the best fit is selected based on the likelihood quality criterion. Lastly, the Gaussian center is used to define a spot integration area from which spot intensity and spot lifetime-based FRET parameters are determined (see Supplementary Fig. 6).

*Step 1: ROI identification.* Spots are identified in the image based on an intensity threshold. First, the data is smoothed with a Gaussian filter to avoid shot noise artifacts. Average background counts range from 0.1 to 0.3 photons / pixel and a threshold of 1 photon / pixel is applied to separate background from signal areas and generate ROIs. The smoothing is only applied for ROI selection, while for further analysis the raw data is used. ROIs that touch image borders or overlay with other ROIs are discarded. ROI identification was done independently for the donor emission upon donor excitation and acceptor emission upon acceptor excitation. The acceptor emission upon donor excitation was not used for localization as it contains additional signal originating from the donor dye due to crosstalk.

*Step 2: Spot localization.* The point spread function (PSF) of the STED microscope is modelled by a 2D Gaussian function:

$$G(x, y)_{x_0\, y_0,\sigma_{\mathrm{PSF}},A,\mathrm{bg}} = A \cdot \exp\left[\frac{(x - x_0)^2 + (y - y_0)^2}{2\sigma_{\mathrm{PSF}}^2}\right] + A_{bg} \qquad (S23)$$

where $A$ denotes the amplitude, $x_0$ and $y_0$ are the center coordinates of the emitter, $\sigma_{\mathrm{PSF}}$ is the width of the spot (determined by the PSF) and $A_{bg}$ a constant background offset. To fit with a model with multiple spots, multiple single 2D Gaussian models are combined in sum while keeping the width identical for all spots:

$$G_n(x, y)_{x_0\, y_0,\sigma_{\mathrm{PSF}},A,\mathrm{bg}} = \sum_{i=1}^{n} A_i \cdot \exp\left[\frac{\left(x - x_{0,i}\right)^2 + \left(y - y_{0,i}\right)^2}{2\sigma_{\mathrm{PSF},i}^2}\right] + A_{bg} \qquad (S24)$$

, where the subscript $n$ indicates the number of 2D Gaussian functions. For the analysis of the double-stranded DNA sample, only a single 2D Gaussian function is fitted per ROI.

To correctly account for the Poissonian statistics of the single photon counting data, we employ a maximum likelihood estimator that provides an unbiased and accurate estimate for the center of a 2D Gaussian spot[20,21]. The probability that a pixel with expectation value $\lambda$ counts $n$ photons is given by the Poissonian distribution, $\mathcal{L}(n|\lambda) = \frac{\lambda^n}{n!} e^{-\lambda}$. The model function generates a set of expectation values $\lambda_i$ for each pixel based on the model parameters $\theta$, $\lambda_i = f(\theta)$. The likelihood that the model describes image is obtained from the product over all pixels

$$\mathcal{L}(\lambda_i|n_i) = e^{-\Sigma_i\lambda_i} \prod_i \frac{\lambda_i^{n_i}}{n_i!}, \qquad (S25)$$

, where $n_i$ is the number of counts in the $i$-th pixel. Rather than calculating the likelihood, the log-likelihood is easier to calculate:

$$\ell(\lambda_i|n_i) = \ln\mathcal{L}(\lambda_i|n_i) = \sum_i (-\lambda_i + n_i \ln \lambda_i - \ln(n_i!)). \qquad (S26)$$

The optimal model parameters are obtained by maximizing the objective function $\ell(\lambda_i|n_i)$. An implementation of the Broyden-Fletcher-Goldfarb-Shanno algorithm is used to find the best solution[22]. This algorithm is based on a quasi-Newtonian that handles many fit parameters well.

The absolute value of the likelihood or log-likelihood is usually not very informative to assess the goodness of the model. A more useful number is obtained by normalizing the likelihood with respect to the likelihood of the best possible model, which is given by the model that is equal to the data itself. This results in the *2I\** value[16,23]:

$$2I^* = -2 \ln\left(\frac{\mathcal{L}(\lambda_i|n_i)}{\mathcal{L}(n_i|n_i)}\right) = -2 \left(\ell(\lambda_i|n_i) - \ell(n_i|n_i)\right). \qquad (S27)$$

Similar to the $\chi^2$ value for Gaussian error distributions, the 2I\* value is always positive, and a lower value is indicative of a better fit. For Gaussian errors, the 2I\* value converges to the $\chi^2$ value. The log-likelihood can be used to obtain confidence intervals for model parameters $\theta$, by calculating the normalized likelihood, or probability density function (pdf):

$$p(\theta) = \frac{\exp(l(\theta))}{\int \exp(l(\theta)) \, d\theta}. \tag{S28}$$

*Step 3: Spot stoichiometry in the multi-spot analysis.* For the DNA origami and hGBP1 measurements, we additionally estimated the spot stoichiometry by counting the number of emitters in the donor and acceptor channels (Supplementary Fig. 2). We consider up to three spots per ROI for the analysis of the donor and acceptor images. To estimate the most likely number of emitters, the best model is selected based on the 2l* value. An absolute 2l* penalty of 0.03 for each spot was imposed to account for the additional fitting parameters. For the DNA origami measurements, we selected only intact constructs carrying all four dyes to remove partially labelled or others dimerized/aggregated structures. For the hGBP1 measurements, only ROIs with at least one donor and acceptor spot were considered for further analysis.

## Predicting the localization precision

MLE supplies an estimate for the uncertainty of the fit parameters. Importantly, the uncertainty in the spot location is given in detail by[20]:

$$Var(x_0) = Var(y_o) = \frac{\sigma_a{}^2}{N} \cdot \left(1 + \int_0^1 \frac{\ln t}{1 + t/\xi} dt\right)^{-1}, \tag{S29}$$

where $\sigma_a{}^2 = \sigma_{PSF}^2 + a^2/12$ is the pixelation broadened spot width, $a$ is the pixel size and $\sigma_{PSF}$ the fitted width of the Gaussian PSF. $\xi$ represents the signal to noise ratio and is given by $\xi = 2\pi \sigma_a{}^2 bg/(N a^2)$, where $N$ is the total amount of photons in the spot. The first term is the general expression for the mean of a distribution sampled $N$ times, while the second term in brackets adds a noise penalty. The computed and the experimental localization precisions are compiled in Supplementary Tab. 5. The experimental data (background and spot sizes) are shown in Supplementary Figs.17 (origami), 25 (DNA ruler) and 29 (hGBP1).

## Model-based analysis of localization-based distance distributions

The distribution of interdye distances $d_{loc}$ between two fluorophores $D$ and $A$ whose position follows a normal distribution with localization precisions $\sigma_{loc,D}$ and $\sigma_{loc,A}$, respectively, is given by a non-central $\chi$-distribution with two degrees of freedom:

$$P_\chi(d_{loc}|R_{mp}^{loc}, \sigma_\chi) = \left(\frac{d_{loc}}{\sigma_\chi^2}\right) \cdot \exp\left(-\frac{d_{loc}^2 + R_{mp}^{loc}{}^2}{2\sigma_\chi^2}\right) \cdot I_0\left(\frac{d_{loc} R_{mp}^{loc}}{\sigma_\chi^2}\right) \tag{S30}$$

where $R_{mp}^{loc}$ is the mean-position distance between the fluorophores and $I_0(x)$ is the modified Bessel function of zero-th order[24,25]. The width parameter $\sigma_\chi$ is determined by the combined localization precision and additional registration error, $\sigma_\chi = \sqrt{\sigma_{loc,D}^2 + \sigma_{loc,A}^2 + \sigma_{reg}^2}$.[24] For large distances compared to the localization uncertainty (i.e., $R_{mp}^{loc} \gg \sigma_\chi$), the distribution approaches a normal distribution. However, if the mean-position distance is on the scale of the localization uncertainty ($R_{mp}^{loc} \approx \sigma_\chi$), the distribution becomes highly asymmetric. In addition, it becomes difficult to fit both the location parameter $R_{mp}^{loc}$ and the scale parameter $\sigma_\chi$ simultaneously because they are highly correlated. As a result, the shape of the distribution is mainly determined by the scale parameter $\sigma_\chi$ but becomes insensitive to $R_{mp}^{loc}$. Niekamp et al. propose that the $\sigma_\chi$ should be fixed to obtain meaningful results for the mean-position distance[24]. As the number of occurrences in a bin is governed by Poisson statistics, maximum likelihood estimation is used to find the optimal model fit parameters[23]. All data are compiled in Supplementary Tab. 12.

### *Alignment and particle averaging for origami measurements*

Particle averaging is a powerful tool to enhance the precision of measurements. However, it requires that molecules from a single molecular species are selected, as the inclusion of other molecular species is detrimental to the result. Here, a single molecular species is obtained in two steps. Firstly, we select only molecular assemblies with two emitters in the donor channel and two emitters in the acceptor channel using the spot stoichiometry. Secondly, we align the four positions to a reference structure and select only those that have a good overall alignment with the reference structure, as described below.

All molecular assemblies were first rotated such that the two donors are at the top. This step enables to identify the NF pair (D1 and A1) on the left and the HF pair (D2 and A2) on the right. In addition, local minima in the fine alignment are avoided where the structures are rotated by 180 degrees. As the biotin anchors are situated on the opposite side of the fluorophores, the classification into the NF and HF pairs was unambiguous. Our alignment procedure can be classified as a case of Procrustes analysis, where the sample is translated and rotated, but not stretched. This problem is linear and can be solved directly using the Kabsch algorithm[26]. An implementation in the Python programming language was used (http://github.com/charnley/rmsd). The Kabsch algorithm minimizes the root-mean-square-displacement (RMSD), which is defined as

$$RMSD_i = \sqrt{\frac{1}{4}\sum_{j \in S}\left(x_{i,j} - x_{ref,j}\right)^2 + \left(y_{i,j} - y_{ref,j}\right)^2} ; \ S = \{D1, D2, A1, A2\} \qquad (S31)$$

Where $x_{i,j}$ and $y_{i,j}$ indicate the position of dye $j$ in construct $i$ and $x_{ref,j}$ and $y_{ref,j}$ indicate the position of the dye in the reference structure. The RMSD is a useful criterion to assess how well the structures align overall (see Supplementary Fig. 9). It is possible to calculate the best possible RMSD when one considers photon noise as the only source of imprecision. For a single localization the precision can be calculated from spot brightness, spot width and signal-to-noise level (see section 'Predicted localization precision'). For completely labelled origamis, the average localization precision for the donors was 1.7 nm and for the acceptors 1.1 nm. The RMSD score takes the Cartesian norm over all four positions, resulting in a lower estimate for the RMSD of 2.8 nm (see also Supplementary Fig. 17). The excess

RMSD is due to surface induced stretching of the origami. Simulations show that the platform is flexible[27,28] matching the variation in platform sizes observed in TEM imaging (Supplementary Fig. 8).

When one has limited pre-knowledge of the structure of the molecular assembly, it is advantageous to select a reference structure from the dataset itself. Here we consider all completely labelled origami platforms with two donor and two acceptor dyes as potential reference structures. For each reference structure, the mean RMSD was calculated over al structures that have RMSD lower than 20 nm and only reference structures where more than 80 structures have RMSD lower than 20 nm are considered. This is done to avoid broken structures polluting the interpretation of the mean RMSD. The lowest scoring candidate was selected as the best reference structure.

The average origami structure is calculated straightforwardly by taking the mean position of all measured structures:

$$\bar{x}_j = \frac{1}{N}\sum_{i=1}^{N} x_{i,j} \qquad \bar{y}_j = \frac{1}{N}\sum_{i=1}^{N} y_{i,j} \qquad (S32)$$

where $j$ indicates the dye label as before and the superscript bar denotes the average structure.

The uncertainty in the average structure is taken as the standard error of the mean (SEM):

$$SEM_{\bar{x}_j} = \sqrt{\frac{1}{N(N-1)}\sum_{i=1}^{N}(x_{i,j}-\bar{x}_j)^2} \qquad SEM_{y_j} = \sqrt{\frac{1}{N(N-1)}\sum_{i=1}^{N}(y_{i,j}-\bar{y}_j)^2} \qquad (S33)$$

The origami model and the mean origami position are aligned such that their center-of-masses (COM) and acceptor-acceptor distance unit vector align. The center-of-mass is calculated by weighing each point equally:

$$x_{com} = \frac{1}{4}\sum_{j\in S} x_j \;; \qquad y_{com} = \frac{1}{4}\sum_{j\in S} y_j \qquad (S34)$$

The coordinate system is rotated such that the x-axis matches the strands direction of the origami platform for Fig. 2a-d of the main text. For Fig. 2e, the mean acceptor-acceptor distance has been aligned with the x-axis such that the Cartesian distances are more spread out over the axis and can be represented in a single plot.

## Assessing the diameter of hGBP1 fibers

To estimate the fiber diameter, we selected structures that showed uniform ring borders from measurements of labelled hGBP1-wt diluted with hGBP1-wt in a ratio of 1:10 (see Fig. 6b, c). Images were deconvolved with Huygens Professional version 20.04 (Scientific Volume Imaging, The Netherlands, http://svi.nl), using the CMLE algorithm, with SNR:10 and 40 iterations. In order to determine the fiber width, multiple line profiles across the fiber were drawn and the full width at half maximum (FWHM) was measured. An average fiber diameter of 73 ± 4 nm was obtained (Supplementary Fig. 28). We observed similar fiber diameters for hGBP1 polymers constituted of labelled Cys8 hGBP1-18-577.

## Transmission electron microscopy

Transmission electron microscopy was performed on a JEM-2100Plus (JEOL Ltd., Tokyo, Japan) operating in bright-field mode at 80 kV acceleration voltage. Samples were prepared by applying 7 microliters of the respective aqueous sample dispersion on carbon-coated copper grids (200 mesh, Science Services) for 1 minute. After blotting, grids were placed on a drop of 3% uranyl acetate solution and immediately blotted. Then, the grid was placed again on another drop of uranyl acetate solution, this time for 30 seconds. After blotting the grid was dried for 20 minutes. The dimensions of the DNA origamis were determined using the image analysis software ImageJ (see Supplementary Fig. 8).

## Atomic force microscopy

AFM measurements were performed on a NanoWizard 4 (JPK Instruments AG, Germany) equipped with a temperature regulated liquid cell and using Sharp Nitrile Lever (Bruker, USA) probes. The softest of the four cantilevers (position D) was used. The spring constant was determined as k = 0.04 N/m using the thermal noise method in both air and PBS buffer. Samples were immersed in PBS buffer at 25 °C and equilibrated for 20 min before approaching the probe. The AFM was operated in the quantitative imaging (QI) mode[29], a high-resolution force mapping method, with a setpoint force of 0.5 nN. The standard scan size was 500 nm × 500 nm at a resolution of 512 × 512 px. To characterize the functionalized surfaces over a wide distance range, we took 5 representative images and logarithmic spacing with respect to the origin of the first image (0 μm) for BSA (1, 10, 40 1000 μm) and for PEG (10, 40 ,1000 and 1010 μm). The JPK SPM Data Processing software (v.6.1.142) was used to process the obtained force curves. Briefly, the curves were smoothed and adjusted for baseline offset and slope. The z-position at a force of 50 pN was defined as height and the slope of the last 4 nm of the extend segment was fitted for hardness information. The generated height and slope maps were flattened by 2nd order line levelling and single pixel outliers were removed by interpolation from the surrounding pixel values. Representative images for all conditions are shown in Supplementary Fig. 23.

## Confocal single-molecule spectroscopy with multiparameter fluorescence detection

Single-molecule FRET experiments with pulsed interleaved excitation (PIE) [30] were performed on a homebuilt confocal fluorescence microscope as described previously[31]. The fluorescent donor molecules were excited by a pulsed white light laser source (SuperK Fianium FIU-15 with spectral filter SuperK Varia, NKT Photonics) at 530 nm, operated at 19.5 MHz using a power of 80 μW at the sample. The acceptor molecules are excited by a pulsed diode laser (LDH-D-C 640) operated at 19.5 MHz with 8 μW. The laser light is guided into the epi-illuminated confocal microscope Olympus IX71 (Olympus, Hamburg, Germany) by dichroic beam splitter F68-532_zt532/640NIRrpo (AHF, Germany) focused by a water immersion objective (UPlanSApo 60x/1.2W, Olympus Hamburg, Germany). The emitted fluorescence is collected through the objective, spatially filtered using a pinhole with 100 μm diameter, split into parallel and perpendicular components via a polarizing beam splitter cube (VISHT11, Gsänger

Optoelektronik, Germany) and spectrally split into donor and acceptor channel by a dichroic mirror (T640LPXR, AHF, Germany). Fluorescence emission was filtered (donor: 47-595/50 ET, acceptor: HQ 730/140, AHF, Germany) and focused on avalanche photodiodes (SPCM-AQRH-14-TR, Excelitas). The detector outputs were recorded by a TCSPC module (HydraHarp 400, PicoQuant, Berlin, Germany), using a time resolution of 2 ps. All samples were measured in Nunc chambers (Lab-Tek, Thermo Scientific) with 500 µL sample volume and a concentration of ~50 pM.

Data analysis was performed using home-written LabView software that was developed in the Seidel lab and is described in[31]. It is available upon request on the homepage of the Seidel group (https://www.mpc.hhu.de/software.html). Single-molecule events were identified using a burst search algorithm according to[32] using a Lee filter, a threshold of 0.2 ms and a minimum of 60 photons per burst. Double labelled species were selected via a stoichiometry cut between S=0.3 to S=0.7. See Supplementary Figs. 12 (origami) and 20 (DNA ruler) and main Fig. 6 (hGBP1).

# Supplementary Notes

## Supplementary Note 1: Detailed analysis workflow

Image and fluorescence spectroscopy analysis was performed using home-built software (ANI software and SEIDEL software, respectively) that follows a joint workflow summarized in Supplementary Fig. 6 (the capital letter references in this note refer to the respective panels in the flowchart). In step A, the raw photon data is separated into the donor channel (detected signal in the donor channel after donor excitation), FRET-sensitized acceptor channel (detected signal in the acceptor channel after donor excitation) and acceptor channel after direct excitation (detected signal in the acceptor channel after acceptor excitation) (see Supplementary Fig. 1a and Supplementary Fig. 3)[33]. The images are then segmented into regions of interest (ROIs) for spot fitting (step B1). A Gaussian filter is applied to smooth shot noise and the image is segmented using an intensity threshold. Both an upper and lower threshold criterion can be applied to reject dirt (higher threshold) or increase sensitivity to weak signals (lower threshold). All further analysis is done using the raw data. Next, the intensity data of each ROI is fitted to 2D Gaussian functions (B2). In the single-spot analysis used for the dsDNA sample, spots are fitted to a single 2D Gaussian function (ANI software) and the spot width and localization precision are estimated (B3-4, left). In the multi-spot analysis, multiple 2D Gaussian functions (SEIDEL software) are fitted and the number of spots is determined based on the log-likelihood value while applying a penalty for overfitting (see section 'Spot stoichiometry', B3-4, right). For both approaches, spot centers are used to define the integration area to obtain the spot fluorescence decay and intensity (C1, see also Supplementary Fig. 4). Localization is performed on time-gated data, whereas FRET-informative photons are collected based on ungated data. Analysis settings for each step are reported in Supplementary Tab. 14 and 15. Spots of donors and acceptors are paired (C2) to gain access to intensity-based FRET indicators, such as the FRET efficiency, stoichiometry, and localization-based interdye distance (C3). Dyes are paired by first selecting the pair with the smallest distance and removing them from the available set. This is repeated until there are either no more donors or acceptors available. Supplementary Tab. 18 lists all available spectroscopic and localization parameters. Available FRET and localization-based indicators are used to select a sub-ensemble for further analysis in the filtering step (D and see *'Filtering procedures'*). For quantitative FRET analysis, the lifetime decay histogram is created by accumulating lifetime decays from spots in the sub-ensemble (F1) and fitted in the Chisurf software package (F2)[15]. For the single-spot analysis, the localization-based interdye distance distributions can be analyzed by two approaches: model-based maximum likelihood fitting (E1) or model-free analysis by the maximum-entropy method (MEM) (E2). For the multi-spot analysis, the structures of interest are aligned and particle averaging is performed (E3). The advantage of the latter approach is that it uses both x- and y-coordinate, whereas the other two use only the norm of the distance vector.

# Supplementary Note 2: Estimation of structural parameters of the DNA origami platform

The origami platform is modelled as a rectangular grid that is parametrized by the rise per base pair along the long axis (y) and the interhelical distance along the short axis (x). To avoid potential errors due to the rotation of the dye attachment point around the helical axis, the dyes are attached such that all of them are located at the same position along the helical turn (Supplementary Fig. 7). Here, we assume no interaction of the dye with the origami platform. Hence, the mean positions of the dyes are assumed to correspond to the xy-positions of the attachment points. The distances between the dyes in terms of the structural parameters of the DNA origami platform can then be expressed as:

$$d_{ij} = \sqrt{\left(n_{\mathrm{bp,ij}}d_{\mathrm{bp}}\right)^2 + \left(n_{\mathrm{h,ij}}d_{\mathrm{h}}\right)^2}; \quad i,j \in \{D1, D2, A1, A2\} \tag{S35}$$

where $d_{\mathrm{ij}}$ is the distance between dyes $i$ and $j$, $d_{\mathrm{bp}}$ and $d_{\mathrm{h}}$ are the rise per base pair and the interhelical distance, and $n_{\mathrm{bp,ij}}$ and $n_{\mathrm{h,ij}}$ are the number of base pairs and helices between the dyes. In the global alignment procedure, we are not only measuring the distance between the dyes but instead resolve their relative displacement as xy-coordinate pairs. In the coordinate frame of the DNA origami nanostructure (Fig. 2b and Supplementary Fig. 7), the displacements are defined as:

$$x'_{\mathrm{ij}} = n_{\mathrm{bp,ij}}d_{\mathrm{bp}} \tag{S36}$$

$$y'_{\mathrm{ij}} = n_{\mathrm{h,ij}}d_{\mathrm{h}} \tag{S37}$$

As the rotation of the whole structure is arbitrary, for display purposes it may be either rotated such that the helices are perpendicular to the x-axis (fig 2b-d) or such that the A1A2 vector lies parallel to the x axis (fig 2e). The angle of rotation is given by the angle $\xi$ of the A1A2 vector with respect to the helical axis (x-axis in the origami frame of reference):

$$\tan \xi = \frac{y_{\mathrm{A1A2}}}{x_{\mathrm{A1A2}}} \tag{S38}$$

The coordinate transformation to the experimental coordinate frame is then performed by counterclockwise rotation:

$$\begin{pmatrix} x \\ y \end{pmatrix} = \begin{pmatrix} \cos \xi & -\sin \xi \\ \sin \xi & \cos \xi \end{pmatrix} \begin{pmatrix} x' \\ y' \end{pmatrix} \tag{S39}$$

The measured distances, xy-displacements, their associated uncertainties, and the values for $n_{\mathrm{bp}}$ and $n_{\mathrm{h}}$ for the different dye pairs are given in Supplementary Tab. 19. The model parameters $d_{bp}$ and $d_h$ are fitted by minimizing the chi-squared value $\chi^2$ assuming normally distributed errors in the x- and y-directions, $\sigma_x$ and $\sigma_y$, defined by:

$$\chi^2 = \sum_{i,j} \frac{\left(x_{ij} - x_{ij,meas}\right)^2}{\sigma_x^2} + \frac{\left(y_{ij} - y_{ij,meas}\right)^2}{\sigma_y^2}; \quad i,j \in \{D1, D2, A1, A2\} \tag{S40}$$

To estimate the confidence intervals of the structural parameters, we performed a parameter scan of the $\chi^2$ surface. The confidence interval was obtained by considering the $\chi^2$ distribution. To this end, the number of degrees of freedom in our fit must be determined. There are four average points, each having an x- and y-coordinate, totaling eight observables. During the alignment first the center of mass of the structure is set equal to the center of mass of the alignment anchor, using two degrees of freedom for (x,y). Next, the rotation of the structure is set to minimize the RMSD. The rotation is parametrized by a single value (i.e., the angle) and uses one degree of freedom. Thus, after the alignment procedure, five independent observables are left. As the fit has two parameters, three degrees of freedom are left. The reduced chi-squared value $\chi^2_{red}$ is then obtained as $\chi^2_{red} = \chi^2/3$.

Supplementary Fig. 11 shows the support plane describing the certainty for interhelical distance and average base pair extension. For three degrees of freedom the statistical variance of a good fit is very large (Supplementary Fig. 11b) such that a good model is 95% likely to have $\chi^2_{red}$ values within [0-2.60]. The $\chi^2_{red}$ value for the fit with the highest probability is 0.73, indicating that the model fits the data within the certainty interval. Consequently, the 95% confidence interval on model parameters is obtained from all models whose $\chi^2_{red}$ falls in the 95% likelihood range.

The obtained interhelical distance is 2.41 nm (68% conf. interval 2.30-2.55 nm) and the average extension of 236 base pairs is 75.0 nm (68% conf. interval 74.2-75.8 nm), corresponding to a rise per base pair of 0.318 nm. We also estimated the rise per base pair solely from the acceptor-acceptor distances (see Supplementary Fig. 10), which yielded a similar distance of 74.8 ± 0.3 nm for 236 base pairs and a corresponding rise per base pair of 0.317 nm.

## Supplementary Note 3: Discussion on error sources in origami samples

We calculate the localization precision for the origami datasets based on photon statistics alone to be O(HF+NF) 2.0 nm, O(NF) 3.0 nm, O(HF) 4.6 nm (see Supplementary Fig. 17). The O(HF+NF) sample has the lowest uncertainty due to photon statistics as more frames were accumulated. The estimated localization precision is smaller than the width of the measured localization distributions (O(HF+NF)$_{NF}$ $_{cut}$ 7.4 nm, O(HF+NF)$_{HF\ cut}$ 3.4 nm, O(NF) 7.4 nm, O(HF) 5.8 nm, see Supplementary Fig. 13), which could be interpreted as additional error contributions due to aberrations. However, we argue that the additional broadening is primarily caused by molecule-to-molecule variations and that the true experimental localization precision is close to the fundamental limit based on the photon statistics. This argument is supported by the occurrence of two distinct populations in the acceptor-to-acceptor distance distributions that consist of two peaks, one narrow and the other broad, centered around the same mean value (Supplementary Fig. 10), indicating that small structural defects occur for a fraction of the origami platforms. The same effect should be expected for smaller distances, however, as the distances are on the order of the localization precision, the structural heterogeneity shows mostly as a peak-broadening. Secondly, we occasionally observe very sharp sub-populations in the distance distributions (Supplementary Fig. 13g), which are unlikely to occur by chance.

## Supplementary Note 4: Discussion of discrepancy between localization and FRET distances for origamis

We observed a discrepancy between the localization-based and FRET-based estimates of the interdye distance for the high-FRET dye pair on the origami nanostructures. The $R_{mp}^{FRET}$ for the origami O(HF) pair has been measured using confocal single-molecule spectroscopy and STED nanoscopy to be in the range 72-75 Å using the intensity and lifetime information (Supplementary Tab. 13). The high amount of consistency between solution-based FRET experiments and FRET nanoscopy indicates the capability of FRET nanoscopy to infer accurate distances from the spectroscopic information. On the other hand, the localization-derived distance $R_{mp}^{loc}$ was found to be 53 ± 7 Å as obtained using the global alignment procedure (Supplementary Tab. 19). Additional localization-based distances can be obtained from fitting non-centered $\chi$-distributions (Supplementary Fig. 13 g and h). However, this data provided much lower accuracy due to contribution of broken constructs which could be removed by the global alignment procedure using the RMSD criterion and the unfavorable properties of the $\chi$-distribution. We thus do not consider these results for the further discussion. From the position of the dye attachment points on the origami structure (Supplementary Fig. 7), it is predicted that the interdye distance corresponds to twice the interhelical distance. Using the value obtained in this work, we thus expect an interdye distance of 48 ± 2 Å (Supplementary Fig. 11). This indicates a clear mismatch between the FRET-derived and the predicted distance.

Single-molecule studies revealed an increased acceptor anisotropy for the origami sample compared to the dsDNA rulers (see Supplementary Fig. 14 and Supplementary Tab. 11), which indicates sticking of the dyes due to specific interactions with DNA backbone or bases. Recently, Hübner et al[34] have shown that the dye Atto647N used in this work tends to stick between the helices of the DNA origami platform. Their study was performed on an origami platform identical to ours, with the only exception that our linker includes an additional unpaired thymine base. The increased anisotropy was only observed for the origami sample with a rotational correlation time of 10 ns, while for the DNA ruler a significantly shorter rotational correlation time of 2 ns was obtained that indicates the absence of sticking interactions (Supplementary Fig. 14). This suggests that the dye-DNA interaction is facilitated by the close proximity of the DNA helices in the origami nanostructure. On the other hand, the donor dye does not exhibit different anisotropies between the dsDNA and origami environment and free rotation can be assumed.

Accurate FRET distances are obtained under the assumption that dyes can freely rotate, such that the $\kappa^2$ value used for the calculation of the Förster radius equals 2/3. However, this assumption no longer holds when the dye is stuck in a specific orientation and the real distance can be larger or smaller depending on the mutual orientation of the transition dipole moments of the donor and acceptor dyes. It is possible to estimate the error on the FRET distance by computing minimal and maximal $\kappa^2$ values given anisotropy values for the donor and acceptor[35] and propagating this error into the uncertainty of $R_{DA}$. Here we integrate over the possible orientations of the donor and acceptor transition dipole moments subject to constraints given by the measured residual anisotropies (Supplementary Tab. 11) to predict the mean and variance of the $\kappa^2$ distribution. These simulations were performed using the ChiSurf software package (https://github.com/Fluorescence-Tools/chisurf)[15]. $\kappa^2$ is found to be in the range [0.33, 1.03] with a mean of 0.63, resulting in an estimated Förster radius of 75.8 Å with an uncertainty of ± 6 Å. We propagate the uncertainty of the Förster radius to the FRET-derived mean positions distance based on the sub-ensemble fluorescence decay analysis of 73 ± 3 Å (Supplementary Fig. 13d, Supplementary Tabs 9 and 13), yielding a final estimate of $R_{mp}^{FRET}$ from the O(HF+NF)$_{highFRET}$ cut sub-ensemble fit of 72 ± 7 Å.

Based on the measured mean-position distances from the localization of $R_{mp}^{loc}$ = 53 ± 7 Å and FRET of $R_{mp}^{FRET}$ = 72 ± 7 Å, corresponding to the projected distance in the xy-plane and the isotropic distance, respectively, we can estimate the corresponding z-displacement between the donor and acceptor fluorophore, $R_z$, using the Pythagorean theorem according to:

$$R_z = \sqrt{R_{mp,FRET}^2 - R_{mp,loc}^2} \qquad (S41)$$

We thus estimate a predicted z-displacement $R_z$ of 50 ± 12 Å from the experimental measurements of the projected and isotropic distances.

The proposed three-dimensional arrangement of the dyes on the origami nanostructure is schematically displayed in Supplementary Fig. 15. The acceptor is positioned between the helices, while the donor is pointing upwards away from the origami surface. The acceptor fluorophore (red) is stuck between two helices pointing away from the donor fluorophore, while the donor fluorophore (orange) is free to diffuse and rotate. The mean position of the donor (orange circle) in its AV (light orange) is assumed to be slightly tilted towards the acceptor position to satisfy the localization-based distance $R_{mp}^{loc}$ of 53 Å, which is shorter than the distance expected for a distance of 2.5 helices of 60 Å based on an interhelical spacing of 24 Å as determined from the global alignment. The estimated z-displacement R$_z$ between the fluorophores of 50 Å matches the length of the dye linker of ~29 Å plus the diameter of the DNA helix of ~20 Å. Correspondingly, the acceptor is assumed to penetrate through the interhelical space, placing its center on the lower side of the origami.

## Supplementary Note 5: Assessment of sticking/unsticking dynamics of dsDNA rulers during acquisition

To assess the occurrence of sticking/unsticking dynamics during the acquisition time (i.e., "jumping" of one end of the DNA rulers), we performed repeated localizations over the course of the measurement. We used the single- and double-biotin dsD(NF) sample due to the absence of energy transfer between the dyes. We recorded 60 frames in total, split the measurement into four intervals 15 frames each, and selected only those molecules, which had sufficient signal in all frame intervals to exclude photobleaching by requiring at least 50 photons both in the donor and acceptor channels. In addition, only molecules with an interdye distance (as estimated from all 60 frames) of less than 30 nm were considered. Each frame interval was processed using the cSTED workflow, yielding the xy-coordinates of the center position of the donor and acceptor and the interdye distance. We visualize the distance vector of each frame interval in a Cartesian plot (Supplementary Fig. 22a). No large jumps were observed and the localizations from the four frame intervals are found to cluster in one region of the plot, indicating that there are no "jumps" during the measurement for both the single-biotin and double-biotin samples.

We additionally quantify the fluctuations of the localization $\sigma_{dyn}$ from the standard deviations of the localizations in the x- and y-directions ($\sigma_{dyn,x}$ and $\sigma_{dyn,y}$) as:

$$\sigma_{dyn} = \sqrt{\sigma_{dyn,x}^2 + \sigma_{dyn,y}^2} \qquad (S42)$$

Similar distributions of $\sigma_{dyn}$ are observed for the single-biotin and double-biotin samples, indicating that no large jumps occur for either immobilization strategy (Supplementary Fig. 22). If jumping were to occur during the measurement, a high value for $\sigma_{dyn}$ should also be correlated with the observation of a seemingly shorter colocalization based interdye distance. However, no correlation is observed between the parameter $\sigma_{dyn}$ and the colocalization-based interdye distance and identical distance distributions are obtained for large and small values of $\sigma_{dyn}$. In the absence of dynamics, $\sigma_{dyn}$ effectively measures the localization precision for an acquisition time of 15 frames. Indeed, the estimated values for $\sigma_{dyn}$ are close to the shot-noise limited localization precision of ~6 nm for the reduced number of photons collected over the shorter acquisition time, indicating that the spread of the repeated localizations is primarily caused by the localization error.

## Supplementary Note 6: Analysis of dsDNA ruler distance distributions by the maximum entropy method

The maximum entropy method (MEM) is an approach to extract the most unbiased distribution of a given parameter that provides a satisfactory fit to the experimental data[36-38]. Instead of minimizing the reduced chi-square, $\chi_r^2$, the following functional is maximized:

$$\Theta = \nu S - \chi_r^2 \qquad (S43)$$

where $\nu$ is a constant scaling factor and $S$ is the entropy functional of the parameter distribution. The entropy is defined by:

$$S = -\sum_i p_i \log \frac{p_i}{m_i} \qquad (S44)$$

where $p_i$ describes the distribution of the parameter of interest and $m_i$ describes the prior knowledge of the distribution. Since we have no prior knowledge, we use a flat prior in the analysis.

We describe the experimental histograms of the colocalization distance $H(d_{loc})$ as a superposition of non-central $\chi$-distributions $P_\chi$ as defined in the section 'Colocalization analysis' (eq. S30) with fixed width parameter $\sigma_\chi$:

$$H(d_{loc}) = \sum_i p\left(R_{mp}^{loc,(i)}\right) \cdot P_\chi\left(d_{loc} \mid R_{mp}^{loc,(i)}, \sigma_\chi\right) \qquad (S45)$$

where the set of kernel functions $\left\{P_\chi\big(d_{\mathrm{loc}}\big|R_{\mathrm{mp}}^{\mathrm{loc,(i)}}, \sigma_\chi\big), \; i = 1 \ldots N\right\}$ is defined over the mean-position distance $R_{\mathrm{mp}}^{\mathrm{loc}}$. Here, we used a fixed value of $\sigma_\chi = 4.4$ nm that was estimated from the distribution of acceptor-acceptor distances for the origami data (see Supplementary Fig. 10).

The reduced chi-squared $\chi_r^2$ is then defined as:

$$\chi_r^2 = \frac{1}{K}\sum_k \frac{1}{w_k^2}\Big(H\big(d_{\mathrm{loc}}^{(\mathrm{k})}\big) - M\big(d_{\mathrm{loc}}^{(\mathrm{k})}\big)\Big)^2 \qquad (S46)$$

where $M$ is the measured histogram, $K$ is the number of bins on the histogram and $w_k$ are the weights of data points given by $w_k = \sqrt{M\big(d_{\mathrm{dloc}}^{(\mathrm{k})}\big)}$ for Poisson counting statistics.

Maximization of $\Theta$ is performed as described in Vinogradov and Wilson[39] over a wide range of values for the regularization parameter $v$. The choice of the regularization parameter $v$ was done by visual inspection of the L-curve plot of the negative entropy $-S$ against the reduced chi-squared $\chi_r^2$, which provided a more robust selection of $v$ compared to corner detection algorithms[40]. The visual analysis yielded a value of $v = 0.1$ that was used for all analyses of the DNA ruler datasets (see Supplementary Fig. 18).

## Supplementary Note 7: Model-based analysis of colocalization distance histograms of dsDNA rulers

In addition to the maximum entropy-based analysis, we also performed a model-based analysis of the distance distribution obtained for the dsDNA rulers (see section 'Colocalization analysis'). However, as no reliable distance estimates could be obtained, we preferred the model-free approach employing the maximum entropy method to infer the underlying distance heterogeneity.

Neither the single-biotin nor the double-biotin samples could be described by a single-component $\chi$-distribution model function using the theoretically predicted localization precision based on the photon statistics, corresponding to a width parameter $\sigma_\chi$ of 4 nm. We thus reasoned that the single-biotin sample experiences excess heterogeneity due to partial sticking to the surface that would lead to a larger effective width of the distance distribution. Indeed, by letting the width $\sigma_\chi$ vary, we could achieve a good fit for all single-biotin samples (Supplementary Tab. 12), with mean-position distances $R_{mp}$ close to zero as would be expected for a singly immobilized DNA ruler. We also tested a single-component model for the description of the broad distance distributions obtained for the double-biotin samples, which provided a poor fit but captured the trend of increasing distances with larger dye separation.

The observation of short colocalization-based distances for the double-biotin sample indicates a potential contamination by singly immobilized molecules, which could be as high as 60% as discussed in Supplementary Note 8. We thus assumed that a fraction of the molecules is immobilized only on one end and behaves like single-biotin molecules. The remaining fraction of molecules are assumed to be immobilized on both ends, thus lying flat on the surface and exhibiting shot-noise limited broadening of the distance distribution ($\sigma_\chi = 4.4$ nm). While the fit provided a good description of the data, the determined distances for the doubly immobilized population showed large deviations from the expected mean-position distances (Supplementary Tab. 12). We also considered an equivalent model where we allowed the width of the doubly immobilized population to vary, which was also unable to provide reliable distances.

## Supplementary Note 8: Estimation of surface roughness and neutravidin density on functionalized surfaces

To immobilize molecules on the surface, we use BSA or PEG-functionalized surfaces and biotin-neutravidin linkage. In addition to the inherent roughness of the BSA or PEG surfaces, neutravidin itself has a size of ~4 nm, comparable to the length of the DNA rulers (18.7 nm, Fig. 4d). The density of neutravidin on the surface is also limited, such that not all doubly biotin labelled DNA rulers are bound on both ends. The DNA origami platforms, on the other hand, possess 8 biotin anchors on their lower side, increasing the chance that at least two biotins are bound, and are much larger and hence less sensitive to local height differences present on the surface. These effects are reflected in our data: while the dsDNA rulers showed shorter than expected distances, indicating that they are angled with respect to the surface, the dye-to-dye distances on the origami platform were found within the expected range, indicating that they lie flat on the surface.

PEG-3000 (average molecular weight of 3 kDa, corresponding to ~70 monomer units) is covalently bound to the surface on one side and functionalized with biotin on the other side. PEG is a flexible polymer chain and considerably smaller than a typical protein. BSA, on the other hand, is a large protein (66 kDa) with a high propensity to stick to glass surfaces. In principle it can build multi-layers of loosely associated proteins. Each BSA molecule is labelled with up to twelve biotin molecules, enabling potential BSA-biotin-BSA crosslinking. In this section, we will investigate the density of neutravidin on the surface using fluorescence and the surface roughness using AFM imaging.

We first estimated the neutravidin concentration on the PEG surface using fluorescently labeled biotin (Supplementary Fig. 24). To assess the fraction of single and double bound DNA molecules, we estimate the surface area that the unbound biotin can explore. To this end, the DNA ruler is considered as a rigid rod of 17 nm length with a flexible 1.5 nm linker with biotin on either side. The linker length is obtained from chemical bond lengths of a C6 linker attached to a nucleic acid[35,41] the size of biotin itself is excluded as it has a fixed conformation in the neutravidin binding pocket. The singly bound molecule can freely rotate around its anchor point. The maximal extension between the terminally attached biotins occurs when both linkers are pointing outward, resulting in 21 nm. The minimal extension occurs when both linkers are pointing inwards, resulting in 13 nm. The area accessible for the unbound biotin is thus:

$$A_{\text{biot}} = \pi(21^2 - 13^2) \text{ nm}^2 = 6.4 \cdot 10^2 \text{ nm}^2 \qquad (S47)$$

To estimate the neutravidin density on the surface, we add Atto647N-biotin in low concentrations to the surface (Supplementary Fig. 24a) to determine the brightness of individual molecules. Next, we titrate the Atto647N concentration until the surface is saturated. Individual spots are no longer visible, so the brightness is used to determine the number of fluorophores. The binding affinity of neutravidin to biotin is high ($K_d$ = 10$^{-15}$ M)[42], hence we assume that the number of fluorophores equals the number of neutravidin binding sites, resulting in a characteristic area per neutravidin molecule $A_{\text{neutr}}$ of $1.1 \cdot 10^3 \ nm^2$

The average number of neutravidin molecules in the accessible area can be calculated as:

$$N_{\text{neutr}} = \frac{A_{\text{biot}}}{A_{\text{neutr}}} = \frac{6.4 \cdot 10^2 \ nm^2}{1.1 \cdot 10^3 \ nm^2} = 0.58 \qquad (S48)$$

When a DNA ruler is observed on the surface, it is given that it has at least a single bond. The probability to find at least one other (second) neutravidin molecule in the accessible area follows Poisson statistics and is given by:

$$p(\text{double bound}) = 1 - e^{-N_{\text{neutr}}} = 44 \ \% \qquad (S49a)$$

The standard error of the neutravidin surface density was estimated as 80 $nm^2$ from multiple independent surface preparations. We additionally consider an uncertainty of the effective linker length of 0.5 nm. By recalculating the accessible area for an effective linker of 1 and 2nm we obtain an estimated uncertainty of the accessible area of 210 $nm^2$. Using standard error propagation, we obtain for the error of the double-bound fraction:

$$\Delta p(\text{double bound}) = N_{\text{neutr}} e^{-N_{\text{neutr}}} \sqrt{\left(\frac{\Delta A_{\text{biot}}}{A_{\text{biot}}}\right)^2 + \left(\frac{\Delta A_{\text{neutr}}}{A_{\text{neutr}}}\right)^2} = 0.11 \qquad (S49b)$$

The likelihood that a DNA ruler is doubly bound is thus obtained as 44 ± 11 %. In the inferred distance distributions of the DNA rulers, however, we found a significant non-zero population also for single-biotin immobilization, and generally observed only a small fraction of short (zero) distances for the double-biotin samples (Fig. 5a-b and Supplementary Fig. 18a-b). This indicates that DNA rulers might also be bound to the surface by unspecific interactions (sticking).

Next, we assessed the surface roughness of the functionalized surfaces by AFM. The PEG-neutravidin surfaces showed height variations that are comparable to the size of individual neutravidin proteins (Supplementary Fig. 23a), causing the DNA rulers to be slightly angled with respect the surface as evidenced by the observed shorter distance in the colocalization analysis (Supplementary Fig. 18). On BSA-functionalized surfaces, we observed height variations of 10-30 nm (Supplementary Fig. 23b), much larger than for the PEG surfaces. Indeed, we observed significantly reduced colocalization distances in the inferred distributions and determined a larger inclination angle of 25° compared to 20° for the PEG surfaces (Fig 5b, main text). For BSA surfaces, a variation of the height profile is observed over distances of 10 nm (Fig. S23b, right column). This indicates that the height profile varies between surface preparations and potentially explains why for some samples a larger difference between the observed and expected distance was found (compare e.g. the dsD(LF) and dsD(NF) in Fig. 5b of the main text). In absence of neutravidin the surfaces are very flat (compare Supplementary Fig. 23 a-c and b-d).

To get the average dimension of surface features, an autocorrelation of the surfaces for the PEG-Neutravidin and BSA-Neutravidin was done (see Fig. 5e and supplementary Fig. 23e, f). Similar to image correlation spectroscopy (ICS) techniques, first the fluctuations around the mean height were obtained,

$$dIm\,(x,y) = Im(x,y) - \langle Im \rangle, \qquad (50a)$$

where $Im(x, y)$ denotes the AFM image and $\langle Im \rangle$ is the average height. The autocorrelation is calculated normally:

$$C(x, y) = \int \int dIm(x', y') dIm(x' - x, y' - y) dx' dy'.$$  $(50b)$

Taking $xc, yc$ as the center of the correlation, and the distance from the center $r = \sqrt{(x - xc)^2 + (y - yc)^2}$, we radially integrate to obtain the 1D correlation function:

$$\tilde{C_{1D}}(r) = \frac{\int \int \delta(r^2 - (x - xc)^2 - (y - yc)^2) C(x, y) dx dy}{2\pi r},$$  $(50c)$

where the division by $2\pi r$ is to divide out the Jacobian of the transformation. Finally, normalization occurs by:

$$C_{1D}(r) = \frac{\tilde{C_{1D}}(r)}{\max\left(\tilde{C_{1D}}(r \neq 0)\right)},$$  $(50d)$

where the $r = 0$ value is skipped as it is not indicative of the average feature size, but rather indicates the variance of the dataset. The characteristic distance is chosen by 1 / e. The average feature size for PEG is found to be 16 nm and for BSA 24 nm. This matches the expectation that the resolution of the AFM images is limited by the AFM tip size, features smaller than this distance cannot be imaged.

## Supplementary Note 9: Estimation of inclination angles of dsDNA rulers

To estimate the inclination angle $\alpha$, we perform a linear regression of the measured localization-based mean-position distance $R_{\text{mp}}^{\text{loc}}$ against the predicted distance between dyes for a flat orientation of the dsDNA $R_{\text{mp}}^{\text{(flat)}}$ according to:

$$R_{\text{mp}}^{\text{loc}} = m \cdot R_{\text{mp}}^{\text{(flat)}}.$$  $(S51)$

From the slope $m$, the inclination angle $\alpha$ is then estimated as:

$$\alpha = arcsin(m),$$  $(S52)$

where $arcsin$ is the inverse sine function.

To reflect the broadening of the observed distance distribution, the uncertainties of the average inclination angles of the single biotin and double biotin samples ($\Delta \alpha$) are obtained by propagating the standard deviation of the distance distributions $\Delta R_{mp}^{loc}$ according to:

$$\Delta \alpha = \frac{d\alpha}{dR_{\text{mp}}^{\text{loc}}} \Delta R_{\text{mp}}^{\text{loc}} = \frac{\Delta R_{\text{mp}}^{\text{loc}}}{R_{\text{mp}}^{\text{(flat)}}} \left( 1 - \left( \frac{R_{\text{mp}}^{\text{loc}}}{R_{\text{mp}}^{\text{(flat)}}} \right)^2 \right)^{-\frac{1}{2}}$$  $(S53)$

$\Delta R_{\text{mp}}^{\text{(exp)}}$ is calculated from the distance distribution $p(R_{\text{mp}}^{\text{loc}})$ obtained from the MEM analysis by:

$$\Delta R_{\text{mp}}^{\text{loc}} = \sqrt{\langle R_{\text{mp}}^{\text{loc}^2} \rangle - \langle R_{\text{mp}}^{\text{loc}} \rangle^2}$$  $(S54)$

where $\langle R_{\mathrm{mp}}^{\mathrm{loc}}{}^2 \rangle = \sum p(R_{mp,i}^{loc}) R_{mp,i}^{loc}{}^2$ and $\langle R_{\mathrm{mp}}^{\mathrm{loc}} \rangle = \sum p(R_{\mathrm{mp,i}}^{\mathrm{loc}}) R_{\mathrm{mp,i}}^{\mathrm{loc}}$. For each measurement series (single or double biotin on BSA or PEG surface), the uncertainty $\Delta\alpha$ is estimated for each measurement of dsD(HF), dsD(MF), dsD(LF) and dsD(NF), and subsequently averaged to obtain the uncertainty of the reported inclination angle. For the estimates of the inclination angles of the double-biotin sample based on the peak distance of the double-biotin population, the reported standard error is obtained from the covariance matrix of the fit using the *curve_fit* function of the SciPy package for Python. See main Fig. 5 and Supplementary Fig. 18.

## Supplementary Note 10: Colocalization analysis of hGBP1

In this section, we describe the model for the interdye distance distributions obtained by cSTED for hGBP1. Two effects need to be accounted for. First, due to arrangement of extended hGBP1 molecules in the ring-like assemblies, a distribution of the projected distance is observed even for a fixed interdye distance. Second, the random distribution of labeled hGBP1 molecules over the fiber rings leads to false-positive pairing of fluorophores attached to different hGBP1 molecules.

For the first part, we consider a uniform distribution of the inclination angle $\alpha$, i.e., $P(\alpha) = const$. The angle $\alpha$ defines the projected interdye distance $R_{\mathrm{mp}}^{\mathrm{loc}}$ as a function of the radius of the disc $R_{\mathrm{disc}}$ by (see schematic in Fig. 6e of the main text):

$$R_{\mathrm{mp}}^{\mathrm{loc}} = R_{\mathrm{disc}} \cos\alpha \qquad (S55)$$

The corresponding distribution of the projected distance in the disc is then obtained as:

$$P_{\mathrm{disc}}(R_{\mathrm{mp}}^{\mathrm{loc}} | R_{\mathrm{disc}}) = \mathrm{P}(\alpha) \frac{\mathrm{d}\alpha(R_{\mathrm{mp}}^{\mathrm{loc}})}{\mathrm{d}R_{\mathrm{mp}}} = \frac{2}{\pi} \left( 1 - \left( \frac{R_{\mathrm{mp}}^{\mathrm{loc}}}{R_{\mathrm{disc}}} \right)^2 \right)^{-\frac{1}{2}} \qquad (S56)$$

The baseline due to random nearest-neighbor pairing of separate donor- or acceptor-only hGBP1 molecules is evaluated based on a Monte-Carlo approach (Supplementary Fig. 29b-c). Considering the diameter and number of localizations of donor and acceptor dyes within each measured ring, the spots are randomly redistributed over the ring and nearest neighbor pairing is performed as for the data. To eliminate stochastic noise and obtain a smooth distribution, the simulation is performed 1000 times.

To obtain the final model function for the colocalization analysis of hGBP1, one has to again consider the inherent distribution of the measured interdye distance $d_{loc}$ due to the localization uncertainty, described by a $\chi$-distribution. The model function is then given by:

$$M_{R_{\mathrm{disc}}, \sigma_\chi}(d_{\mathrm{loc}}) = \int_0^\infty \left( A_{\mathrm{BL}} BL(R_{\mathrm{mp}}^{\mathrm{loc}}) + A_{\mathrm{disc}} P_{\mathrm{disc}}(R_{\mathrm{mp}}^{\mathrm{loc}} | R_{\mathrm{disc}}) \right) P_\chi(R_{\mathrm{loc}}, \sigma_\chi | R_{\mathrm{mp}}^{\mathrm{loc}}) dR_{\mathrm{mp}}^{\mathrm{loc}} \qquad (S57)$$

where $BL(R_{mp})$ describes the shape of the baseline of randomly distributed spots on the ring, $P_\chi(R_{\mathrm{loc}}, \sigma_{chi} | R_{\mathrm{mp}}^{\mathrm{loc}})$ is the $\chi$-distribution with two degrees of freedom with the central parameter $R_{mp}$ and width parameter $\sigma_\chi$, and $A_{\mathrm{BL}}$ and $A_{\mathrm{disc}}$ are the corresponding amplitudes.

To generate the reported histogram of the colocalization distances in Fig. 6h, only those spots with a proximity ratio (uncorrected FRET efficiency calculated from the raw signals) above 0.35 were selected (Supplementary Fig. 29a). The amplitude of the baseline was fitted only for long distances (> 50 nm) and fixed for the fit over the whole distance range using the full model function (Fig. 6h, main text). 68% confidence intervals were determined from the marginal probability distribution of the disc radius $R_{\mathrm{disc}}$ obtained from a two-dimensional parameter scan of the fit parameters $R_{\mathrm{disc}}$ and $\sigma_\chi$ (Fig. 6i).

# Supplementary Tables

## Sample information

***Supplementary Tab. 1: Overview of DNA sequences for the DNA rulers (dsD).*** The donor strand (D-strand) is labeled with Alexa594 dye and acceptor strand (A-strand) with Atto647N dye (for details see 'Sample preparation, dsDNA). The labeling sites of the donor and acceptor are shown by green and in red bases in the sequence, respectively.

| Sample | Base position (Linker), strand | # Biotin-anchor | Dyes (Donor/Acceptor) | Sequence |
|---|---|---|---|---|
| dsD(HF) | T 6 (C6), D-strand C 30(C6), A-strand | 1 | Alexa594 / Atto647N | 5´-Biotin - CGT ACT GAT TAA TCT CCG CAA ATG TGA ACG CGT ACT GAT TAA TCT CCG CAA ATG T - 3´<br>5' - A CAT TTG CGG AGA TTA ATC AGT ACG CGT TCA CAT TTG CGG AGA TTA ATC AGT ACG - 3´ |
| dsD(MF) | T 6 (C6), D-strand C 24(C6), A-strand | 1 | Alexa594 / Atto647N | 5´-Biotin - CGT ACT GAT TAA TCT CCG CAA ATG TGA ACG CGT ACT GAT TAA TCT CCG CAA ATG T - 3´<br>5' - ACA TTT GCG GAG ATT AAT CAG TAC GCG TTC ACA TTT GCG GAG ATT AAT CAG TAC G - 3' |
| dsD(LF) | T 6 (C6), D-strand T 19(C6), A-strand | 1 | Alexa594 / Atto647N | 5´-Biotin - CGT ACT GAT TAA TCT CCG CAA ATG TGA ACG CGT ACT GAT TAA TCT CCG CAA ATG T - 3´<br>5' – ACA TTT GCG GAG ATT AAT CAG TAC GCG TTC ACA TTT GCG GAG ATT AAT CAG TAC G - 3´ |
| dsD(NF) | T 6 (C6), D-strand T 6 (C6), A-strand | 1 | Alexa594 / Atto647N | 5´-Biotin - CGT ACT GAT TAA TCT CCG CAA ATG TGA ACG CGT ACT GAT TAA TCT CCG CAA ATG T - 3´<br>5' – ACA TTT GCG GAG ATT AAT CAG TAC GCG TTC ACA TTT GCG GAG ATT AAT CAG TAC G - 3´ |
| dsD(HF) | T 6 (C6), D-strand C 30(C6), A-strand | 2 | Alexa594 / Atto647N | 5´-Biotin - CGT ACT GAT TAA TCT CCG CAA ATG TGA ACG CGT ACT GAT TAA TCT CCG CAA ATG T - 3´-Biotin<br>5' - A CAT TTG CGG AGA TTA ATC AGT ACG CGT TCA CAT TTG CGG AGA TTA ATC AGT ACG - 3´ |
| dsD(MF) | T 6 (C6), D-strand C 24(C6), A-strand | 2 | Alexa594 / Atto647N | 5´-Biotin - CGT ACT GAT TAA TCT CCG CAA ATG TGA ACG CGT ACT GAT TAA TCT CCG CAA ATG T – 3'-Biotin<br>5' - ACA TTT GCG GAG ATT AAT CAG TAC GCG TTC ACA TTT GCG GAG ATT AAT CAG TAC G - 3' |
| dsD(LF) | T 6 (C6), D-strand T 19(C6), A-strand | 2 | Alexa594 / Atto647N | 5´-Biotin - CGT ACT GAT TAA TCT CCG CAA ATG TGA ACG CGT ACT GAT TAA TCT CCG CAA ATG T - 3´-Biotin<br>5' – ACA TTT GCG GAG ATT AAT CAG TAC GCG TTC ACA TTT GCG GAG ATT AAT CAG TAC G - 3´ |
| dsD(NF) | T 6 (C6), D-strand T 6 (C6), A-strand | 2 | Alexa594 / Atto647N | 5´-Biotin - CGT ACT GAT TAA TCT CCG CAA ATG TGA ACG CGT ACT GAT TAA TCT CCG CAA ATG T - 3´-Biotin<br>5' – ACA TTT GCG GAG ATT AAT CAG TAC GCG TTC ACA TTT GCG GAG ATT AAT CAG TAC G - 3´ |

**Supplementary Tab. 2: Sequences for staple strands of the DNA origami platform**

| Name | Sequence | Modification |
|------|----------|--------------|
| A1 | TTTTCACTCAAAGGGCGAAAAACCATCACC | |
| A2 | GTCGACTTCGGCCAACGCGCGGGGTTTTTC | |
| A3 | TGCATCTTTCCCAGTCACGACGGCCTGCAG | |
| A4 | TAATCAGCGGATTGACCGTAATCGTAACCG | |
| A5 | AACGCAAAATCGATGAACGGTACCGGTTGA | |
| A6 | AACAGTTTTGTACCAAAAACATTTTATTTC | |
| A7 | TTTACCCCAACATGTTTTAAATTTCCATAT T | 3' Atto647N |
| A8 | TTTAGGACAAATGCTTTAAACAATCAGGTC T | 3' Atto594 |
| A9 | CATCAAGTAAAACGAACTAACGAGTTGAGA | |
| A10 | AATACGTTTGAAAGAGGACAGACTGACCTT | |
| A11 | AGGCTCCAGAGGCTTTGAGGACACGGGTAA | |
| A12 | AGAAAGGAACAACTAAAGGAATTCAAAAAA | |
| B1 | CAAATCAAGTTTTTTGGGGTCGAAACGTGGA | |
| B2 | CTCCAACGCAGTGAGACGGGCAACCAGCTGCA | |
| B3 | TTAATGAACTAGAGGATCCCCGGGGGGTAACG | |
| B4 | CCAGGGTTGCCAGTTTGAGGGGACCCGTGGGA | |
| B5 | ACAAACGGAAAAGCCCCAAAAACACTGGAGCA | |
| B6 | AACAAGAGGGATAAAAATTTTTAGCATAAAGC | |
| B7 | TAAATCGGGATTCCCAATTCTGCGATATAATG | |
| B8 | CTGTAGCTTGACTATTATAGTCAGTTCATTGA | |
| B9 | ATCCCCCTATACCACATTCAACTAGAAAAATC | |
| B10 | TACGTTAAAGTAATCTTGACAAGAACCGAACT | |
| B11 | GACCAACTAATGCCACTACGAAGGGGGTAGCA | |
| B12 | ACGGCTACAAAAGGAGCCTTTAATGTGAGAAT | |
| C1 | AGCTGATTGCCCTTCAGAGTCCACTATTAAAGGGTGCCGT | |
| C2 | ATTAAGTTTACCGAGCTCGAATTCGGGAAACCTGTCGTGC | 5' biotin |
| C4 | GTATAAGCCAACCCGTCGGATTCTGACGACAGTATCGGCCGCAAGGCG | |
| C5 | TATATTTTGTCATTGCCTGAGAGTGGAAGATT | |
| C6 | GATTTAGTCAATAAAGCCTCAGAGAACCCTCA | |
| C7 | CGGATTGCAGAGCTTAATTGCTGAAACGAGTA | |
| C8 | ATGCAGATACATAACGGGAATCGTCATAAATAAAGCAAAG | |
| C9 | ATAAGGGAACCGGATATTCATTACGTCAGGACGTTGGGAA | 5' biotin |
| C11 | TTTATCAGGACAGCATCGGAACGACACCAACCTAAAACGAGGTCAATC | |
| C12 | ACAACTTTCAACAGTTTCAGCGGATGTATCGG | |
| D1 | AAAGCACTAAATCGGAACCCTAATCCAGTT | |
| D2 | TGGAACAACCGCCTGGCCCTGAGGCCCGCT | |
| D3 | TTCCAGTCGTAATCATGGTCATAAAAGGGG | |
| D4 | GATGTGCTTCAGGAAGATCGCACAATGTGA | |
| D5 | GCGAGTAAAAATATTTAAATTGTTACAAAG | |
| D6 | GCTATCAGAAATGCAATGCCTGAATTAGCA | |
| D7 | AAATTAAGTTGACCATTAGATACTTTTGCG | |
| D8 | GATGGCTTATCAAAAAGATTAAGAGCGTCC | |
| D9 | AATACTGCCCAAAAGGAATTACGTGGCTCA | |
| D10 | TTATACCACCAAATCAACGTAACGAACGAG | |
| D11 | GCGCAGACAAGAGGCAAAAGAATCCCTCAG | |
| D12 | CAGCGAAACTTGCTTTCGAGGTGTTGCTAA | |
| E1 | AGCAAGCGTAGGGTTGAGTGTTGTAGGGAGCC | |
| E2 | CTGTGTGATTGCGTTGCGCTCACTAGAGTTGC | |
| E3 | GCTTTCCGATTACGCCAGCTGGCGGCTGTTTC | |
| E4 | ATATTTTGGCTTTCATCAACATTATCCAGCCA | |
| E5 | TAGGTAAACTATTTTTGAGAGATCAAACGTTA | |
| E6 | AATGGTCAACAGGCAAGGCAAAGAGTAATGTG | |
| E7 | CGAAAGACTTTGATAAGAGGTCATATTTCGCA | |
| E8 | TAAGAGCAAATGTTTAGACTGGATAGGAAGCC | |
| E9 | TCATTCAGATGCGATTTTAAGAACAGGCATAG | |
| E10 | ACACTCATCCATGTTACTTAGCCGGAAAGCTGC | |
| E11 | AAACAGCTTTTTGCGGGATCGTCAACACTAAA | |
| E12 | TAAATGAATTTTCTGTATGGGATTAATTTCTT | |
| F1 | CCCGATTTAGAGCTTGACGGGGAAAAAGAATA | |
| F2 | GCCCGAGAGTCCACGCTGGTTTGCAGCTAACT | |
| F3 | CACATTAAAATTGTTATCCGCTCATGCGGGCC | |
| F4 | TCTTCGCTGCACCGCTTCTGGTGCGGCCTTCC | |
| F5 | TGTAGCCATTAAAATTGCGATTAAATGCCGGA | |
| F6 | GAGGGTAGGATTCAAAAGGGTGAGACATCCAA | |
| F7 | TAAATCATATAACCTGTTTAGCTAACCTTTAA | |
| F8 | TTGCTCCTTTCAAATATCGCGTTTGAGGGGGT | |
| F9 | AATAGTAAACACTATCATAACCCTCATTGTGA | |
| F10 | ATTACCTTTGAATAAGGCTTGCCCAAATCCGC | |
| F11 | GACCTGCTCTTTGACCCCCAGCGAGGGAGTTA | |
| F12 | AAGGCCGCTGATACCGATAGTTGCGACGTTAG | |
| G1 | CCCAGCAGGCGAAAAATCCCTTATAAATCAAGCCGGCG | |
| G2 | GCGATCGGCAATTCCACACAACAGGTGCCTAATGAGTG | 5' biotin |
| G4 | TAAATCAAAATAATTCGCGTCTCGGAAACCAGGCAAAGGGAAGG | |
| G5 | GAGACAGCTAGCTGATAAATTAATTTTTGT | |
| G6 | TTTGGGGATAGTAGTAGCATTAAAAGGCCG | |
| G7 | GCTTCAATCAGGATTAGAGAGTTATTTTCA | |
| G8 | CGTTTACCAGACGACAAAGAAGTTTTGCCATAATTCGA | |
| G9 | TTGTGTCGTGACGAGAAACACCAAATTTCAACTTTAAT | 5' biotin |
| G11 | TGACAACTCGCTGAGGCTTGCATTATACCAAGCGCGATGATAAA | |
| G12 | TCTAAAGTTTTGTCGTCTTTCCAGCCGACAA | |
| H1 | TCAATATCGAACCTCAAATATCAATTCCGAAA T | |
| H2 | GCAATTCACATATTCCTGATTATCAAAGTGTA | |
| H3 | AGAAAACAAAGAAGATGATGAAACAGGCTGCG | |
| H4 | ATCGCAAGTATGTAAATGCTGATGATAGGAAC | |
| H5 | GTAATAAGTTAGGCAGAGGCATTTATGATATT | |
| H6 | CCAATAGCTCATCGTAGGAATCATGGCATCAA | |
| H7 | AGAGAGAAAAAATGAAAATAGCAAGCAAACT | |
| H8 | TTATTACGAAGAACTGGCATGATTGCGAGAGG | |
| H9 | GCAAGGCCTCACCAGTAGCACCATGGGCTTGA | |
| H10 | TTGACAGGCCACCACCAGAGCCGCGATTTGTA | |
| H11 | TTAGGATTGGCTGAGACTCCTCAATAACCGAT | |
| H12 | TCCACAGACAGCCCTCATAGTTAGCGTAACGA | |
| AA1 | AACGTGGCGAGAAAGGAAGGGAAACCAGTAA | |
| AA2 | TCGGCAAATCCTGTTTGATGGTGGACCCTCAA | |
| AA3 | AAGCCTGGTACGAGCCGGAAGCATAGATGATG | |
| AA4 | CAACTGTTGCGCCATTCGCCATTCAAACATCA | |
| AA5 | GCCATCAAGCTCATTTTTTAACCACAAATCCA | |
| AA6 | CAACCGTTTCAAATCACCATCAATTCGAGCCA | |
| AA7 | TTCTACTACGCGAGCTGAAAAGGTTACCGCGC | |
| AA8 | CCAACAGGAGCGAACCAGACCGGAGCCTTTAC | |
| AA9 | CTTTTGCAGATAAAAACCAAAATAAAGACTCC | |
| AA10 | GATGGTTTGAACGAGTAGTAAATTTACCATTA | |
| AA11 | TCATCGCCAACAAAGTACAACGGACGCCAGCA | |
| AA12 | ATATTCGGAACCATCGCCCACGCAGAGAAGGA | |
| BB1 | TAAAAGGGACATTCTGGCCAACAAAGCATC | |
| BB2 | ACCTTGCTTGGTCAGTTGGCAAAGAGCGGA | |
| BB3 | ATTATCATTCAATATAATCCTGACAATTAC | |
| BB4 | CTGAGCAAAAATTAATTACATTTTGGGTTA | |
| BB5 | TATAACTAACAAAGAACGCGAGAACGCCAA | |
| BB6 | CATGTAATAGAATATAAAGTACCAAGCCGT | |
| BB7 | TTTTATTTAAGCAAATCAGATATTTTTTGT | |
| BB8 | TTAACGTCTAACATAAAAACAGGTAACGGA | |
| BB9 | ATACCCAACAGTATGTTAGCAAATTAGAGC | |
| BB10 | CAGCAAAAGGAAACGTCACCAATGAGCCGC | |
| BB11 | CACCAGAAAGGTTGAGGCAGGTCATGAAAG | |
| BB12 | TATTAAGAAGCGGGGTTTTGCTCGTAGCAT | |
| CC1 | TCAACAGTTGAAAGGAGCAAATGAAAAATCTAGAGATAGA | |
| CC2 | ATTCATTTTTGTTTGGATTATACTAAGAAACCACCAGAAG | 5' biotin |
| CC4 | TCAAATATAACCTCCGGCTTAGGTAACAATTTCATTTGAAGGCGAATT | |
| CC5 | GTAAAGTAATCGCCATATTTAACAAAACTTTT | |
| CC6 | TATCCGGTCTCATCGAGAACAAGCGACAAAAG | |
| CC7 | TTAGACGGCCAAATAAGAAACGATAGAAGGCT | |
| CC8 | CGTAGAAAATACATACCGAGGAAACGCAATAAGAAGCGCA | |
| CC9 | CACCCTCAGAAACCATCGATAGCATTGAGCCATTTGGGAA | 5' biotin |
| CC11 | GCGGATAACCTATTATTCTGAAACAGACGATTGGCCTTGAAGAGCCAC | |
| CC12 | TCACCAGTACAAACTACAACGCCTAGTACCAG | |
| DD1 | ACCCTTCTGACCTGAAAGCGTAAGACGCTGAG | |
| DD2 | AGCCAGCAATTGAGGAAGGTTATCATCATTTT | |
| DD3 | GCGGAACATCTGAATAATGGAAGGTACAAAAT | |
| DD4 | CGCGCAGATTACCTTTTTTAATGGGAGAGACT | |
| DD5 | ACCTTTTTATTTTAGTTAATTTCATAGGGCTT | |
| DD6 | AATTGAGAATTCTGTCCAGACGACTAAACCAA | |
| DD7 | GTACCGCAATTCTAAGAACGCGAGTATTATTT | |
| DD8 | ATCCCAATGAGAATTAACTGAACAGTTACCAG | |
| DD9 | AAGGAAACATAAAGGTGGCAACATTATCACCG | |
| DD10 | TCACCGACGCACCGTAATCAGTAGCAGAACCG | |
| DD11 | CCACCCTCTATTCACAAACAAATACCTGCCTA | |
| DD12 | TTTCGGAAGTGCCGTCGAGAGGGTGAGTTTCG | |
| EE1 | CTTTAGGGCCTGCAACAGTGCCAATACGTG | |
| EE2 | CTACCATAGTTTGAGTAACATTTAAAATAT | |
| EE3 | CATAAATCTTTGAATACCAAGTGTTAGAAC | |
| EE4 | CCTAAATCAAAATCATAGGTCTAAACAGTA | |
| EE5 | ACAACATGCCAACGCTCAACAGTCTTCTGA | |
| EE6 | GCGAACCTCCAAGAACGGGTATGACAATAA | |
| EE7 | AAAGTCACAAAATAAACAGCCAGCGTTTTA | |
| EE8 | AACGCAAAGATAGCCGAACAAACCCTGAAC | |
| EE9 | TCAAGTTTCATTAAAGGTGAATATAAAAGA | |
| EE10 | TTAAAGCCAGAGCCGCCACCCTCGACAGAA | |
| EE11 | GTATAGCAAACAGTTAATGCCCAATCCTCA | |
| EE12 | AGGAACCCATGTACCGTAACACTTGATATAA | |
| FF1 | GCACAGACAATATTTTTGAATGGGGTCAGTA | |
| FF2 | TTAACACCAGCACTAACAACTAATCGTTATTA | |
| FF3 | ATTTTAAAATCAAAATTATTTGCACGGATTCG | |
| FF4 | CCTGATTGCAATATATGTGAGTGATCAATAGT | |
| FF5 | GAATTTATTTAATGGTTTGAAATATTCTTACC | |
| FF6 | AGTATAAAGTTCAGCTAATGCAGATGTCTTTC | |
| FF7 | CTTATCATTCCCGACTTGCGGGAGCCTAATTT | |
| FF8 | GCCAGTTAGAGGGTAATTGAGCGCTTTAAGAA | |
| FF9 | AAGTAAGCAGACACCACGGAATAATATTGACG | |
| FF10 | GAAATTATTGCCTTTAGCGTCAGACCGGAACC | |
| FF11 | GCCTCCCTCAGAATGGAAAGCGCAGTAACAGT | |
| FF12 | GCCCGTATCCGGAATAGGTGTATCAGCCCAAT | |
| GG1 | AGATTAGAGCCGTCAAAAAACAGAGGTGAGGCCTATTAGT | |
| GG2 | AACAATAACGTAAAACAGAAATAAAAATCCTTTGCCCGAA | 5' biotin |
| GG4 | GTGATAAAAAGACGCTGAGAAGAGATAACCTTGCTTCTGTTCGGGAGA | |
| GG5 | GTTTATCAATATGCGTTATACAAACCGACCGT | |
| GG6 | GCCTTAAACCAATCAATAATCGGCACGCGCCT | |
| GG7 | GAGAGATAGAGCCGTCTTTCCAGAGGTTTTGAA | |
| GG8 | GTTTATTTTGTCACAATCTTACCGAAGCCCTTTAATATCA | |
| GG9 | AGCCACCACTGTAGCGCGTTTTCAAGGGAGGGAAGGTAAA | 5' biotin |
| GG11 | CAGGAGGTGGGGTCAGTGCCTTGAGTCTCTGAATTTACCGGGGAACCAG | |
| GG12 | CCACCCTCATTTTCAGGGATAGCAACCGTACT | |
| HH1 | CTTTAATGCGCGAACTGATAGCCCCACCAG | |
| HH2 | CAGAAGATTAGATAATACATTTGTCGACAA | |
| HH3 | CTCGTATTAGAAATTGCGTAGATACAGTAC | |
| HH4 | CTTTTACAAAATCGTCGCTATTAGCGATAG T | 3' Atto647N |
| HH5 | CTTAGATTTAAGGCGTTAAATAAAGCCTGT | |
| HH6 | TTAGTATCACAATAGATAAGTCCACGAGCA | |
| HH7 | TGTAGAAATCAAGATTAGTTGCTCTTACCA T | 3' Atto594 |
| HH8 | ACGCTAACACCCACAAGAATTGAAAATAGC | |
| HH9 | AATAGCTATCAATAGAAAATTCAACATTCA | |
| HH10 | ACCGATTGTCGGCATTTCGGTCATAATCA | |
| HH11 | AAATCACCTTCCAGTAAGCGTCAGTAATAA | |
| HH12 | GTTTTAACTTAGTACCGCCACCCAGAGCCA | |
| Scaffold | TTCCCTTCCTTTCTCGCCACGTTCGCCGGCTTTCCCCGTCAAGCTCTAAATCGGGGGCTCCCT TTAGGGTTCCGATTTAGTGCTTTACGGCACCTCGACCCCAAAAAACTTGATTTGGGTGATGGTT CACGTAGTGGGCCATCGCCCTGATAGACGGTTTTTCGCCCTTTGACGTTGGAGTCCACGTTCT TTAATAGTGGACTCTTGTTCCAAACTGGAACAACACTCAACCCTATCTCGGGCTATTCTTTTGA TTTATAAGGGATTTTGCCGATTTCGGAACCACCATCAAACAGGATTTTCGCCTGCTGGGGCAA ACCAGCGTGGACCGCTTGCTGCAACTCTCTCAGGGCCAGGCGGTGAAGGGCAATCAGCTGTT GCCCGTCTCACTGGTGAAAAGAAAAACCACCCTGGCGCCCAATACGCAAACCGCCTCTCCCC GCGCGTTGGCCGATTCATTAATGCAGCTGGCACGACAGGTTTCCCGACTGGAAAGCGGGCAG TGAGCGCAACGCAATTAATGTGAGTTAGCTCACTCATTAGGCACCCCAGGCTTTACACTTTATG CTTCCGGCTCGTATGTTGTGTGGAATTGTGAGCGGATAACAATTTCACACAGGAAACAGCTAT GACCATGATTACGAATTCGAGCTCGGTACCCGGGGATCCTCTAGAGTCGACCTGCAGGCATG CAAGCTTGGCACTGGCCGTCGTTTTACAACGTCGTGACTGGGAAAACCCTGGCGTTACCCAA CTTAATCGCCTTGCAGCACATCCCCCTTTCGCCAGCTGGCGTAATAGCGAAGAGGCCCGCAC CGATCGCCCTTCCCAACAGTTGCGCAGCCTGAATGGCGAATGGCGCTTTGCCTGGTTTCCGG CACCAGAAGCGGTGCCGGAAAGCTGGCTGGAGTGCGATCTTCCTGAGGCCGATACTGTCGTC GTCCCCTCAAACTGGCAGATGCACGGTTACGATGCGCCCATCTACACCAACGTGACCTATCCC ATTACGGTCAATCCGCCGTTTGTTCCCACGGAGAATCCGACGGGTTGTTACTCGCTCACATTT AATGTTGATGAAAGCTGGCTACAGGAAGGCCAGACGCGAATTATTTTTGATGGCGTTCCTATT GGTTAAAAAATGAGCTGATTTAACAAAAATTTAATGCGAATTTTAACAAAATATTAACGTTTACA ATTTAAATATTTGCTTATACAATCTTCCTGTTTTTGGGGCTTTTCTGATTATCAACCGGGGTACA TATGATTGACATGCTAGTTTTACGATTACCGTTCATCGATTCTCTTGTTTGCTCCAGACTCTCAG GCAATGACCTGATAGCCTTTGTAGATCTCTCAAAAATAGCTACCCTCTCCGGCATTAATTTATC AGCTAGAACGGTTGAATATCATATTGATGGTGATTTGACTGTCTCCGGCCTTTCTCACCCTTTT GAATCTTTACCTACACATTACTCAGGCATTGCATTTAAAATATATGAGGGTTCTAAAAATTTTTA TCCTTGCGTTGAAATAAAGGCTTCTCCCGCAAAAGTATTACAGGGTCATAATGTTTTTGGTACA ACCGATTTAGCTTTATGCTCTGAGGCTTTATTGCTTAATTTTGCTAATTCTTTGCCTTGCCTGTA TGATTTATTGGATGTTAATGCTACTACTATTAGTAGAATTGATGCCACCTTTTCAGCTCGCGCCC CCAAATGAAAATATAGCTAAACAGGTTATTGACCATTTGCGAAATGTATCTAATGGTCAAACTAA ATCTACTCGTTCGCAGAATTGGGAATCAACTGTTATATGGAATGAAACTTCCAGACACCGTACT TTAGTTGCATATTTAAAACATGTTGAGCTACAGCATTATATTCAGCAATTAAGCTCTAAGCCATC CGCAAAAATGACCTCTTATCAAAAGGAGCAATTAAAGGTACTCTCTAATCCTGACCTGTTGGAG TTTGCTTCCGGTCTGGTTCGCTTTGAAGCTCGAATTAAAACGCGATATTTGAAGTCTTTCGGGC | Note that the start according to the convention in the Picasso software [3,43]. |

```
TTCCTCTTAATCTTTTTGATGCAATCCGCTTTGCTTCTGACTATAATAGTCAGGGTAAAGACCTG
ATTTTTGATTTATGGTCATTCTCGTTTTCTGAACTGTTTAAAGCATTTGAGGGGGATTCAATGAA
TATTTATGACGATTCCGCAGTATTGGACGCTATCCAGTCTAAACATTTTACTATTACCCCCTCTG
GCAAAACTTCTTTTGCAAAAGCCTCTCGCTATTTTGGTTTTTATCGTCGTCTGGTAAACGAGGG
TTATGATAGTGTTGCTCTTACTATGCCTCGTAATTCCTTTTGGCGTTATGTATCTGCATTAGTTG
AATGTGGTATTCCTAAATCTCAACTGATGAATCTTTCTACCTGTAATAATGTTGTTCCGTTAGTT
CGTTTTATTAACGTAGATTTTTCTTCCCAACGTCCTGACTGGTATAATGAGCCAGTTCTTAAAAT
CGCATAAGGTAATTCACAATGATTAAAGTTGAAATTAAACCATCTCAAGCCCAATTTACTACTCG
TTCTGGTGTTTCTCGTCAGGGCAAGCCTTATTCACTGAATGAGCAGCTTTGTTACGTTGATTTG
GGTAATGAATATCCGGTTCTTGTCAAGATTACTCTTGATGAAGGTCAGCCAGCCTATGCGCCT
GGTCTGTACACCGTTCATCTGTCCTCTTTCAAAGTTGGTCAGTTCGGTTCCCTTATGATTGACC
GTCTGCGCCTCGTTCCGGCTAAGTAACATGGAGCAGGTCGCGGATTTCGACACAATTTATCAG
GCGATGATACAAATCTCCGTTGTACTTTGTTTCGCGCTTGGTATAATCGCTGGGGGTCAAAGA
TGAGTGTTTTAGTGTATTCTTTTGCCTCTTTCGTTTTAGGTTGGTGCCTTCGTAGTGGCATTAC
GTATTTTACCCGTTTAATGGAAACTTCCTCATGAAAAAGTCTTTAGTCCTCAAAGCCTCTGTAG
CCGTTGCTACCCTCGTTCCGATGCTGTCTTTCGCTGCTGAGGGTGACGATCCCGCAAAAGCG
GCCTTTAACTCCCTGCAAGCCTCAGCGACCGAATATATCGGTTATGCGTGGGCGATGGTTGTT
GTCATTGTCGGCGCAACTATCGGTATCAAGCTGTTTAAGAAATTCACCTCGAAAGCAAGCTGA
TAAACCGATACAATTAAAGGCTCCTTTTGGAGCCTTTTTTTTGGAGATTTTCAACGTGAAAAAT
TATTATTCGCAATTCCTTTAGTTGTTCCTTTCTATTCTCACTCCGCTGAAACTGTTGAAAGTTGT
TTAGCAAAATCCCATACAGAAAATTCATTTACTAACGTCTGGAAAGACGACAAAACTTTAGATC
GTTACGCTAACTATGAGGGCTGTCTGTGGAATGCTACAGGCGTTGTAGTTTGTACTGGTGACG
AAACTCAGTGTTACGGTACATGGGTTCCTATTGGGCTTGCTATCCCTGAAAATGAGGGTGGTG
GCTCTGAGGGTGGCGGTTCTGAGGGTGGCGGTTCTGAGGGTGGCGGTACTAAACCTCCTGA
GTACGGTGATACACCTATTCCGGGCTATACTTATATCAACCCTCTCGACGGCACTTATCCGCC
TGGTACTGAGCAAAACCCCGCTAATCCTAATCCTTCTCTTGAGGAGTCTCAGCCTCTTAATACT
TTCATGTTTCAGAATAATAGGTTCCGAAATAGGCAGGGGCATTAACTGTTTATACGGGCACT
GTTACTCAAGGCACTGACCCCGTTAAAACTTATTACCAGTACACTCCTGTATCATCAAAAGCCA
TGTATGACGCTTACTGGAACGGTAAATTCAGAGACTGCGCTTTCCATTCTGGCTTTAATGAGGA
TTTATTTGTTTGTGAATATCAAGGCCAATCGTCTGACCTGCCTCAACCTCCTGTCAATGCTGGC
GGCGGCTCTGGTGGTGGTTCTGGTGGCGGCTCTGAGGGTGGTGGCTCTGGTTCCGGTGATTTTGA
CTGAGGGTGGCGGCTCTGAGGGAGGCGGTTCCGGTGGTGGCTCTGGTTCCGGTGATTTTGA
TTATGAAAAGATGGCAAACGCTAATAAGGGGGCTATGACCGAAAATGCCGATGAAAACGCGCT
ACAGTCTGACGCTAAAGGCAAACTTGATTCTGTCGCTACTGATTACGGTGCTGCTATCGATGG
TTTCATTGGTGACGTTTCCGGCCTTGCTAATGGTAATGGTGCTACTGGTGATTTTGCTGGCTCT
AATTCCCAAATGGCTCAAGTCGGTGACGGTGATAATTCACCTTTAATGAATAATTTCCGTCAAT
ATTTACCTTCCCTCCCTCAATCGGTTGAATGTCGCCCTTTTGTCTTTGGCGCTGGTAAACCATA
TGAATTTTCTATTGATTGTGACAAAATAAACTTATTCCGTGGTGTCTTTGCGTTTCTTTTATATGTT
GCCACCTTTATGTATGTATTTTCTACGTTTGCTAACATACTGCGTAATAAGGAGTCTTAATCAT
GCCAGTTCTTTTGGGTATTCCGTTATTATTGCGTTTCCTCGGTTTCCTTCTGGTAACTTTGTTCG
GCTATCTGCTTACTTTTCTTAAAAAGGGCTTCGGTAAGATAGCTATTGCTATTTCATTGTTTCTT
GCTCTTATTATTGGGCTTAACTCAATTCTTGTGGGTTATCTCTCTGATATTAGCGCTCAATTACC
TTCTCTCTGTAAAGGCTGCTATTTTCATTTTTGACGTTAAACAAAAAATCGTTTCTTATTTGGATT
GGGATAAATAATATGGCTGTTTATTTTGTAACTGGCAAATTAGGCTCTGGAAAGACGCTCGTTA
GCGTTGGTAAGATTCAGGATAAAATTGTAGCTGGGTGCAAAATAGCAACTAATCTTGATTTAAG
GCTTCAAAACCTCCCGCAAGTCGGGAGGTTCGCTAAAACGCCTCGCGTTCTTAGAATACCGGA
TAAGCCTTCTATATCTGATTTGCTTGCTATTGGGCGCGGTAATGATTCCTACGATGAAAATAAA
AACGGCTTGCTTGTTCTCGATGAGTGCGGTACTTGGTTTAATACCCGTTCTTGGAATGATAAG
GAAAGACAGCCGATTATTGATTGGTTTCTACATGCTCGTAAATTAGGATGGGATATTATTTTTCT
TGTTCAGGACTTATCTATTGTTGATAAACAGGCGCGTTCTGCATTAGCTGAACATGTTGTTTATT
GTCGTCGTCTGGACAGAATTACTTTACCTTTTGTGGTGCGGCTATCTTTATATTCTCTTATTACTGGCTCG
AAAATGCCTCTGCCTAAATTACATGTTGGCGTTGTTAAATATGGCGATTCTCAATTAAGCCCTA
CTGTTGAGCGTTGGCTTTATACTGGTAAGAATTTGTATAACGCATATGATACTAAACAGGCTTT
TTCTAGTAATTATGATTCCGGTGTTTATTCTTATTTAACGCCTTATTTATCACACGGTCGGTATTT
CAAACCATTAAATTTAGGTCAGAAGATGAAATTAACTAAAATATATTTGAAAAAGTTTTCTCGCG
TTCTTTGTCTTGCGATTGGATTTGCATCAGCATTTACATATAGTTATATAACCCAACCTAAGCCG
GAGGTTAAAAAGGTAGTCTCTCAGACCTATGATTTTGATAAATTCACTATTGACTCTTCTCAG
GTCTTAATCTAAGCTATCGCTATGTTTTCAAGGATTCTAAGGGAAAATTAATTAATAGCGACGAT
TTACAGAAGCAAGGTTATTCACTCACATATATTGATTTATGTACTGTTTCCATTAAAAAAGGTAA
TTCAAATGAAATTGTTAAATGTAATTAATTTTGTTTTCTTGATGTTTGTTTCATCATCTTCTTTTGC
TCAGGTAATTGAAATGAATAATTCGCCTCTGCGCGATTTTGTAACTTGGTATTCAAAGCAATCA
GGCGAATCCGTTATTGTTTCTCCCGATGTAAAAGGTACTGTTACTGTATATTCATCTGACGTTA
AACCTGAAAATCTACGCAATTTCTTTATTTCTGTTTTACGTGCAAAATAAGGGTTGATATGGTGGT
TCTAACCCTTCCATTATTCAGAAGTATAATCCAAACAATCAGGATTATATTGATGAATTGCCATC
ATCTGATAATCAGGAATATGATGATAATTCCGCTCCTTCTGGTGGTTTCTTTGTTCCGCAAAAT
GATAATGTTACTCAAACTTTTAAAATTAATAACGTTCGGGCAAAGGATTTAATACGAGTTGTCGA
ATTGTTTGTAAAGTCTAATACTTCTAAATCCTCAAATGTATTATCTATTGACGGCTCTAATCTATT
AGTTGTTAGTGCTCCTAAAGATATTTTAGATAACCTTCCTCAATTCCTTTCAACTGTTGATTTGC
CAACTGACCAGATATTGATTGAGGGTTTGATATTTGAGGTTCAGCAAGGTGATGCTTTAGATTT
TTCATTTGCTGCTGGCTCTCAGCGTGGCACTGTTGCAGGCGGTGTTAATACTGACCGCCTCAC
CTCTGTTTTATCTTCTGCTGGTGGTTCGTTCGGTATTTTTAATGGCGATGTTTTAGGGCTATCA
GTTCGCGCATTAAAGACTAATAGCCATTCAAAAATATTGTCTGTGCCACGTATTCTTACGCTTT
CAGGTCAGAAGGGTTCTATCTCTGTTGGCCAGAATGTCCCTTTTATTACTGGTCGTGTGACTG
GTGAATCTGCCAATGTAAATAATCCATTTCAGACGATTGAGCGTCAAAATGTAGGTATTTCCAT
GAGCGTTTTTCCTGTTGCAATGGCTGGCGGTAATATTGTTCTGGATATTACCAGCAAGGCCGA
TAGTTTG
```

# Results

**Supplementary Tab. 3: FRET correction parameters.** List of correction factors for intensity-based FRET efficiencies, including the additional parameters required for the correction under STED conditions. The meaning and the computation is described in chapter '*Spectroscopy and image analysis*', in the sections '*Determination of intensity-based correction factors* ', '*Intensity-based spectroscopic parameters*', '*Determination of intensity-based FRET efficiencies under STED conditions*' and '*Determination of average intensity-based FRET efficiencies*'.

| Sample | Surface | $G$ (eq. S15) | Correction Factor | | | | | | | | | | | | | | | | $\frac{g_{D\mid D}}{g_{A\mid A}}$ |
| | | | $\alpha$ | $\beta$ | $\gamma$ | $\delta$ | $\gamma'$ | $x_D^0$ | $x_A^0$ | $x_D^d$ | $x_A^d$ | $\Phi_{F,D}^0$ | $\Phi_{F,A}^0$ | $\Phi_{F,D}^d$ | $\Phi_{F,A}^d$ | $I_{Dem\mid Dex}^{(BG)}$ (kHz) | $I_{Aem\mid Dex}^{(BG)}$ | $I_{Aem\mid Aex}^{(BG)}$ | |
|---|---|---|---|---|---|---|---|---|---|---|---|---|---|---|---|---|---|---|---|
| Single Biotin dsD | BSA-Biotin | 0.84 | 0.493 ±5 | 1.33 ±1 | 1.37 ±2 | 0.12 ±2 | 1.6 ± 1 | 0.14 ± 1 | 0.12± 1 | 0.86± 2 | 0.88± 2 | 0.84± 2 | 0.69± 2 | 0.06 ± 5 | 0.04 ±1 | 0.66±1 | 2.55± 10 | 2.35± 10 | 0.6 |
| Double Biotin dsD | BSA-Biotin | 0.84 | 0.493 ±5 | 1.33 ±1 | 1.37 ±2 | 0.12 ±2 | 1.6± 1 | 0.14 ± 1 | 0.12± 1 | 0.86± 2 | 0.88± 2 | 0.84± 2 | 0.69± 2 | 0.06 ± 5 | 0.04 ±1 | 0.67±1 | 2.05± 10 | 2.03± 10 | 0.6 |
| Single Biotin dsD | NHS-PEG-Biotin | 0.84 | 0.493 ±5 | 0.48 ±1 | 1.37± 2 | 0.23 ±2 | 1.6 ± 1 | 0.14± 1 | 0.12± 1 | 0.86± 2 | 0.88± 2 | 0.84± 2 | 0.69± 2 | 0.06 ± 5 | 0.04 ±1 | 0.65±1 | 1.63± 10 | 1.35± 10 | 0.6 |
| Double Biotin dsD | NHS-PEG-Biotin | 0.84 | 0.493 ±5 | 0.77 ±1 | 1.0 ±2 | 0.12 ±2 | 1.6 ± 1 | 0.14± 1 | 0.12± 1 | 0.86± 2 | 0.88± 2 | 0.84± 2 | 0.69± 2 | 0.06 ± 5 | 0.04 ±1 | 0.60 ±1 | 1.55± 10 | 1.41± 10 | 0.6 |
| O(HF+NF) | NHS-PEG-Biotin | 0.89 | 0.52 ± 2 | 1.2 ± 1 | 2.3 ± 2 | 0.07 ± 2 | 1.5 ± 1 | 0.21 ± 1 | 0.14 ± 1 | 0.80 ± 2 | 0.86 ± 2 | 0.81 ± 2 | 0.78 ± 2 | 0.060 ± 5 | 0.048 ± 5 | 0.20 ± 2 | 0.65 ± 10 | 0.54 ± 10 | 0.42 ± 6 |
| O(NF) | NHS-PEG-Biotin | 0.77 | 0.52 ± 2 | 2.1 ± 1 | 2.3 ± 2 | 0.07 ± 2 | 1.3 ± 1 | 0.29 ± 1 | 0.17 ± 1 | 0.71 ± 2 | 0.83 ± 2 | 0.81 ± 2 | 0.78 ± 2 | 0.060 ± 5 | 0.048 ± 5 | 0.20 ± 2 | 0.65 ± 10 | 0.54 ± 10 | 0.42 ± 6 |
| O(HF) | NHS-PEG-Biotin | 0.77 | 0.52 ± 2 | 1.5 ± 1 | 2.3 ±2 | 0.07 ± 2 | 2.8 ± 2 | 0.17 ± 1 | 0.20 ± 1 | 0.83 ± 2 | 0.80 ± 2 | 0.81 ± 2 | 0.78 ± 2 | 0.060 ± 5 | 0.048 ± 5 | 0.20 ± 2 | 0.65 ± 10 | 0.54 ± 10 | 0.42 ± 6 |
| dsD (sm) | in solution | 0.87 | 0.34 | 0.71 | 0.92 | 0.11 | - | - | - | - | - | 0.67 | 0.65 | - | - | | | | 0.95 |
| O (sm) | in solution | 0.87 | 0.75 | 1.54 | 1.23 | 0.13 | - | - | - | - | - | 0.85 | 0.65 | - | - | | | | 0.45 |
| hGBP1 | | 0.84 | 0.56 ±5 | 1.8± 1 | 1.13 ±2 | 0.12 ±2 | 1.6 ± 1 | 0.17± 1 | 0.12± 1 | 0.83 ± 2 | 0.88 ± 2 | 0.85± 2 | 0.69 | 0.69± 2 | 0.04 ±1 | 4.67 | 5.41 | 6.4 | 0.72 |

**Supplementary Tab. 4: Determination of the Förster radius for dsDNA and DNA origami.** List of the parameters for two dye pairs for calculating the Förster radii $R_0$. The absorption and fluorescence spectra as well as the spectral overlap spectrum $J$ are shown in Supplementary Fig. 26.

| Dye pairs | $\kappa^2$ | $n_{im}$ | $\Phi_{F,D}$ | $\varepsilon_{A,max}$ [M$^{-1}$cm$^{-1}$] | $J$ [cm$^{-1}$M$^{-1}$nm$^4$] | $R_0$ [Å] |
|---|---|---|---|---|---|---|
| Alexa594-Atto647N (for DNA rulers) | 2/3 | 1.40 | 0.80 | 150000 | $1.008 \cdot 10^{16}$ | 71 |
| Atto594-Atto647N (for origami) | 2/3 | 1.40 | 0.85[10] | 150000 | $1.56 \cdot 10^{16}$ | 76.5 |

**Supplementary Tab. 5: Microscope resolution and predicted precision.** The theoretical precision was calculated using measured values for each dataset according to equation S29. The full width at half maximum of the point spread function was obtained from the average $\sigma_{PSF}$ of all molecules. See also Supplementary Fig. 17 (origami) and Supplementary Fig. 25 (DNA ruler).

| Sample | Immob. | # biotin anchors | Predicted localization precision [nm] | Donor FWHM [nm] | Acceptor FWHM [nm] |
|---|---|---|---|---|---|
| dsD(HF) | BSA | 1 | 3.17 | 65 | 50 |
| dsD(HF) | BSA | 2 | 3.47 | 65 | 49 |
| dsD(MF) | BSA | 1 | 2.97 | 68 | 51 |
| dsD(MF) | BSA | 2 | 3.42 | 68 | 52 |
| dsD(LF) | BSA | 1 | 3.03 | 67 | 53 |
| dsD(LF) | BSA | 2 | 2.97 | 83 | 63 |
| dsD(NF) | BSA | 1 | 3.94 | 68 | 53 |
| dsD(NF) | BSA | 2 | 3.45 | 67 | 50 |
| dsD(HF) | PEG | 1 | 4.54 | 56 | 67 |
| dsD(HF) | PEG | 2 | 5.04 | 63 | 50 |
| dsD(MF) | PEG | 1 | 4.25 | 52 | 67 |
| dsD(MF) | PEG | 2 | 3.34 | 68 | 52 |
| dsD(LF) | PEG | 1 | 4.89 | 54 | 67 |
| dsD(LF) | PEG | 2 | 3.76 | 68 | 52 |
| dsD(NF) | PEG | 1 | 4.53 | 71 | 58 |
| dsD(NF) | PEG | 2 | 4.41 | 72 | 56 |
| O(NF) | PEG | 6 | 2.9 | 75 | 62 |
| O(HF) | PEG | 6 | 3.4 | 93 | 74 |
| O(HF-NF) | PEG | 6 | 2.0 | 70 | 56 |

**Supplementary Tab. 6: Accessible volume parameters.** Dye parameters for the AV simulations of Alexa594 and Atto647N on dsDNA with Förster radius $R_0$ [Å] = 71, and of Atto594 and Atto647N on the origami nanostructures with a Förster radius $R_0$ [Å] = 76.5. On the origami, an extension of the linker length due to the addition of an unpaired thymine base of 8.3 Å is considered based on the length of a phosphate-sugar-phosphate fragment in dsDNA. See main text '*Spectroscopy and image analysis, Accessible Volume Simulations*'..

| Dye | linker length [Å] | linker width [Å] | dye radius [Å] |
|---|---|---|---|
| Alexa594 (dsDNA) | 20.5 | 4.5 | 3.5 |
| Atto647N (dsDNA) | 21.0 | 4.5 | 3.5 |
| Atto594 (Origami) | 28.8 | 4.5 | 3.5 |
| Atto647N (Origami) | 29.3 | 4.5 | 3.5 |

**Supplementary Tab. 7: Expected FRET efficiencies and distances from accessible volume simulations.** Recommended dye parameters for the AV simulations for high-FRET (HF)), mid-FRET (MF), low-FRET(LF) and no-FRET (NF) dsDNA (dsD) sample. $R_0$ = 71 Å. See also Supplementary Tabs. 6 (AV parameters) and 10 (corresponding experimental data). The simulated FRET related distances were converted into $R_{mp}^{FRET}$ as shown in Supplementary Fig. 5 using the equation given in Supplementary Tab. S8.

| Sample | $\langle E \rangle$ | $\langle R_{DA} \rangle_E$ [Å] | $\langle R_{DA} \rangle$ [Å] | $R_{mp}^{FRET}$ [Å] |
|---|---|---|---|---|
| dsD(HF) | 0.556 | 68.4 | 68.5 | 66.3 |
| dsD(MF) | 0.254 | 85.0 | 87.1 | 85.5 |
| dsD(LF) | 0.058 | 112.9 | 115.5 | 114.2 |
| dsD(NF) | 0.015 | 143.2 | 145.4 | 144.4 |

**Supplementary Tab. 8: Conversion of FRET-related distances into mean-position distances**. List of polynomial coefficients describing conversion into $R_{mp}^{FRET}$. The corresponding polynomials are given by $R_{mp}^{FRET} = a_0 + a_1 \langle R_{DA} \rangle_E + a_2 \langle R_{DA} \rangle_E^2 + a_3 \langle R_{DA} \rangle_E^3 + a_4 \langle R_{DA} \rangle_E^4 + a_5 \langle R_{DA} \rangle_E^5$

and $R_{mp}^{FRET} = a_0 + a_1 \langle R_{DA} \rangle + a_2 \langle R_{DA} \rangle^2 + a_3 \langle R_{DA} \rangle^3 + a_4 \langle R_{DA} \rangle^4 + a_5 \langle R_{DA} \rangle^5$

The polynomials were determined based on AV simulations on dsDNA using a Förster radius of $R_0$ = 71 Å for the dye pair Alexa594-Atto647N and $R_0$ = 76.5 Å for dye pair Alexa594-Atto647N. The polynomials were determined as described in Kalinin et al.[18]. The corresponding graphs are shown in Supplementary Fig 5.

| Conversion | sample | $a_0$ | $a_1$ | $a_2$ | $a_3$ | $a_4$ | $a_5$ |
|---|---|---|---|---|---|---|---|
| $\langle R_{DA} \rangle_E \rightarrow R_{mp}^{FRET}$ | Alexa594-Atto647N | -68.1 | 4.48 | -0.08 | 1.1E-03 | -6.84E-06 | 1.67E-08 |
| | Atto594-Atto647N | -153.4 | 8.09 | -0.15 | 1.7E-03 | -9.62E-06 | 2.09E-08 |
| $\langle R_{DA} \rangle \rightarrow R_{mp}^{FRET}$ | Alexa594-Atto647N | -64.9 | 5.34 | -0.15 | 1.4E-03 | -8.76E-06 | 2.02E-05 |
| | Atto594-Atto647N | -161.4 | 10.27 | -0.21 | 2.3E-03 | -1.23E-05 | 2.50E-08 |

| Sample | Immob. | # biotin-anchors | $x_{STED}$ | $\tau_{STED}$ [ns] | $x_{FRET}$ | $\tau_{F, Donly}$ [ns] | $R_0$ [Å] | $\langle R_{DA}\rangle$ [Å]; $\sigma = 6$ Å | $\chi^2_r$ |
|---|---|---|---|---|---|---|---|---|---|
| **FRET nanoscopy** | | | | | | | | | |
| dsD(HF) | BSA | 1 | 0.76 | 0.3 | 0.22 | 3.84 | 71 | 72.9+5 \ -4 | 1.18 |
| dsD(HF) | BSA | 2 | 0.77 | 0.3 | 0.22 | 3.84 | 71 | 71.9+4 \ -3 | 1.04 |
| dsD(MF) | BSA | 1 | 0.76 | 0.33 | 0.17 | 3.84 | 71 | 88.1+6 \ -8 | 1.07 |
| dsD(MF) | BSA | 2 | 0.72 | 0.35 | 0.28 | 3.84 | 71 | 87.9+4 \ -8 | 1.12 |
| dsD(LF) | BSA | 1 | 0.7 | 0.33 | 0.3 | 3.84 | 71 | 109.2+6 \ -11 | 1.09 |
| dsD(LF) | BSA | 2 | 0.75 | 0.33 | 0.25 | 3.84 | 71 | 110.1+ 6\ -6 | 1.09 |
| dsD(NF) | BSA | 1 | 0.67 | 0.34 | 0.33 | 3.84 | 71 | 115 | 0.99 |
| dsD(NF) | BSA | 2 | 0.69 | 0.38 | 0.31 | 3.84 | 71 | 124.5 | 1.04 |
| dsD(HF) | PEG | 1 | 0.74 | 0.36 | 0.26 | 3.84 | 71 | 74.9+4 \ -8 | 1.06 |
| dsD(HF) | PEG | 2 | 0.77 | 0.28 | 0.23 | 3.84 | 71 | 75.5+3 \ -10 | 1.01 |
| dsD(MF) | PEG | 1 | 0.69 | 0.39 | 0.31 | 3.84 | 71 | 89.6+4 \ -2 | 1.43 |
| dsD(MF) | PEG | 2 | 0.65 | 0.41 | 0.35 | 3.84 | 71 | 87.6+7 \ -22.6 | 1.09 |
| dsD(LF) | PEG | 1 | 0.67 | 0.34 | 0.33 | 3.84 | 71 | 102+18 \ -9 | 1.00 |
| dsD(LF) | PEG | 2 | 0.66 | 0.38 | 0.34 | 3.84 | 71 | 104.3+12 \ -6 | 1.3 |
| dsD(NF) | PEG | 1 | 0.72 | 0.36 | 0.28 | 3.84 | 71 | 107.8 | 1.04 |
| dsD(NF) | PEG | 2 | 0.69 | 0.38 | 0.31 | 3.84 | 71 | 115 | 1.17 |
| O(NF)[1] | PEG | 6 | 0.80 | 0.23 ± 5 | - | 3.69± 12[1] | 76.5 | - | 1.02 |
| O(HF)[1] | PEG | 6 | 0.84 | 0.29 | 0.15 | 3.69± 12[1] | 76.5 | 75+6 \ -6 | 1.59 |
| O(HF-NF)$_{NF\ cut}$[2] | PEG | 6 | 0.84 | 0.25 ± 3 | 0[f] | 3.74 ± 10[2] | 76.5 | - | 1.33 |
| O(HF-NF)$_{HF\ cut}$[2] | PEG | 6 | 0.88 | 0.27 | 0.1 | 3.74 ± 10[2] | 76.5 | 76+3 \-3 | 1.00 |
| **Confocal single-molecule spectroscopy** | | | | | | | | | |
| dsD(HF) | - | 1 | - | - | 1 | 3.96 | 71 | 70 | 1.2 |
| dsD(MF) | - | 1 | - | - | 1 | 3.96 | 71 | 90; σ = 24 Å | 1.2 |
| dsD(LF) | - | 1 | - | - | 1 | 3.96 | 71 | 115 | 1.5 |
| dsD(NF) | - | 1 | - | - | 1 | 3.96 | 71 | 133 | 1.1 |
| O(NF) | - | 6 | - | - | 0[f] | 3.753 | 76.5 | -- | 0.86 |
| O(HF) | - | 6 | - | - | 1[f] | 3.753 | 76.5 | 78.3 | 1.2 |
| O(HF-NF) | - | 6 | - | - | 0.4 | 3.753 | 76.5 | 78.3 | 2.5 |

**Supplementary Tab. 10: Intensity-based FRET efficiencies of all measurements.** FRET efficiencies are determined from the center of the plots of the FRET efficiency versus fluorescence-weighted lifetime, and are converted into interdye distances as described in the section '*Accessible volume simulations*' and Supplementary Fig. 5. *E-τ* plots for origamis are shown Supplementary Figs. 12a and 13a for confocal single-molecule spectroscopy and FRET nanoscopy, respectively. Corresponding plots for the dsDNA rulers are shown in Supplementary Figs. 19 and 20 for FRET nanoscopy and confocal single-molecule spectroscopy, respectively.

| Sample | Surface | # Biotin-anchor | $\langle E \rangle$ | $\langle R_{\mathrm{DA}} \rangle_{\mathrm{E}}$ [Å] | $R_{mp}^{FRET}$ [Å] |
|---|---|---|---|---|---|
| **FRET nanoscopy** | | | | | |
| **dsD(HF)** | BSA | 1 | 0.47 | 72 ±2.6 | 71 ±3 |
| **dsD(HF)** | BSA | 2 | 0.44 | 74 ±2.6 | 73 ±3 |
| **dsD(MF)** | BSA | 1 | 0.22 | 88 ±3.8 | 89 ±4 |
| **dsD(MF)** | BSA | 2 | 0.21 | 89 ±3.8 | 89 ±4 |
| **dsD(LF)** | BSA | 1 | 0.1 | 103 ±9.8 | 106 ±10 |
| **dsD(LF)** | BSA | 2 | 0.09 | 104± 11.3 | 106 ±11 |
| **dsD(HF)** | PEG | 1 | 0.46 | 73 ±2.5 | 72 ±3 |
| **dsD(HF)** | PEG | 2 | 0.52 | 70 ±2.5 | 68 ±3 |
| **dsD(MF)** | PEG | 1 | 0.23 | 87 ±3.7 | 88 ±4 |
| **dsD(MF)** | PEG | 2 | 0.19 | 90 ±4.0 | 91 ±4 |
| **dsD(LF)** | PEG | 1 | 0.05 | 115 ±8.2 | 117 ±8 |
| **dsD(LF)** | PEG | 2 | 0.07 | 111 ±7.8 | 113 ±8 |
| **O(NF)** | PEG | 8 | 0.02± 0.02 | >120 | - |
| **O(HF)** | PEG | 8 | 0.53± 0.02 | 76 ±2.0 | 74 ±2.0 |
| **O(HF-NF)**NF cut | PEG | 8 | 0.02± 0.02 | >120 | - |
| **O(HF-NF)**HF cut | PEG | 8 | 0.45± 0.02 | 79 ±2.0 | 77 ±2.0 |
| **hGBP1** | - | - | 0 | - | - |
| **Confocal single molecule spectroscopy** | | | | | |
| **dsD(HF)** | - | | 0.5 | 70 | - |
| **dsD(MF)** | - | | 0.22 | 86 | - |
| **dsD(LF)** | - | | 0.06 | 111 | - |
| **dsD(NF)** | - | | 0.02 | 134 | - |
| **O(NF)** | - | 8 | 0 | >120 | - |
| **O(HF)** | - | 8 | 0.53 | 75 | 73 |
| **O(HF-NF)** | - | 8 | - | - | - |
| **hGBP1** | - | - | >0.5- | - | - |

***Supplementary Tab. 11: Steady state anisotropies for origami and dsDNA ruler.*** Donor ($r_D$) and acceptor steady state anisotropies ($r_A$) are determined from single-molecule measurements (see Supplementary Fig. 14). Only FRET-active molecules carrying both dyes were considered. Similarly, the rotational correlation times for donor and acceptor ($\rho_D$, $\rho_A$), respectively, are obtained by a graphical fit to the Perrin equation (eq. S21c) using a fundamental anisotropy, $r_0 = 0.374$, for the donor and acceptor.

| Sample | $r_D$ | $\rho_D$ [ns] | $r_A$ | $\rho_A$ [ns] |
|---|---|---|---|---|
| O(HF+NF) | 0.07 | 0.8 | 0.25 | 10 |
| O(NF) | 0.06 | 0.7 | 0.25 | 9 |
| O(HF) | 0.09 | 0.7 | 0.25 | 10 |
| dsD(HF) | 0.11 | 0.9 | 0.13 | 2.2 |
| dsD(MF) | 0.1 | 1.1 | 0.18 | 4 |
| dsD(LF) | 0.08 | 1.1 | 0.12 | 1.9 |
| dsD(NF) | 0.08 | 1.1 | 0.13 | 2.2 |

**Supplementary Tab. 12: Model-based analysis of distance distributions of dsDNA rulers.** Summary of fit results for surface immobilization by BSA-Biotin and NHS-PEG-Biotin. The procedure is described in Supplementary Note 7 and in the section 'Colocalization analysis' (see eq. S30). For each surface immobilization the single biotin distance distribution is fitted by single-component $\chi$-distribution fit (I). For double biotin distribution two independent fit models were applied, assuming a single biotin component with fixed $\sigma_{\chi,1}$ from single biotin fit (II), and additional fixing of double biotin component $\sigma_{\chi,2}$ = 4 nm (III). Fixed values are highlighted in grey. Further parameters are described in the chapter *'Spectroscopy and Image analysis' in the section 'Model-based analysis of localization-based distance distributions'*.

| Sample | Surface | Number of biotin anchor | Model component | Amplitude $A_1$ | $R_{mp,1}$ [nm] | $\sigma_{\chi,1}$ | Amplitude $A_2$ | $R_{mp,2}$ [nm] | $\sigma_{\chi,2}$ |
|---|---|---|---|---|---|---|---|---|---|
| **I. Single biotin – all parameters free** | | | | | | | | | |
| dsD(HF) | BSA | 1 | 1 | 1 | 0 | 5.21 | | | |
| dsD(MF) | BSA | 1 | 1 | 1 | 0 | 5.03 | | | |
| dsD(LF) | BSA | 1 | 1 | 1 | 4.95 | 4.41 | | | |
| dsD(NF) | BSA | 1 | 1 | 1 | 5.8 | 4.95 | | | |
| **II. Double biotin** | | | | | | | | | |
| **Single biotin parameters fixed, double biotin distance and sigma free** | | | | | | | | | |
| dsD(HF) | BSA | 2 | 2 | 0 | 0 | 5.21 | 1 | 4.13 | 5.17 |
| dsD(MF) | BSA | 2 | 2 | 0 | 0 | 5.03 | 1 | 7.54 | 5 |
| dsD(LF) | BSA | 2 | 2 | 0.29 | 4.95 | 4.41 | 0.71 | 10.4 | 3.59 |
| dsD(NF) | BSA | 2 | 2 | 0.12 | 5.8 | 4.95 | 0.88 | 10.72 | 6.03 |
| **III. Double biotin** | | | | | | | | | |
| **Single biotin parameters fixed, double biotin distance free, but sigma fixed** | | | | | | | | | |
| dsD(HF) | BSA | 2 | 2 | 0.73 | 0 | 5.21 | 0.27 | 8.89 | 4 |
| dsD(MF) | BSA | 2 | 2 | 0.29 | 0 | 5.03 | 0.71 | 9.61 | 4 |
| dsD(LF) | BSA | 2 | 2 | 0.18 | 4.95 | 4.41 | 0.82 | 9.7 | 4 |
| dsD(NF) | BSA | 2 | 2 | 0.48 | 5.8 | 4.95 | 0.52 | 13.99 | 4 |
| Sample | Immob. | Number of biotin anchor | Model component | Amplitude $A_1$ | $R_{mp,1}$ [nm] | $\sigma_{\chi,1}$ | Amplitude $A_2$ | $R_{mp,2}$ [nm] | $\sigma_{\chi,2}$ |
| **I. Single biotin – all parameters free** | | | | | | | | | |
| dsD(HF) | PEG | 1 | 1 | 1 | 0 | 5.77 | 1 | 0 | 5.77 |
| dsD(MF) | PEG | 1 | 1 | 1 | 0 | 6.33 | 1 | 0 | 6.33 |
| dsD(LF) | PEG | 1 | 1 | 1 | 0 | 7.08 | 1 | 0 | 7.08 |
| dsD(NF) | PEG | 1 | 1 | 1 | 7.18 | 4.99 | 1 | 7.18 | 4.99 |
| **II. Double biotin** | | | | | | | | | |
| **Single biotin parameters fixed, double biotin distance and sigma free** | | | | | | | | | |
| dsD(HF) | PEG | 2 | 2 | 0.27 | 0 | 5.77 | 0.73 | 0.03 | 7.7 |
| dsD(MF) | PEG | 2 | 2 | 0 | 0 | 6.33 | 1.0 | 7.2 | 6.2 |
| dsD(LF) | PEG | 2 | 2 | 0.5 | 0 | 7.08 | 0.5 | 12.92 | 3.36 |
| dsD(NF) | PEG | 2 | 2 | 0.67 | 7.18 | 4.99 | 0.33 | 15.62 | 3.86 |
| **III. Double biotin** | | | | | | | | | |
| **Single biotin parameters fixed, double biotin distance free, but sigma fixed** | | | | | | | | | |
| dsD(HF) | PEG | 2 | 2 | 0.77 | 0 | 5.77 | 0.23 | 12.01 | 4 |
| dsD(MF) | PEG | 2 | 2 | 0.64 | 0 | 6.33 | 0.37 | 12.4 | 4 |
| dsD(LF) | PEG | 2 | 2 | 0.43 | 0 | 7.08 | 0.57 | 12.26 | 4 |
| dsD(NF) | PEG | 2 | 2 | 0.66 | 7.18 | 4.99 | 0.34 | 15.47 | 4 |

**Supplementary Tab. 13: Overview of distances.** All distances are mean-position distances $R_{mp}$. Where applicable, the FRET related distances were converted into $R_{mp}^{FRET}$ as shown in Supplementary Fig. 5 using by using equation given in Supplementary Tab. S8. The experimental origin is described as follows: FN: FRET nanoscopy, SM: confocal single-molecule spectroscopy, int: intensity-based ($\langle R_{DA} \rangle_E$ in Supplementary Tab. 10), lt: lifetime-based ($\langle R_{DA} \rangle$, in Supplementary Tab. 9) loc: localization-based ($R_{mp}^{loc}$). For the accessible volume simulations, we used an atomistic model of the origami generated by the CanDo software[44,45], based on an assumed inter-helical distance of 2.25 nm for the accessible volume simulations. The superscript [a] denotes that values are obtained from aligned structures, which is in general more accurate due to the better filtering of broken constructs. See Supplementary Fig. 13a (origami) and Supplementary Figs. 19 and 21 (DNA ruler).

| Sample | Surface | # Biotin-anchor | $R_{mp}^{loc}$ | $R_{mp}^{FRET}$ | | | | |
| | | | FN-loc | AV sim. | FN-int | FN-lt | SM-int | SM-lt |
|---|---|---|---|---|---|---|---|---|
| **O(NF)** | PEG | 8 | 165 ± 5 | 144 | >120 | >120 | >120 | >120 |
| **O(HF)** | PEG | 8 | 94 ± 10 | 48 | 74 ± 2 | 72 ± 6 | 73 | 75 |
| **O(HF-NF)**$_{NF cut}$ | PEG | 8 | 147 ± 7[a] | 144 | >120 | >120 | -- | >120 |
| **O(HF-NF)**$_{HF cut}$ | PEG | 8 | 53 ± 7[a] | 48 | 77 ± 2 | 73 ± 3 | -- | 75 |
| **dsD(HF)** | BSA | 1 | 0 | 66.3 | 71 ± 3 | 70.5 | 70 | 70 |
| **dsD(HF)** | BSA | 2 | 49.5 | 66.3 | 73 ± 3 | 69.5 | 70 | 70 |
| **dsD(MF)** | BSA | 1 | 19.5 | 85.5 | 89 ± 4 | 88.9 | 86 | 90 |
| **dsD(MF)** | BSA | 2 | 73.5 | 85.5 | 89 ± 4 | 88.8 | 86 | 90 |
| **dsD(LF)** | BSA | 1 | 54 | 114.2 | 106 ± 10 | 108.3 | 111 | 115 |
| **dsD(LF)** | BSA | 2 | 90 | 114.2 | 106 ± 11 | 109.2 | 111 | 115 |
| **dsD(NF)** | BSA | 1 | 72 | 144.4 | - | - | - | - |
| **dsD(NF)** | BSA | 2 | 150 | 144.4 | - | - | - | - |
| **dsD(HF)** | PEG | 1 | 0 | 66.3 | 72 ±3 | 72.6 | 70 | 70 |
| **dsD(HF)** | PEG | 2 | 64.5 | 66.3 | 68 ±3 | 73.2 | 70 | 70 |
| **dsD(MF)** | PEG | 1 | 31.5 | 85.5 | 88 ±4 | 88.2 | 86 | 90 |
| **dsD(MF)** | PEG | 2 | 75 | 85.5 | 91 ±4 | 86.0 | 86 | 90 |
| **dsD(LF)** | PEG | 1 | 70.5 | 114.2 | 117 ±8 | 101.1 | 111 | 115 |
| **dsD(LF)** | PEG | 2 | 112.5 | 114.2 | 113 ±8 | 103.5 | 111 | 115 |
| **dsD(NF)** | PEG | 1 | 79.5 | 144.4 | - | - | - | - |
| **dsD(NF)** | PEG | 2 | 133.5 | 144.4 | - | - | - | - |

# Filtering and analysis settings

**Supplementary Tab. 14: Overview of used analysis settings.** Values between square brackets indicate corresponding software settings as listed is Supplementary Tab. 16. Software settings that are not explicitly mentioned all have default values. The procedures are described in the chapter '*Spectroscopy and Image analysis' in the section 'Filtering procedures*' and Supplementary Figs. 3 and 6.

| Parameter | O(HF-NF) | O(NF) | O(HF) | dsD (all) | hGBP1 |
|---|---|---|---|---|---|
| Analysis workflow | multi-spot | | | single-spot | multi-spot |
| **Localization analysis** | | | | | |
| # images taken | 2036 (100%) | 1500 (100%) | 2992 (100%) | -- | 258 (rings) |
| # frames for localization [Framestop_localization] | 60 | 20 | 20 | 60 | 100 |
| # frames for FRET [Framestop_FRET] | 20 | 20 | 20 | 30 | 20 |
| gate for localization [Ggate_loc, Rgate_loc, Ygate_loc] | 3.7 ns - 19.2 ns | 3.7 ns - 19.2ns | 0.0 ns - 19.2 ns | 3.8 ns -15.4 ns | 3.8 ns -15.4 ns |
| gate for spot lifetime tail fit [Ggate_lt, Rgate_lt, Ygate_lt] | 3.7 ns - 19.2 ns | 3.7 ns - 19.2 ns | 3.7 ns - 19.2 ns | 3.8 ns -15.4 ns | 3.2 ns -15.4 ns |
| ROI size [ROIsize] | 30 x 30 pixels | 20 x 20 pixels | 20 x 20 pixels | 21 x 21 pixels | 17x17 pixels |
| absolute threshold [ROI_threshold_abs] | 1 | 1 | 1 | 2 | 1 |
| minimal brightness per spot [garbageBrightness] | 20 | 20 | 20 | -- | 40 |
| # FRET pairs | 1394 (67%) | 724 (48%) | 987 (33%) | | 1533 (spots) |
| **Data filtering for particle averaging** | | | | | |
| spot stoichiometry | 2 Donors, 2 Acceptors | -- | -- | -- | -- |
| RMSD (alignment score) | < 10 nm | -- | -- | -- | -- |
| No. of platforms | 101 (7%) | -- | -- | -- | -- |
| **Data filtering for fluorescence spectroscopy** | | | | | |
| spot stoichiometry | ≥1 Donor, ≥1 Acceptor | ≥1 Donor, ≥1 Acceptor | 1 Donor, 1 Acceptor | -- | ≥1 Donor, ≥1 Acceptor |
| acceptor intensity | -- | -- | >150 counts | -- | -- |
| background for lifetime fitting [kHz] | 0.1 | 0.5 | 0.6 | -- | 0.4 |
| # FRET pairs | 1384 (68 %) | 677 (45%) | 327 (11%) | -- | 1533 |
| **Data filtering for sub-ensemble lifetime decay** | | | | | |
| spot stoichiometry | 1, 2 or 3 Donors and 1, 2 or 3 Acceptors | | | -- | 1, 2 or 3 Donors and 1, 2 or 3 Acceptors |
| efficiency for highFRET cut | E > 0.31 | -- | E > 0.31 | -- | - |
| # FRET pairs for highFRET cut | 372 (18 %) | -- | 694 (23 %) | -- | - |
| efficiency for noFRET cut | E < 0.2 | E < 0.2 | -- | -- | - |
| # FRET pairs for noFRET cut | 729 (36 %) | 450 (30%) | -- | -- | - |
| proximity ratio cut, PR | | | | | P > 0.35 |

**Supplementary Tab. 15: Selection criteria for dsDNA rulers.** The number of spots (# spots) corresponds to the remaining spots after each selection step. The procedures are described in '*Spectroscopy and Image analysis, Filtering procedures*' and Supplementary Figs. 3 and 6.

| Sample | Surface | # biotin anchors | Total number of spots | Image indicator | | | FRET indicator | |
|---|---|---|---|---|---|---|---|---|
| | | | | Selection $\sigma_{PSF,D}$ [nm] | Selection $\sigma_{PSF,A}$ [nm] | # spots | Selection Stoichiometry | # spots |
| dsD(HF) | BSA | 1 | 2116 | 20- 35 | 15-30 | 1316 | 0.4-0.6 | 583 |
| dsD(HF) | BSA | 2 | 1461 | 20- 35 | 15-30 | 825 | 0.4-0.6 | 346 |
| dsD(MF) | BSA | 1 | 2044 | 20- 35 | 15-30 | 1360 | 0.4-0.6 | 554 |
| dsD(MF) | BSA | 2 | 2784 | 20- 35 | 15-30 | 1669 | 0.4-0.6 | 705 |
| dsD(LF) | BSA | 1 | 2064 | 20- 35 | 15-30 | 1337 | 0.4-0.6 | 553 |
| dsD(LF) | BSA | 2 | 3819 | 20- 35 | 15-30 | 882 | 0.4-0.6 | 321 |
| dsD(NF) | BSA | 1 | 2382 | 20- 35 | 15-30 | 1533 | 0.4-0.6 | 652 |
| dsD(NF) | BSA | 2 | 3246 | 20- 35 | 15-30 | 1800 | 0.4-0.6 | 712 |
| dsD(HF) | PEG | 1 | 2299 | 20- 35 | 15-30 | 1491 | 0.35-0.65 | 1195 |
| dsD(HF) | PEG | 2 | 1120 | 20- 35 | 15-30 | 708 | 0.35-0.65 | 554 |
| dsD(MF) | PEG | 1 | 1188 | 20- 35 | 15-30 | 858 | 0.35-0.65 | 717 |
| dsD(MF) | PEG | 2 | 653 | 20- 35 | 15-30 | 378 | 0.35-0.65 | 233 |
| dsD(LF) | PEG | 1 | 1082 | 20- 35 | 15-30 | 602 | 0.4-0.6 | 428 |
| dsD(LF) | PEG | 2 | 1410 | 20- 35 | 15-30 | 811 | 0.35-0.65 | 495 |
| dsD(NF) | PEG | 1 | 1151 | 20- 35 | 15-30 | 689 | 0.35-0.65 | 520 |
| dsD(NF) | PEG | 2 | 3884 | 20- 35 | 15-30 | 2450 | 0.35-0.65 | 1845 |

**Supplementary Tab. 16: Used chemicals and hardware for 'Sample Preparation'**.

| Chemicals | | |
|---|---|---|
| **Description** | **Full name** | **Supplier** |
| Trolox | 6-Hydroxy-2,5,7,8-tetramethylchroman-2-carboxylic acid | Sigma-Aldrich |
| Neutravidin | | Thermo Scientific |
| BSA-biotin | Albumin, biotin labelled bovine | Sigma-Aldrich |
| Helmanex lll | | Sigma-Aldrich |
| Ethanolamine hydrochlorid | | Sigma-Aldrich |
| DMSO | Dimethyl sulphoxide | VWR chemicals |
| Biotin-PEG-NHS, M = 3145 g / mol | | Iris Biotech GmbH |
| Chloroform > 99.9% | | Sigma-Aldrich |
| Acetonitrile | Methyl cyanide | Sigma-Aldrich |
| PBS | | GIBCO |
| **Hardware and consumables** | | |
| **Description** | **Article name** | **Supplier** |
| Sonificator | RK510 H | Bandelin electronic |
| Plasma Cleaner | Femto | Diener Electronic |
| Plasma Cleaner for AFM data, operated at 100W and ambient gas | PlasmaFlecto 10 | plasma technology GmbH |
| Cover glasses, 24mm x 60 mm, 170 +-5 µm thick | No. 1.5H | Marienfeld |

**Supplementary Tab. 17: Complete description of all analysis parameters for the SEIDEL software**. See '*Spectroscopy and image analysis, Filtering procedures',* Supplementary Figs. 3 and 6.

| Parameter | Description |
|---|---|
| **Read/Write settings (A)** | |
| File directory | Folder containing the raw photon (.ptu) files |
| Output folder | Location to save output files. |
| **Lifetime image loading (B1)** | |
| Ggate_loc, Rgate_loc, Ygate_loc | Donor, FRET sensitized acceptor, direct excited acceptor localization gate range. Localization may be performed on gated or ungated data. |
| Ggate_lt, Rgate_lt, Ygate_lt | Donor, FRET sensitized acceptor, direct excited acceptor lifetime gate range. The spot-wise lifetime is always fitted on gated data. |
| Ggate_I, Rgate_I, Ygate_I | Donor, FRET sensitized acceptor, direct excited acceptor intensity gate range. The spot-wise intensity is always obtained from ungated data. |
| Framestop_FRET | Starting from the first frame, how many frames are merged to collect statistics for FRET analysis. Not all frames are used to exclude bleaching which is more likely to occur as more frames are accumulated. |
| Ntacs | Number of TAC channels in lifetime decay. Default 256. |
| Rebin | If given, the image is rebinned by an integer factor. Default False. |
| Framestop_localization | How many frames to collect for localization. Default all frames are taken |
| **Fit region Selection (B2)** | |
| ROI_threshold_abs | Minimal amplitude for ROI selection. Default 1 count / pixel. |
| ROI_threshold_relative | Minimal amplitude for ROI selection relative to image max. Default 0.3 (norm. int.) |
| min_distance | Minimal distance between the centers of two ROIs in pixels. Default 15 pixels. |
| smoothIntensitySigma | Standard deviation of Gaussian filter to smooth the image and identify regions of interest. Default 2 pixels. |
| ROIsize | Side of square region of interest. Default 20 pixels |
| spot integration area | Integration area around a spot center used for spectroscopy information. Default 7 x 7 pixels. |
| **Fit parameters (B3)** | |
| DTwoIstar | Minimal improvement in 2I* after an additional spot has been added to the fit model for the fit to be accepted as better. Default 0.03 normalized likelihood) |
| garbageBrightness | Spots with a lower intensity are discarded. Default 20 counts integrated over a spot. |
| junkIstar | Pure-noise data sometime have very low 2I*. Fits with a 2I* below are not considered good. Default value 0.30 (normalized likelihood) |
| fitbg | fit background when fitting Gaussian spots. Default True |
| setbg | Initial estimate for background. Default 0.2 counts / pixel. |
| Ellipt_circ | If True, allows for elliptical spots. Default False. |
| **Correction parameters (C)** | |
| max_dist (post-processing) | All pair distances more than max_dist away from the mean displacement are considered junk and kicked out. Default 100nm. |
| **Correction and selection parameters (D)** | |
| Image_stoichiometry | Image stoichiometry lists the number of green and yellow spots that are located in an image. It is different from stoichiometry obtained from photon counts. One can chose to incorporate only images with e.g. 1 green and 1 yellow spot. |
| Channel shift correction | Correct for channel shift. Per default a (0nm, 5nm) shift in (x, y) is applied as the acceptor direct excitation image is shifted by half a pixel because of the line interleaved donor and acceptor direct excitation and constant scanner velocity in y. |
| Lifetime background | Background level for lifetime tail fits. Default: 0 kHz. |

***Supplementary Tab. 18: List of filtering parameters.*** Some parameters are defined for a single spot and others are defined only for a donor-acceptor spot pair (multi-spot precedure). Acceptor (PIE) is used to indicate acceptor under direct excitation, acceptor (FRET) is used to indicate FRET sensitized acceptor. Additional intensity-derived parameters are also available but not listed explicitly. [1]Anisotropy is only available in AnI-3SF program. [2]Lifetime decay histogram is an array of values and therefore is separately exported (Supplementary Fig. 6, panel F1). [3]Correction parameters are used to calculate these values in a separate program (Margarita). See '*Spectroscopy and image analysis, Filtering procedures',* Supplementary Figs 3 and 6.

| FRET indicator | Type |
|---|---|
| Donor / acceptor (PIE) position | Single spot |
| Donor / acceptor (PIE) localization photons | Single spot |
| Donor / acceptor (PIE) spot width | Single spot |
| Donor / acceptor (PIE) spot ellipticity | Single spot |
| Donor / acceptor (PIE) background | Single spot |
| Number of donor / acceptor (PIE) spots | Single spot |
| Donor / acceptor (FRET) / acceptor (PIE) lifetime | Single spot |
| Donor / acceptor (FRET) / acceptor (PIE) intensity | Single spot |
| Anisotropy[1] | Single spot |
| Donor / acceptor (FRET) / acceptor (PIE) lifetime decay histogram[2] | Single spot |
| Localization distance | Spot pair |
| Proximity ratio (uncorrected Efficiency) | Spot pair |
| Stoichiometry[3] | Spot pair |
| Efficiency (corrected)[3] | Spot pair |

*Supplementary Tab. 19: Position of dyes on the origami platform determined by localization analysis.* The estimated positions are shown in the main text in Fig. 2b and the respective interdye distances in Fig. 2e. The positions are aligned according to the origami frame of reference such that the D1 point is in the origin and that the x axis is parallel to the helical axis, as indicated by the superscript [D1]. The distribution of alignment scores is displayed in Supplementary Fig. 9. For the interdye distances, the experimental coordinate system is used which is rotated by an angle of 18 degrees with respect to the origami frame of reference such that both acceptor positions lie on the x axis, as indicated by the superscript [AA]. For the distances, the spacing between the dyes in terms of the structural parameters of the origami nanostructure (number of helices $n_h$ and base pairs $n_{bp}$) is given in addition. x, y: measured x- and y-displacement; x model, y model: x- and y-displacement based on the structural model described in Supplementary Note 2 (eq. S35) with d: interdye distance, d model: interdye distance based on the modeled x- and y-displacements..

| | x [nm] | x model [nm] | y [nm] | y model [nm] | d [nm] | d model [nm] | $n_h$ | $n_{bp}$ |
|---|---|---|---|---|---|---|---|---|
| **D1 position** | -0.5± 0.4 | $0^{D1}$ | 0.1±0.3 | $0^{D1}$ | | | | |
| **D2 position** | 75.0±0.4 | 75 | 2.7±0.4 | 2.4 | | | | |
| **A1 position** | 0.35±0.4 | $0^{D1}$ | -14.6±0.3 | -14.4 | | | | |
| **A2 position** | 75.1±0.4 | 75 | -2.6±0.4 | -2.4 | | | | |
| **D1-A1 distance (NF)** | -1.8±0.7 | -2.3 | 14.5±0.7 | 14.2 | 14.6±0.7 | 14.4 | -6 | 0 |
| **D1-D2 distance** | 74.5±0.7 | 74.0 | 9.3±1.0 | 9.5 | 75.1±0.7 | 74.6 | 1 | 236 |
| **D1-A2 distance** | 76.4±0.7 | 76.3 | -5.3±0.7 | -4.7 | 76.5±0.7 | 76.4 | -1 | 236 |
| **A1-D2 distance** | 74.5±0.7 | 73.3 | 14.5±0.7 | 14.2 | 75.9±0.7 | 74.7 | 7 | 236 |
| **A1-A2 distance** | 75.9±0.6 | 75.6 | $0^{AA}$±0 | $0^{AA}$ | 75.9±0.6 | 75.6 | 5 | 236 |
| **D2-A2 distance (HF)** | -0.3±0.7 | -0.8 | 5.3±0.7 | 4.7 | 5.3±0.7 | 4.8 | -2 | 0 |

**Supplementary Tab. 20: Polynomial coefficients for linker corrected static FRET line**.

Linker-corrected static FRET–lines are approximated by a fourth-order polynomial as given in equation 10 in the main text. See Supplementary Figs. 13a , 12 (origami) and 21, 29 (DNA Ruler).

| | sample | $\tau_{D(0)}$ [ns] | $a_1$ | $a_2$ | $a_3$ | $a_4$ | $a_5$ |
|---|---|---|---|---|---|---|---|
| **Imaging** | **dsDNA** | 3.84 | -0.0125 | 0.5910 | 0.3127 | -0.0800 | 0.0069 |
| | **Origami** | 3.6500 | -0.0067 | 0.7172 | 0.2383 | -0.0678 | 0.0065 |
| | **hGBP1** | 3.9 | -0.0126 | 0.5925 | 0.3070 | -0.0773 | 0.0066 |
| **confocal sm spectr.** | **dsDNA** | 3.9 | -0.0131 | 0.5815 | 0.3138 | -0.0786 | 0.0066 |
| | **Origami** | 3.9 | -0.0109 | 0.6352 | 0.2726 | -0.0684 | 0.0058 |
| | **hGBP1** | 3.84 | -0.0125 | 0.5910 | 0.3127 | -0.0800 | 0.0069 |

**Method figures**

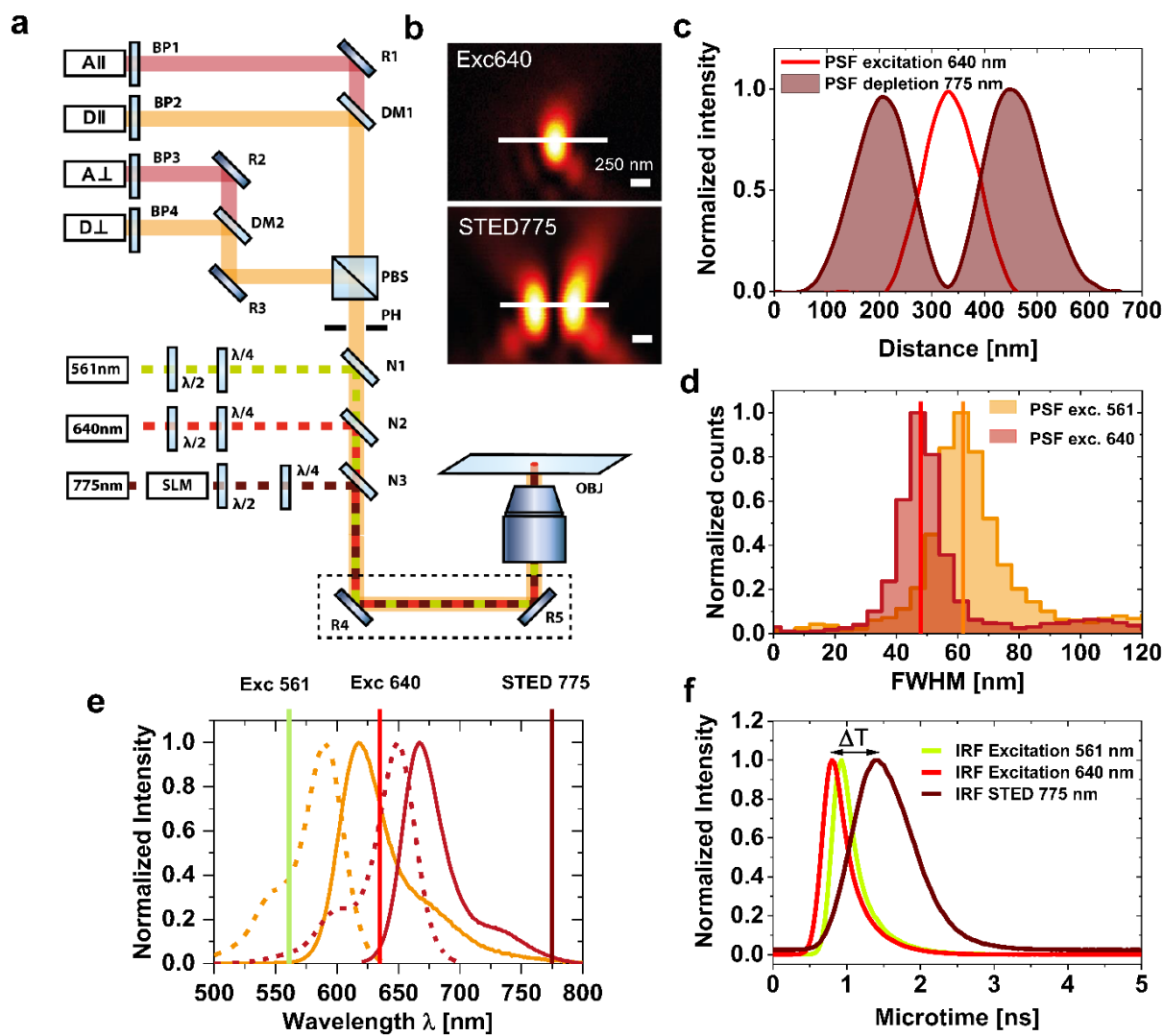

***Supplementary Fig. 1: STED-FRET microscope. a)*** Scheme of the microscope with pulsed laser excitation at 561 nm and 635 nm, and depletion at 775 nm. The STED 775 laser is modulated by a spatial light modulator (SLM) to generate 2D donut transverse mode. The excitation and depletion lasers are overlaid by notch filters (N1-N3). The emitted light is guided back through a pinhole (PH), split by polarization using a polarizing beam splitter (PBS), split spectrally by dichroic mirrors (DM), and filtered by band pass filters (BP) in front of the detectors. A detailed description is given in Supplementary Material and Methods - STED-FRET microscope. **b-c)** Recorded point spread function (PSF) of excitation laser 640 nm and depletion laser 775nm measured on gold beads (size: 150nm). **d)** Histogram of determined full width at half maximum (FWHM) of PSF measured on single biotin dsD(BF) sample. **e)** Excitation and emission spectra of used dyes attached to dsDNA. Vertical lines illustrate the used excitation and depletion wavelength. **f)** Recorded instrument response functions (IRF) of used excitation/ depletion laser (561nm, 640nm and 775nm) showing the time delay of 0.5 ns for the depletion laser 775nm.

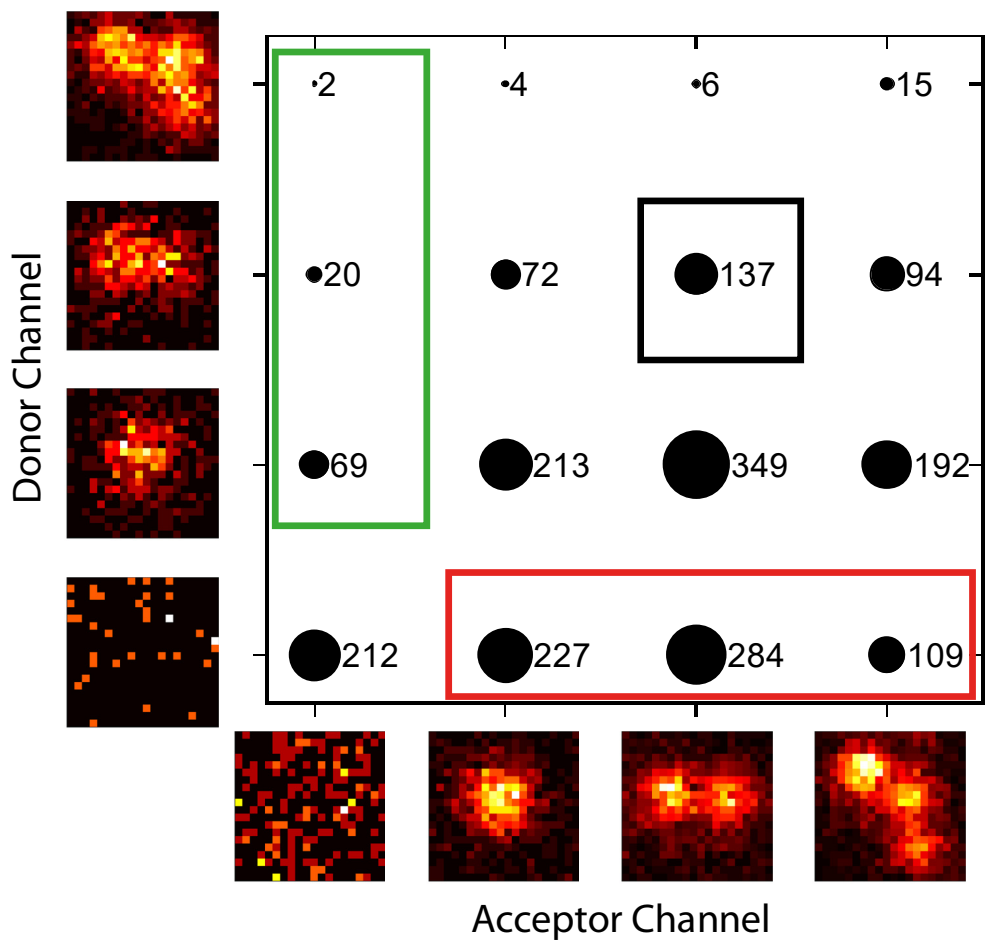

**Supplementary Fig. 2: Spot stoichiometry for DNA origami.** Data corresponds to the O(HF+NF) sample. Black circles with numbers indicate the number of constructs corresponding to that spot stoichiometry species. The green box indicates constructs without acceptors (donor-only), used to determine the crosstalk and donor only lifetime. Similarly, the red box indicates constructs without donors (acceptor-only), used to determine direct excitation and acceptor lifetime. The black box indicates fully labelled structures with two FRET pairs, which are used to obtain an average structure. Partially labelled structures (1 donor - 1 acceptor, 1 donor – 2 acceptors, 2 donors – 1 acceptor) each contain one FRET pair and all available FRET pairs are included in Fig. 2b of the main text. An emitter stoichiometry of >2 can occur due to crowding or aggregation. Overall, more acceptor spots than donor spots are detected, attributed to the weaker donor signal due to a lower detection efficiency and quenching by FRET. See '*Spectroscopy and image analysis, Filtering procedures*', Supplementary Figs. 3 and 6.

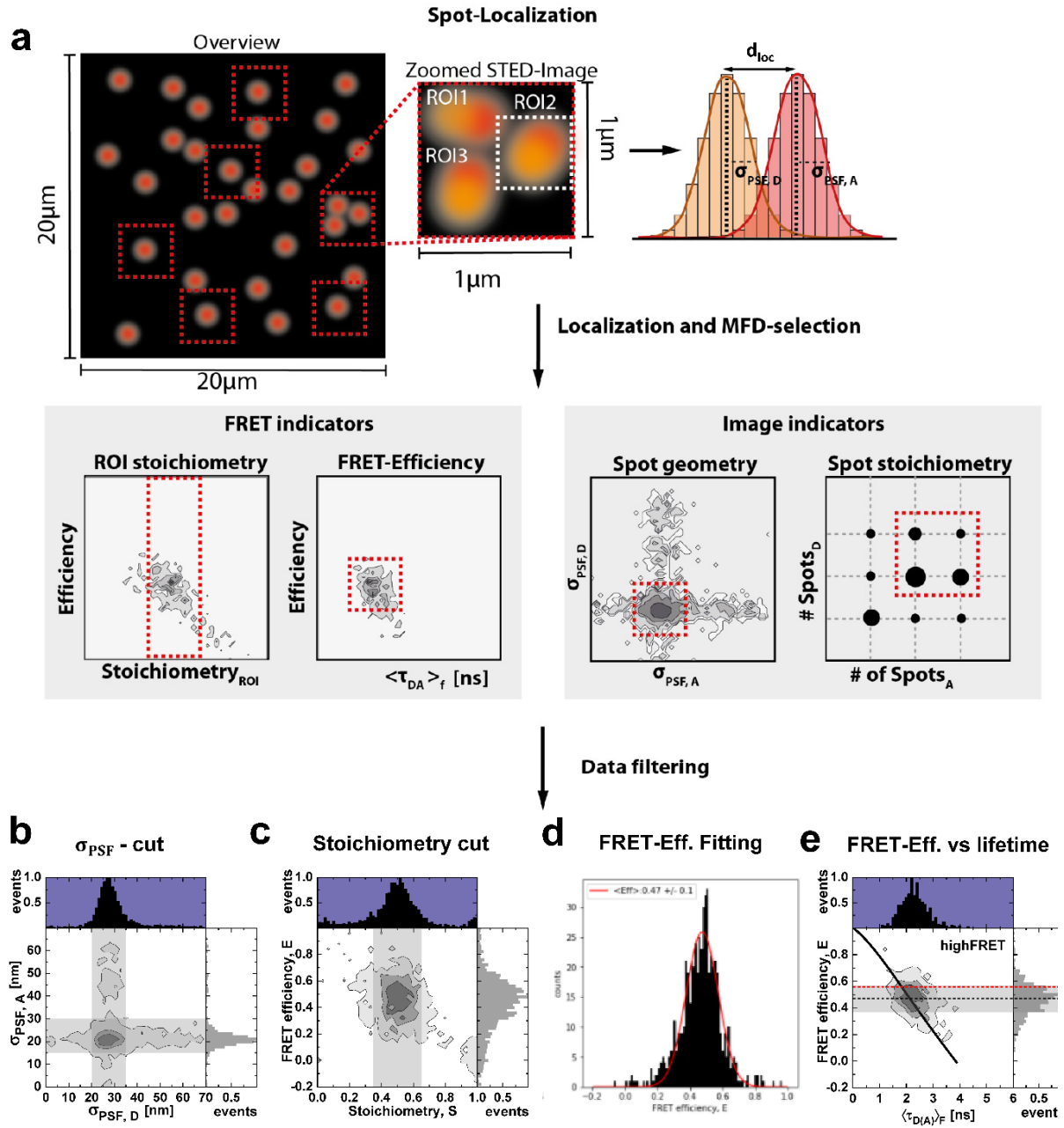

**Supplementary Fig. 3: Schematic data filtering workflow.** a) Schematic illustration of filtering procedure for dsDNA (dsD) and origami samples (O). b-e: Applied data filtering shown on single biotin labeled dsD(HF) sample. Intensity based FRET-derived distances are determined by b) filtering of 2D Gaussian fitted spots by sigma selection c) only donor acceptor pairs are selected with stoichiometry of 0.5 d) FRET efficiency histogram is fitted by single Gaussian model to determine the mean and sigma (grey area). e) 2D histogram of FRET Efficiency vs fluorescence weighted donor lifetime. Greyed area marked $1\sigma$ fitted FRET efficiency histogram and is used to calculate the mean FRET efficiency (horizontal, dotted, black line). The predicted FRET efficiency by AV simulation is shown as horizontal red dotted line. See also Supplementary Tabs 14, 15, 17, 18 and Supplementary Fig. 6 for details

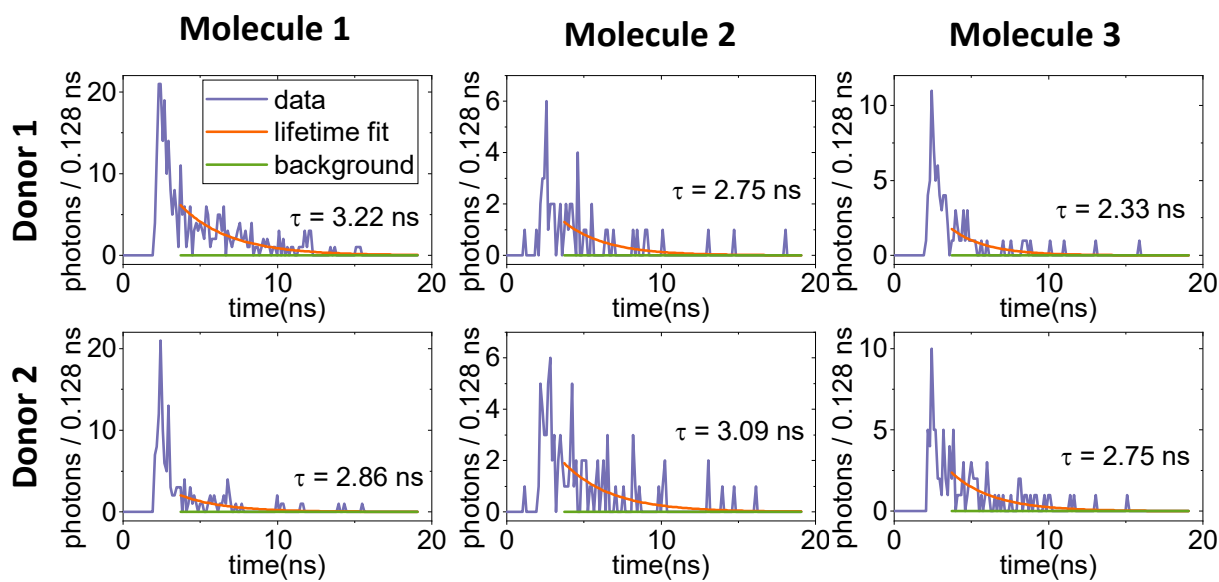

**Supplementary Fig. 4: Exemplary ROI-wise donor fluorescence decay fits.** Three origami molecules carrying two FRET pairs O(HF+NF) were randomly selected. See '*Spectroscopy and image analysis, Determination of spot-integrated fluorescence lifetimes*'.

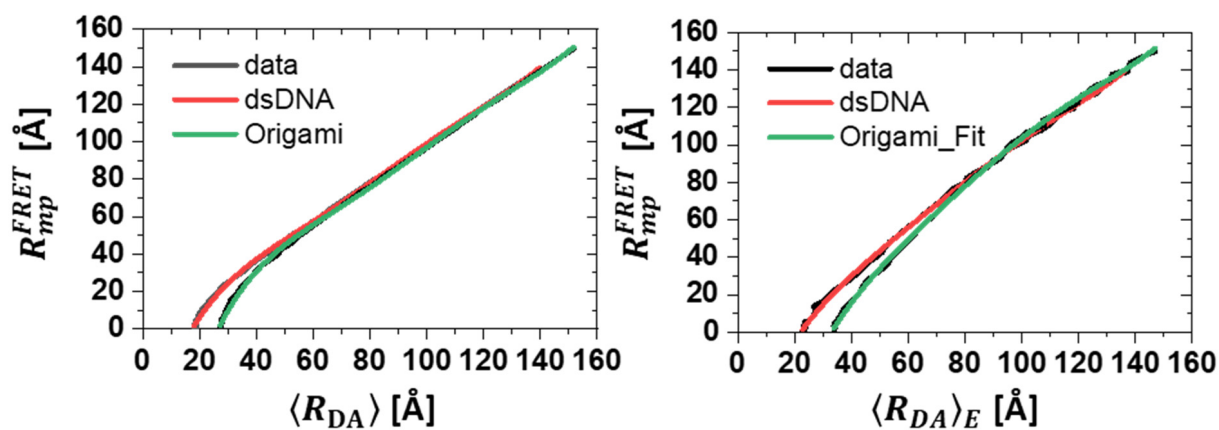

**Supplementary Fig. 5: Conversion between FRET-averaged distances and mean-position distances.** Two FRET pairs were analyzed: Alexa594-Atto647N (for DNA rulers) and Atto594-Atto647N (for origami). The relation between the mean position distance $R_{mp}^{FRET}$ and the mean donor-acceptor distance $\langle R_{DA} \rangle$ (left) or the FRET-averaged interdye distance $\langle R_{DA} \rangle_E$ was approximated by a fifth-order polynomial. The polynomial coefficients and the corresponding equation are given in Supplementary Tab. 8. The polynomials were determined as described in Kalinin et al.[18]

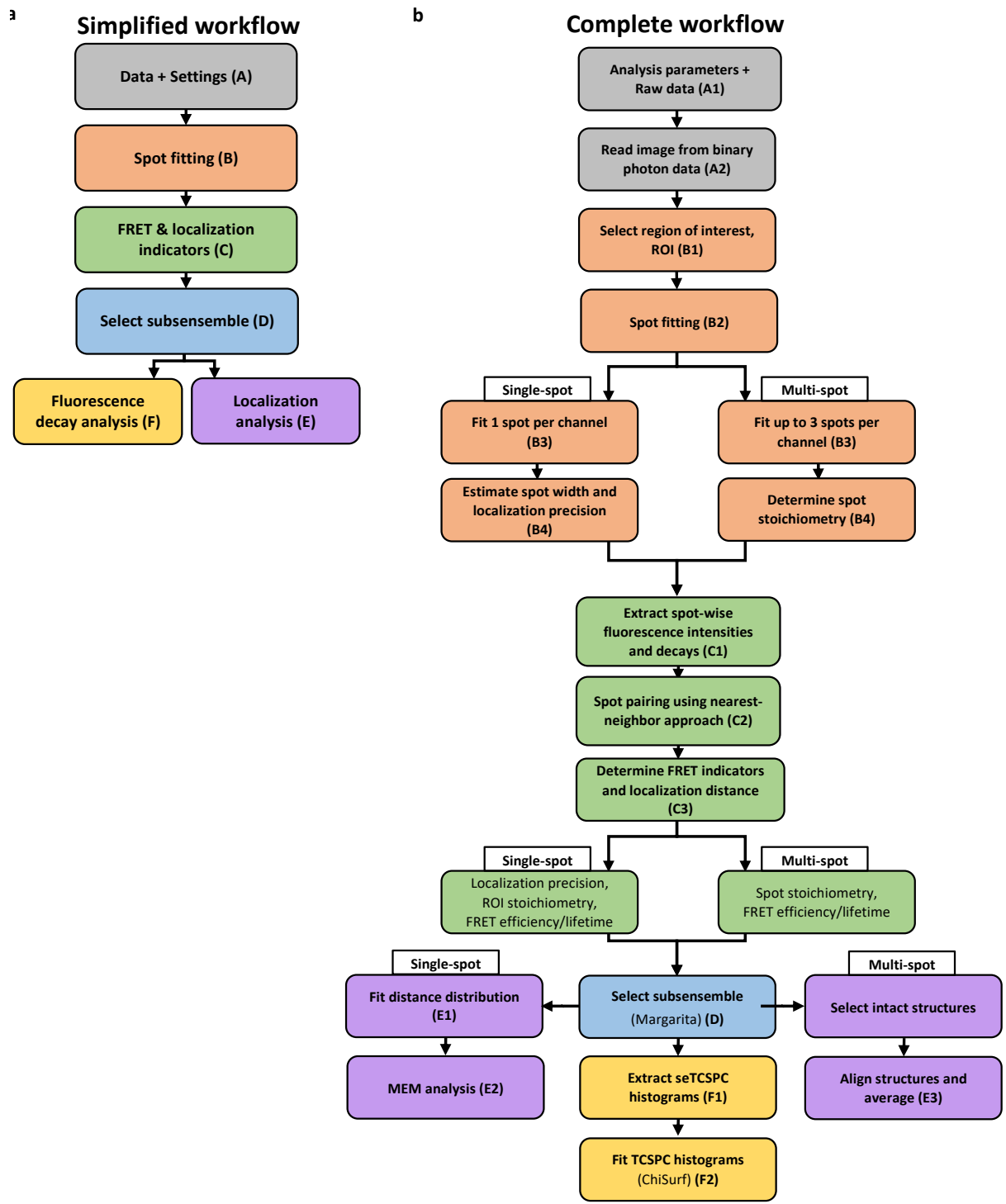

**Supplementary Fig. 6: Analysis workflow. a**) Simplified workflow. **b**) Complete workflow. See Supplementary Note 1 for a detailed description of the workflow.

**Origami figures**

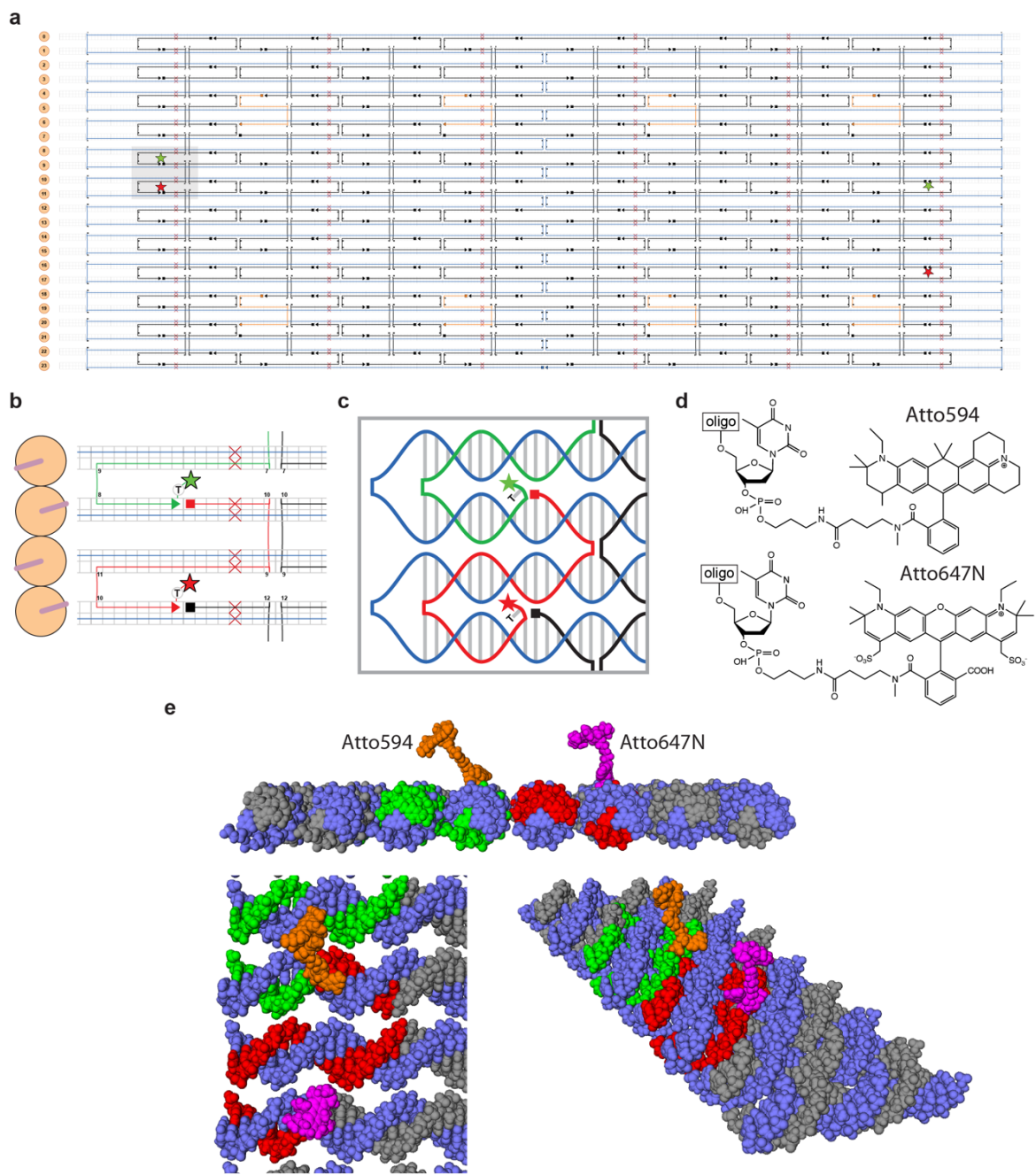

***Supplementary Fig. 7: DNA origami design.* a)** Strand diagram of the single-layer DNA origami rectangle. Green and red stars indicate the labeling positions of the donor (Atto594) and acceptor (Atto647N) dyes, respectively. The schematic was generated using the caDNAno software[5]. Squares signify the 5' end and arrow heads the 3' end of the staple strands. **b)** Zoom-in of the high FRET dye pair (gray shaded area in panel a). The scaffold strand is colored in blue, staple strands in black and the strands carrying the donor and acceptor are colored green and red, respectively. 5' ends of staple strands are indicated by squares, 3' ends by arrow tips. The dyes are linked to the 3' end with an unpaired thymine base pair as a spacer. A side-view of the four displayed helices on the left indicates the orientation of the orange shaded base pairs. Base pairs used for the dye attachment point upwards

from the plane are defined by the origami rectangle. Red crosses indicate deletions in the structure used to reduce twisting of the single-layer sheet. **c)** Cartoon representation of the fragment shown in panel b, providing a more detailed view of the routing of the scaffold and staple strands. **d)** Chemical structures of the dyes and the labeling chemistry used for the attachment to the 3' end. A C3-aminolink group on the terminal phosphate is reacted with a N-hydroxysuccinimidyl residue on the dyes. **e)** 3D model of the dyes linked to the origami structure in a side view (top), top view (bottom left) and perspective view (bottom right). The strands are colored as in panels b and c. Dyes are colored orange (Atto594) and magenta (Atto647N). The single-stranded scaffold overhangs are not displayed. See '*Sample preparation, DNA origami*', Supplementary Tab. 2 and Supplementary Fig. 15.

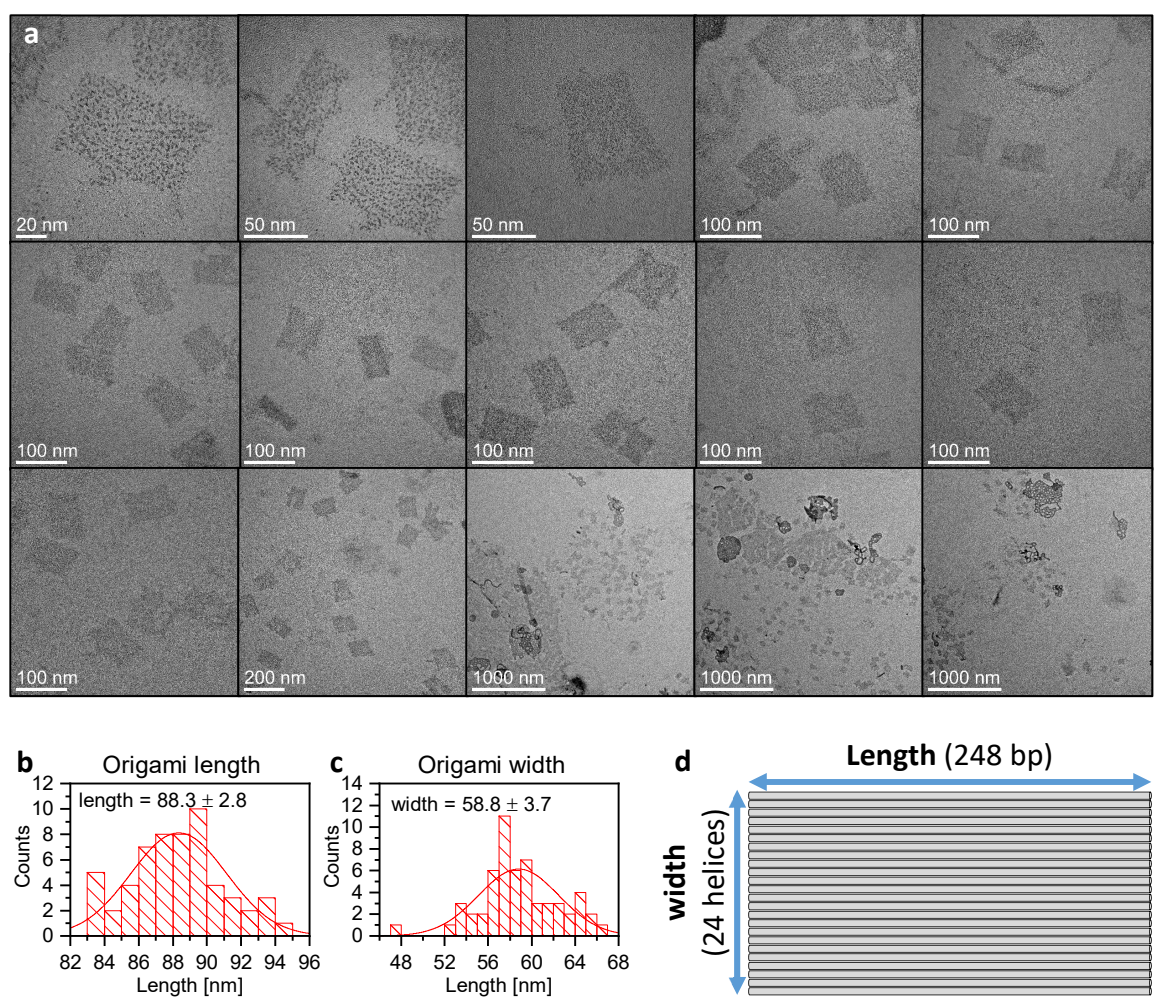

**Supplementary Fig. 8: TEM imaging of DNA origami platforms. a)** Overview of recorded TEM images. Images are sorted from higher magnification to lower magnification. **b-c)** The dimensions were measured manually using ImageJ image analysis software and fitted to Gaussian distribution to determine the mean. **d)** Schematic of the origami platform. The dimension of the origami platform is the same regardless which labels are attached to it. The applied procedures are described in '*Transmission electron microscopy*'.

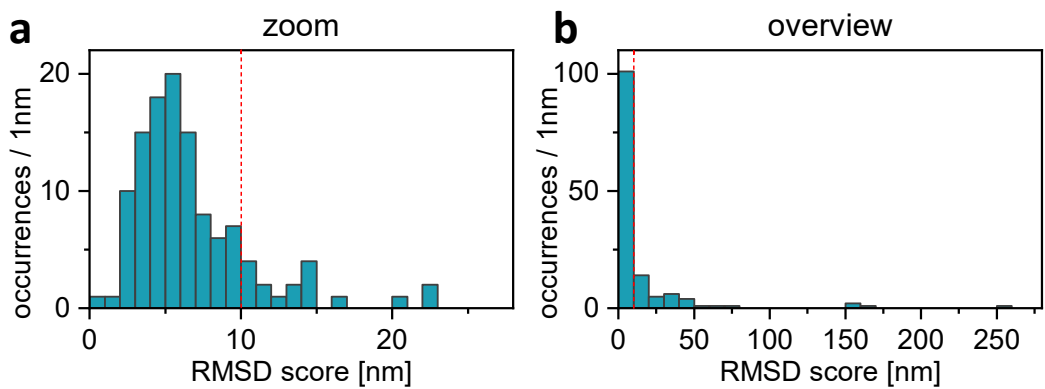

**Supplementary Fig. 9: Distribution of alignment RMSD scores obtained for origamis. a)** The distribution of the root-mean-square-displacement (RMSD) scores as calculated with respect to the best reference structure for the O(HF+NF) sample. It shows a single peak around 5 nm, close to the lowest possible localization precision based on photon statistics of 2.8 nm. Origamis with an RMSD score higher than 10 nm (red line) are discarded for further analysis. **b)** Extreme outliers (RMSD score > 100 nm) are attributed to aggregation or random placement of two partially labelled structure in close proximity. Origami selection criteria are listed in Supplementary Tab. S14. See also *'Spectroscopy and image analysis, Alignment and particle averaging for origami measurements'*.

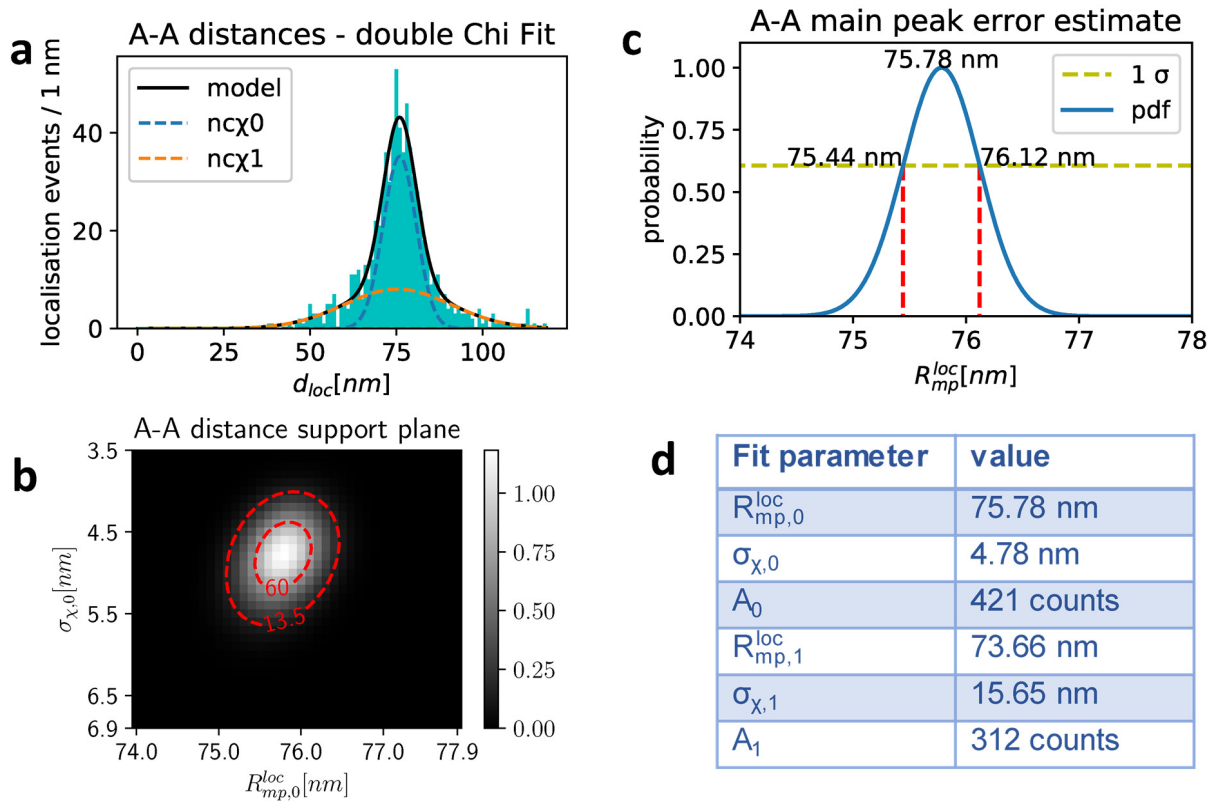

| Fit parameter | value |
|---|---|
| $R_{mp,0}^{loc}$ | 75.78 nm |
| $\sigma_{\chi,0}$ | 4.78 nm |
| $A_0$ | 421 counts |
| $R_{mp,1}^{loc}$ | 73.66 nm |
| $\sigma_{\chi,1}$ | 15.65 nm |
| $A_1$ | 312 counts |

**Supplementary Fig. 10: Acceptor-to-acceptor distance distribution of origamis.** Constructs with 2 acceptors and 0, 1, 2 or 3 donor spots were selected from the O(HF+NF) sample. **a)** The colocalization distance histogram was fitted with maximum likelihood optimization and two non-central $\chi$-distributions, each with fitting parameters A, $R_{mp}^{loc}$ and $\sigma_{\chi}$. A single $\chi$-distribution model was rejected based on the Aikaike information criterion (AIC=478, not shown), which was significantly better for the two-component fit (AIC=340). **b)** Probability density function (pdf) for $R_{mp}^{loc}$ and $\sigma_{\chi}$ calculated from the likelihood. *p*-values are normalized to indicate the likelihood of the true value being in a 1 x 1 nm area. Dashed lines indicate 60% and 13.5% probability relative to the highest *p*-value. The optimum is well defined and no covariance between $R_{mp}^{loc}$ and $\sigma_{\chi}$ is observed, as is expected when $R_{mp}$ / $\sigma_{\chi}$ >> 1. **c)** Pdf for $R_{mp,0}^{loc}$ where the highest *p*-value is normalized to one. The distribution is Gaussian and 60% *p*-values indicate the 1$\sigma$ confidence interval, yielding an acceptor-acceptor distance of 75.8 ± 0.3 nm. The acceptors are separated by 5 helices, or 12nm along the short direction. Using the Pythagorean theorem, we obtain the distance along the helical strands to be 74.8 ± 0.3 nm. **d)** Fit parameters corresponding to **a** (see eq.30). The presence of two peaks indicates two species of origamis. The defined population with narrow width (index 0) is attributed to correctly folded origamis as the width is close to the predicted localization precision (see Supplementary Tab. 5). The broad distribution (index 1) is attributed to incorrectly folded or broken origamis. For further details see Supplementary Note 2.

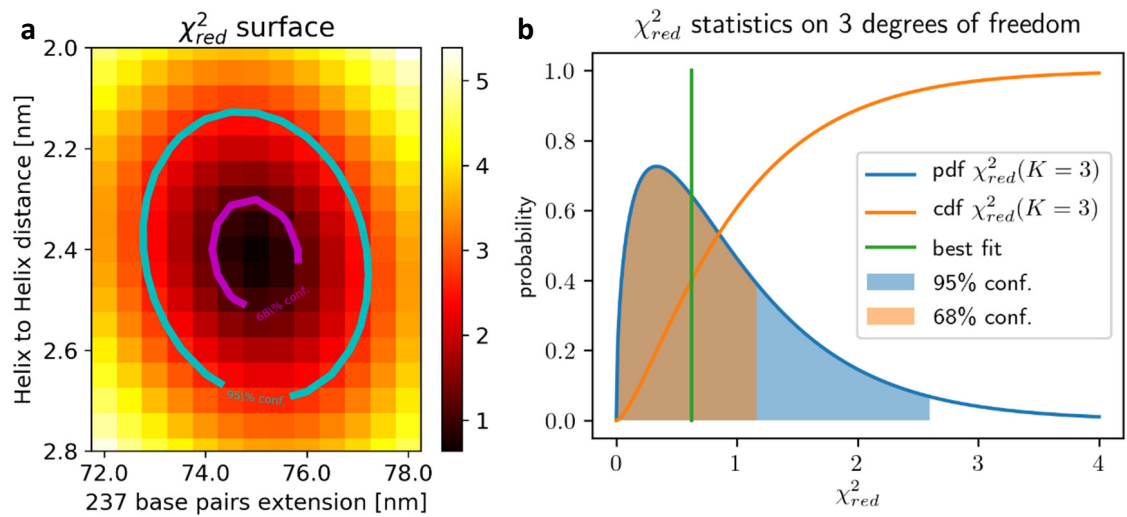

**Supplementary Fig. 11: Error analysis on origami lattice constants.** **a)** $\chi^2_{\mathrm{red}}$ surface as a function of the base pair extension and average helix to helix distance. Confidence interval limits are reported in Supplementary note 2. **b)** $\chi^2_{\mathrm{red}}$ distribution for three free parameters. Limiting $\chi^2_{\mathrm{red}}$ values for confidence intervals are obtained from integrating the curve from the left (shaded areas) For further details see Supplementary Note 2.

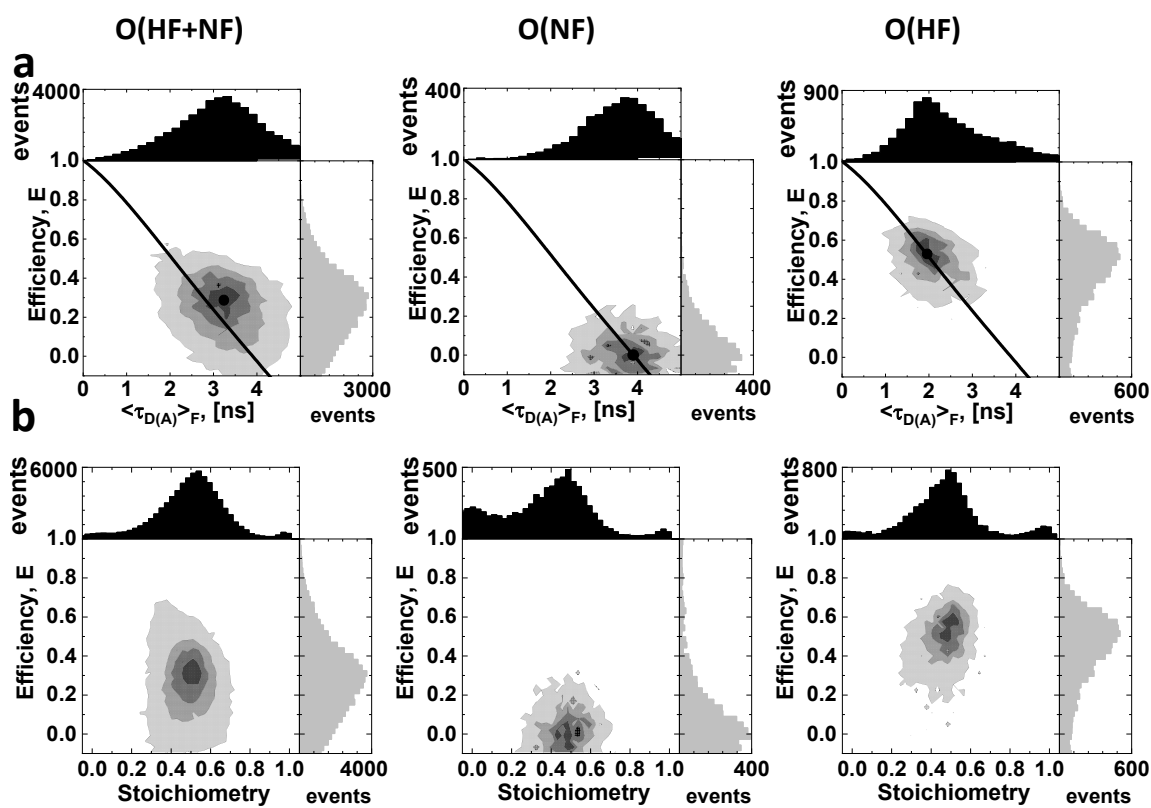

**Supplementary Fig. 12: Confocal single-molecule FRET measurements of three origami samples: O(HF+NF), O(NF) and O(HF).** Confocal single-molecule measurements using multiparameter fluorescence detection were performed to characterize the O(HF+NF), O(NF) and O(HF). **a)** Two-dimensional frequency histogram of the ROI-integrated intensity-based FRET efficiency, E, and the fluorescence-weighted average donor fluorescence lifetime, $\langle\tau_{D(A)}\rangle_F$. The ROI populations lie on a static FRET line including the contribution of the flexible linker (solid line, given by eq.10 in the Methods section of the main text using the parameters in Supplementary Tab. 20). The population centers are reported in Supplementary Tab. 10. O(HF+NF) contains a mixture of two FRET species and lies above the static FRET line. **b)** Two-dimensional frequency histogram of the ROI-integrated intensity-based FRET efficiency, E, and stoichiometry. The necessary setup correction factors for the computation of the plotted corrected parameters are compiled in Supplementary Tab. 3. The stoichiometry histograms (b) show a shoulder towards lower a stoichiometry of 0.33, indicating constructs where two acceptors and a single donor is observed. This observation correlates with longer diffusion times (not shown), indicating aggregation of origami platforms, e.g. via single strand loops protruding from the short side of the platform. See '*Confocal single-molecule spectroscopy with multiparameter fluorescence detection*', and compare with Supplementary Fig. 13.

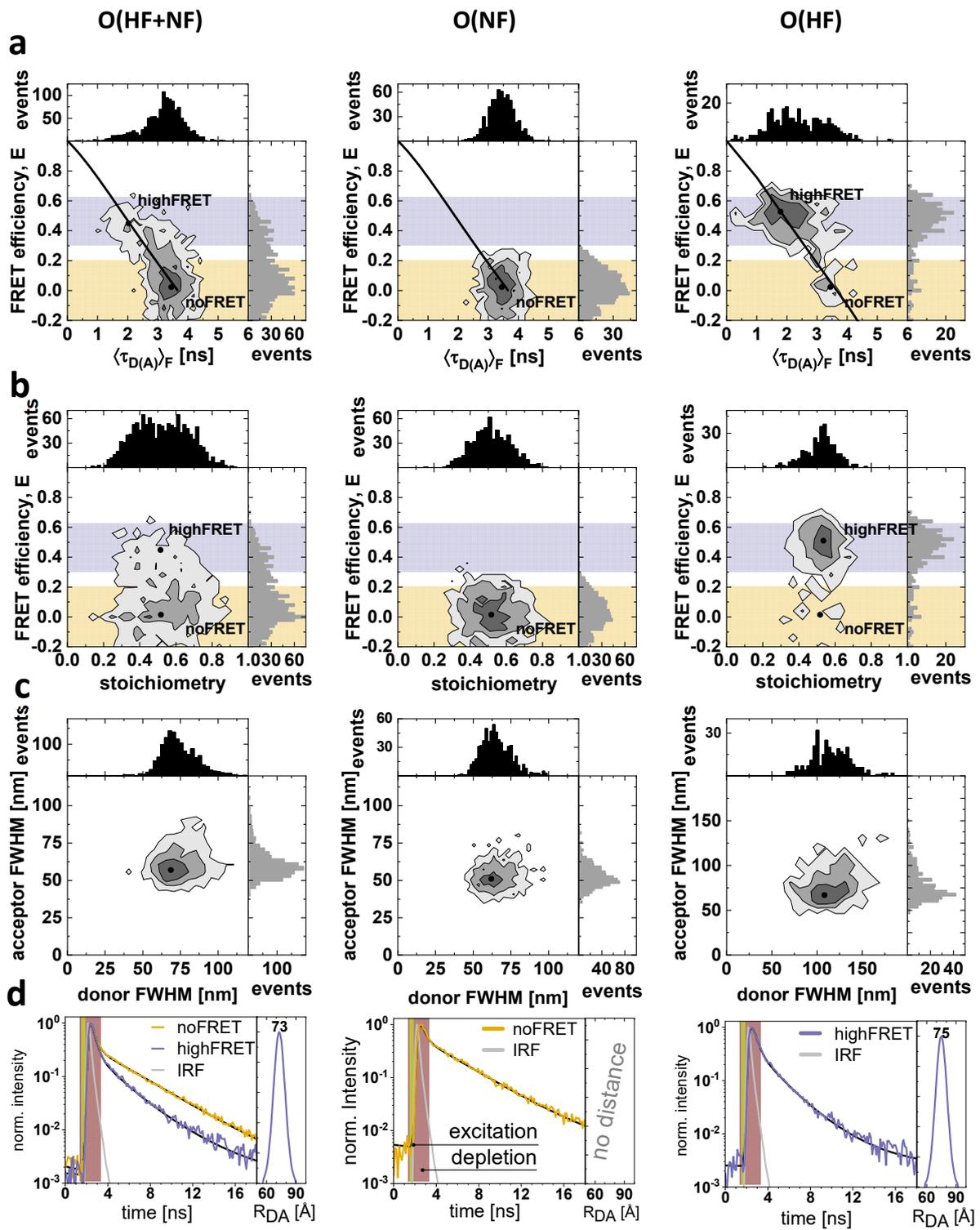

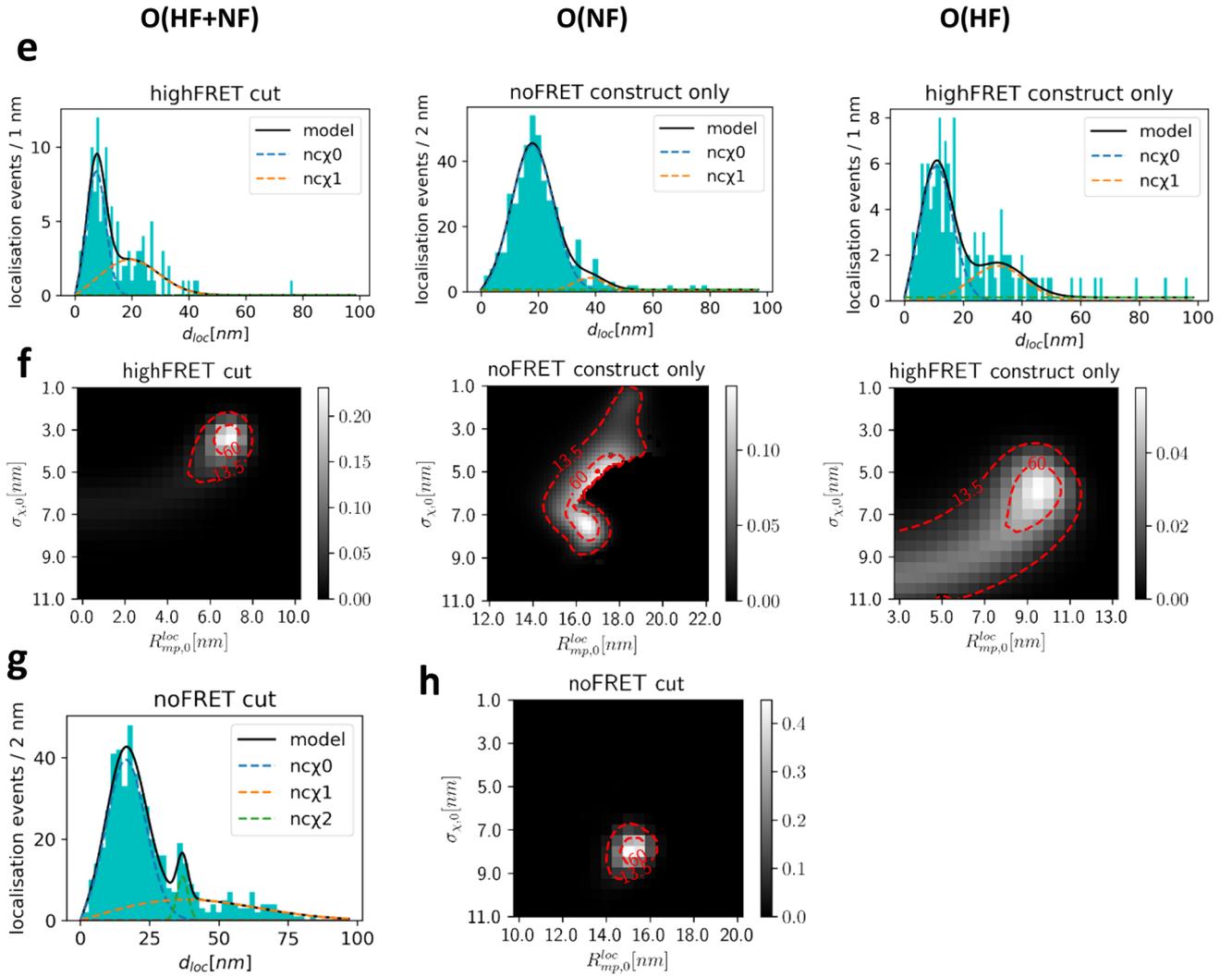

**Supplementary Fig. 13: FRET nanoscopy of three origami constructs: O(HF+NF), O(NF) and O(HF). a)** Two-dimensional frequency histogram of the ROI-integrated intensity-based FRET efficiency, E, and the fluorescence-weighted average donor fluorescence lifetime, $\langle \tau_{D(A)} \rangle_F$. All ROI populations lie on a static FRET line including the contribution of the flexible linker (solid line, given by eq.10 in the Methods section of the main text using the parameters in Supplementary Tab. 20). Populations lie on the static FRET line. For the O(HF+NF) and O(HF) tailing of the high FRET population is visible due to acceptor bleaching. Numeric values for the population centers are reported in Supplementary Tab. 10. **b)** Corresponding two-dimensional frequency histogram of the ROI-integrated intensity-based FRET efficiency, E, and the fluorescence-stoichiometry. The stoichiometry for O(HF+NF) is broader as the dataset originates from two measurement days with different laser power. The FRET populations lie in the violet shaded areas. **c)** Full-width-at-half-max (FWHM) of the donor and acceptor as determined for Gaussian spot fitting. The resolution in the donor channel and acceptor channels are 60 nm and 50 nm for the O(NF) sample, 67 nm and 57 nm for the O(HF+NF) sample and 105 nm and 70 nm for the O(HF) sample, respectively. For the O(HF) data, FRET shortened the lifetime of the donor, reducing the number of photons available for localization on gated data. Consequently, localization was done on ungated data to better identify spots. **d)** Sub-ensemble donor decay histograms and fits with obtained distance distributions (side panel). Yellow bar indicates the duration of the excitation pulse and light red bar indicates duration of the STED pulse. Fit models are described in the methods. All fit results are compiled in Supplementary Tab. 9. **e,g)** Localization histograms and fits. Localizations outside the main peak are attributed to incomplete filtering of broken structures and an additional $\chi$-distribution (see eq. S30) was used to describe the data. The green line for O(NF) and O(HF) indicates a constant background offset. **f,h)** 2D Probability density functions (pdf) for the principal peak in (e, g). *p*-values are normalized to indicate the likelihood of the true value being in a 1 x 1 nm area. Dashed lines indicate 60% and 13.5% probability relative to the highest *p*-value. Purple and Yellow area in a, b indicate efficiency cuts used in d-h. Additional filtering criteria are reported in Supplementary Tab. 14. Comparison of columns in spectroscopic indicators (a, b, d) show that the O(HF+NF) parameters consist of the sum of O(NF) and O(HF), indicating that accurate spectroscopic parameters can be determined from the two dyes on the same platform (75nm separation) as good as if they were isolated. This result is confirmed by comparison of localization histograms where main populations for O(HF+NF) (NF: 15.0 ± 0.7 nm, HF: 6.8 ± 1 nm) match the populations of O(NF) (NF: 16.5 ± 1 nm) and O(HF) (HF: 9.5 ± 1.5 nm). While the localization histograms can be used to determine any distance, incomplete filtering of broken constructs is detrimental to the accuracy and it represents a challenge to improve filtering further in future work. In comparison, filtering by aligning structures is able to completely reject broken structures. The former is more widely applicable as it can use probe with at least two labels, whereas the latter is restricted to constructs with at least 3 labels.

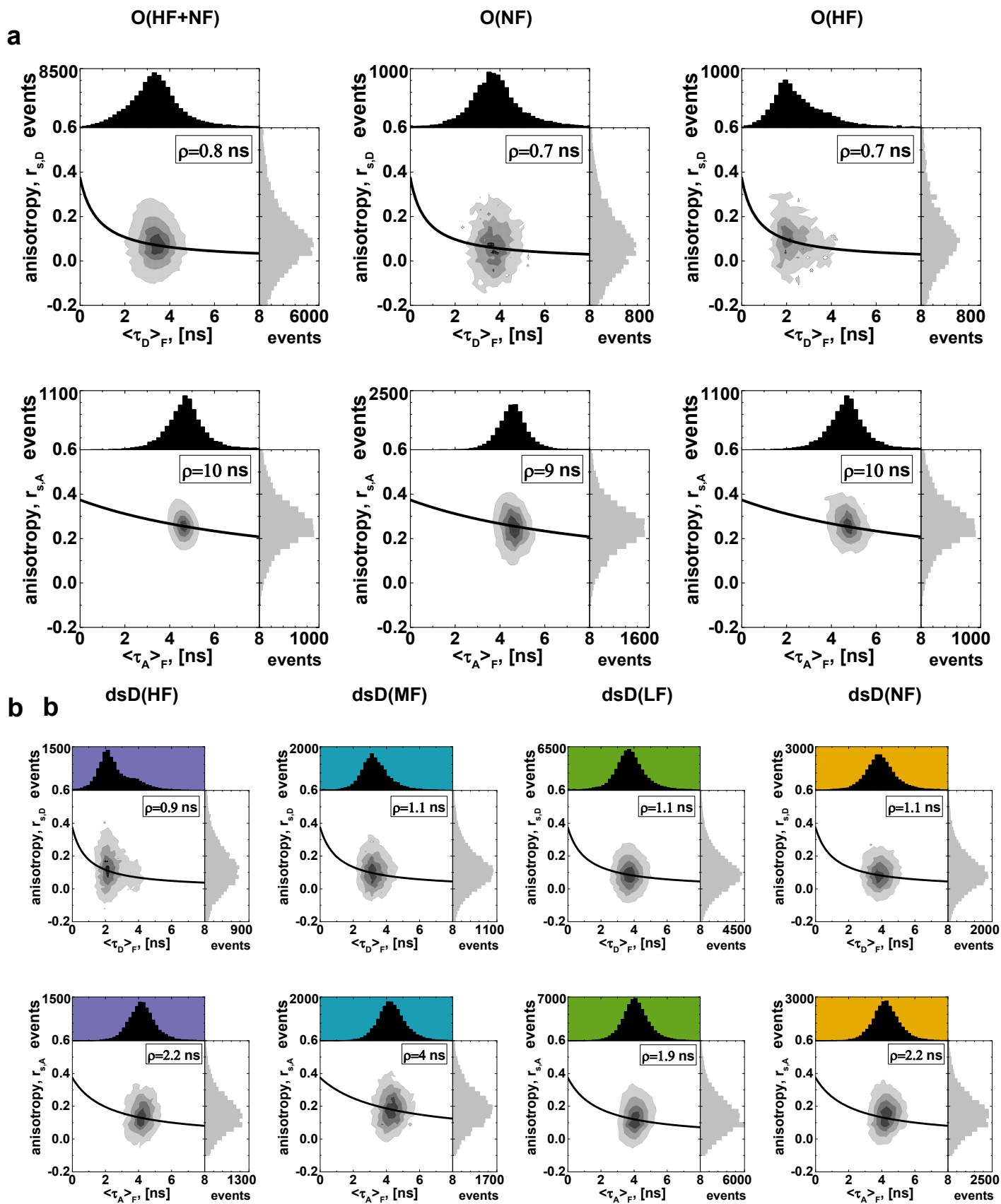

**Supplementary Fig. 14: Confocal single-molecule measurements of anisotropies and fluorescence lifetimes for origami and dsDNA samples.** a) Two-dimensional frequency histograms for the steady-state anisotropies of the Atto594 donor (rs,D, top row) and Atto647N acceptor ($r_{s,A}$, bottom

row) and the fluorescence lifetime for origami samples obtained by confocal single-molecule measurements. Each plot is overlaid with the Perrin equation (eq. S21c) using a fundamental anisotropy, $r_0 = 0.374$, of the donor and acceptor. The corresponding rotational correlation time is indicated in the graph. For the O(HF+NF) sample, the weighted averages of the two donor, respectively two acceptor dyes are reported. The rotational correlation time for acceptor Atto647N is high, indicating sticking. This effect was also observed by others[34,46]. The donor rotational correlation time is fast (< 2.5ns), indicating that the dye can move freely. b) Same as for a, but for the dsDNA ruler samples which are labelled with Alexa594 as donor and Atto647N as acceptor. All rotational correlation times are fast (<2.5ns) except for the medium FRET acceptor. This indicates that the sticking of the acceptor dye is specific to the origami sample. This observation matches the origami FRET distances, which over-estimate the predicted distance due to $\kappa^2$ effects. As a control, dsDNA rulers yield the correct predicted distance. Values are reported in Supplementary Tab. 11. The procedures are described in the section *'Confocal single-molecule spectroscopy with multiparameter fluorescence detection'*.

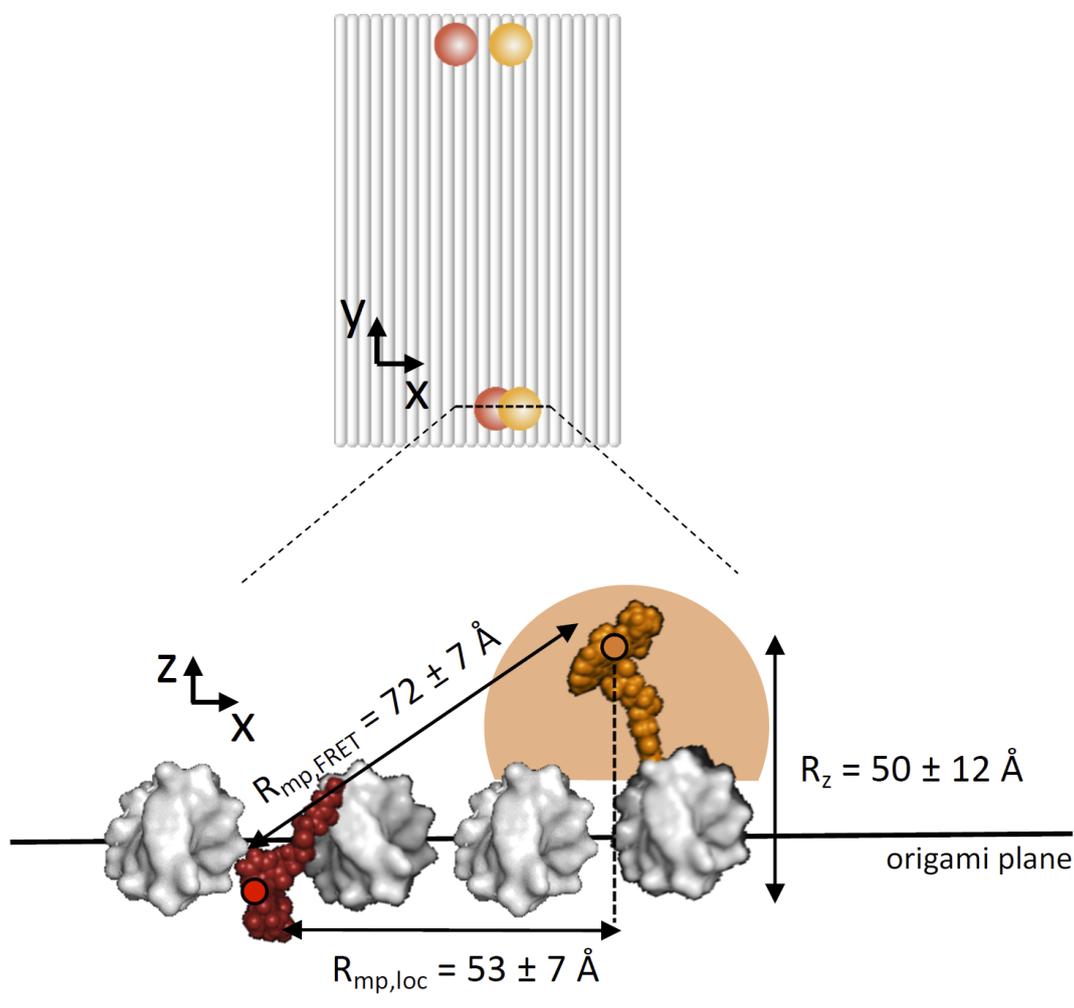

**Supplementary Fig. 15: Proposed model for the three-dimensional positioning of the dyes on the origami platform.**

The potential reasons for the discrepancy between localization and FRET distances for origami are discussed in Supplementary Note 4.

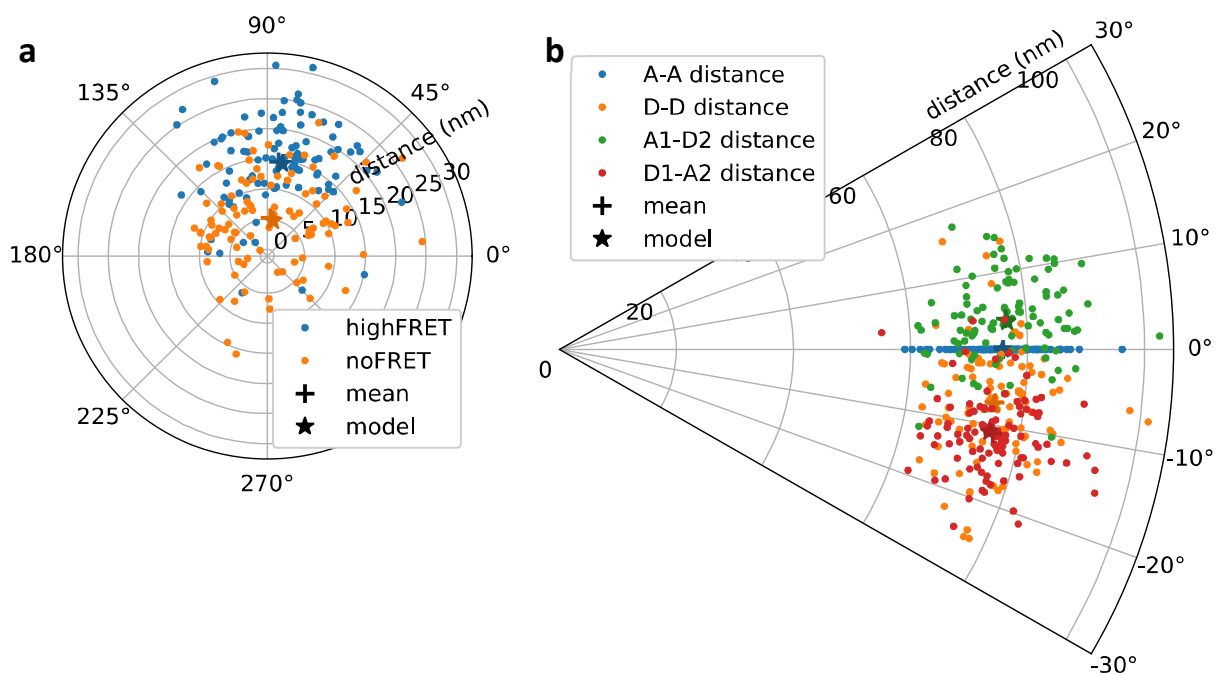

**Supplementary Fig. 16: Alternative origami alignment on acceptor-acceptor distance.**

The origamis are aligned such that the acceptor-acceptor distance is oriented along the x-axis and the NF acceptor is placed in the origin. **a)** Interpair distances **b)** Intrapair distances. This alignment approach is advantageous as one may use any sample that has two acceptors, for example 1 donor – 2 acceptors, and obtain better statistics. However, it does not use all information available and has sub-optimal alignment. We may rotate all structures without loss of information, hence it is natural to display distances in polar coordinates. The same dye labelling conventions as in Fig. 2 (main text) have been used. The mean and predicted distances match closely, such that the symbols overlap (refer to Fig. 2d of the main text for a zoomed figure).

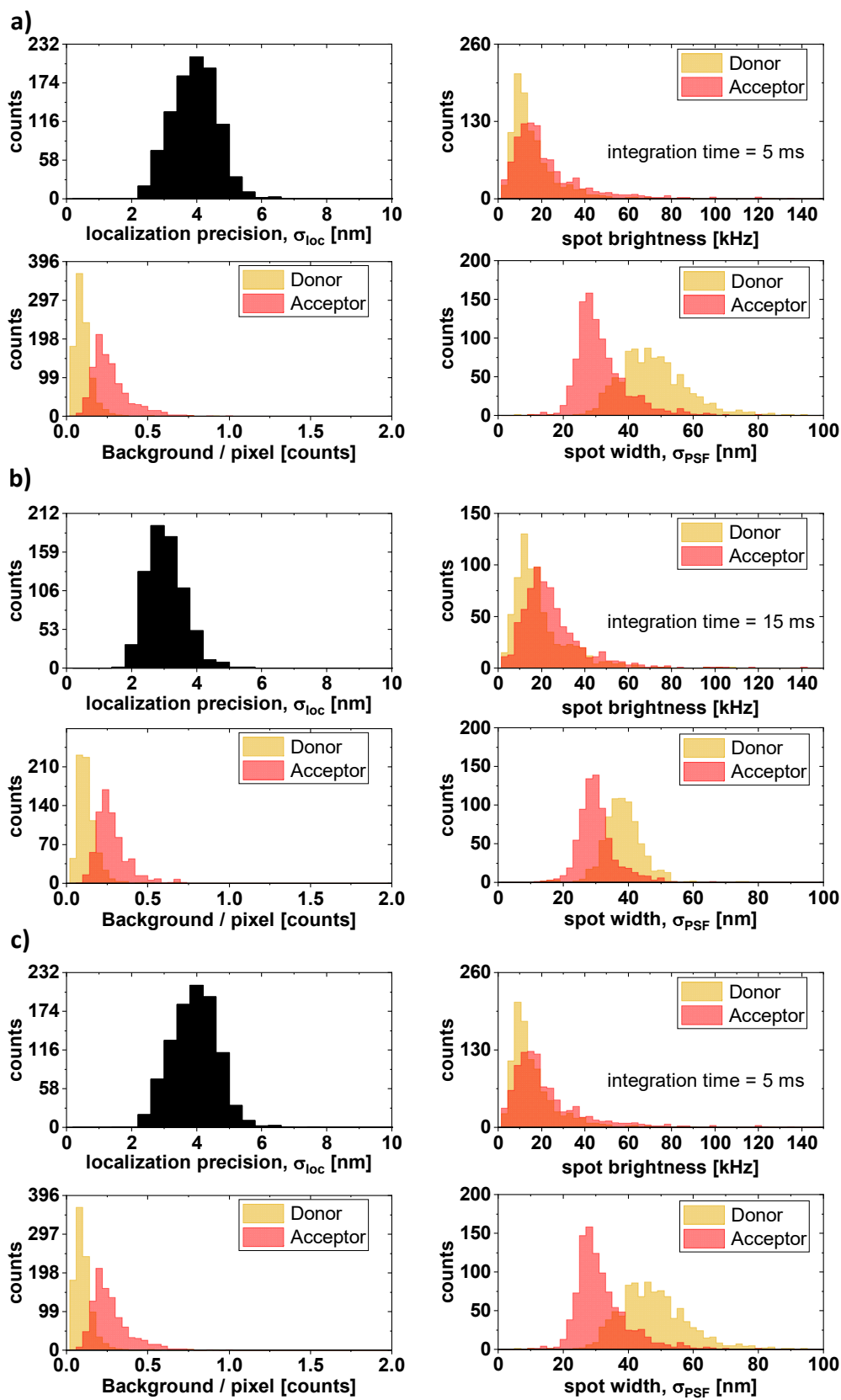

**Supplementary Fig. 17: Localization precision of origamis. a)** Localization characteristics for the O(HF+NF) sample. The localization precision (top-left) is calculated from equation S29 and depends on the spot brightness (top-right), the background level (bottom-left) and the spot width (bottom-right). **b)** The same is shown for the O(NF) and the **c)** O(HF) sample.

# DNA ruler figures

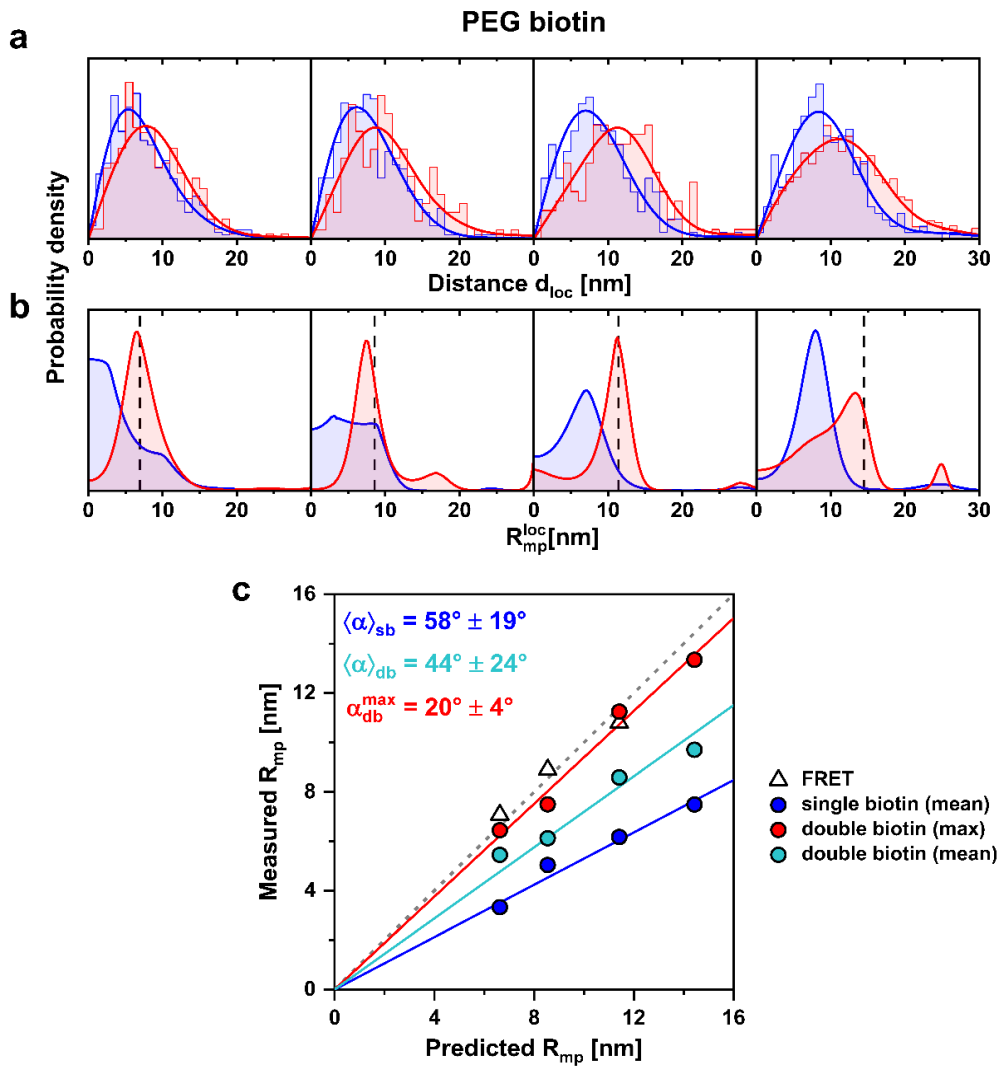

**Supplementary Fig. 18: Colocalization analysis of DNA rulers on PEG surfaces. a)** Distributions of the measured donor-acceptor distances from localization analysis for the single-biotin (blue) and double-biotin (red) samples (from left to right: dsD(HF), dsD(MF), dsD(LF), dsD(NF)) on the PEG surface. Fitted distributions based on maximum entropy analysis are shown as solid lines. **b)** The inferred distributions of the center distance of the $\chi$-distribution from maximum entropy analysis of the distance distributions shown in **a** (from left to right: dsD(HF), dsD(MF), dsD(LF), dsD(NF). For the MEM analysis, the width of the $\chi$-distributions was set to $\sigma_\chi$ = 4.4 nm. Expected distances between the mean positions of the fluorophores based on the AV model are shown as dashed lines. **c)** Distance-distance plots of the measured mean or peak (maximum) values of the inferred distance distributions against the predicted distances. Solid lines are linear fits to the data. The slopes of the fits define the inclination angle $\alpha$ which is calculated based on the mean and width of the distance distribution, $\langle\alpha\rangle$, or the peak value of the double-biotin population, $\alpha_{db}^{max}$. For the mean angles, the error is estimated from the width of the distance distribution. Compare Fig. 5a-c of the main text for the corresponding analysis of measurements performed on BSA-functionalized surfaces. See Supplementary Note 6.

# Surface immobilization: BSA- Biotin

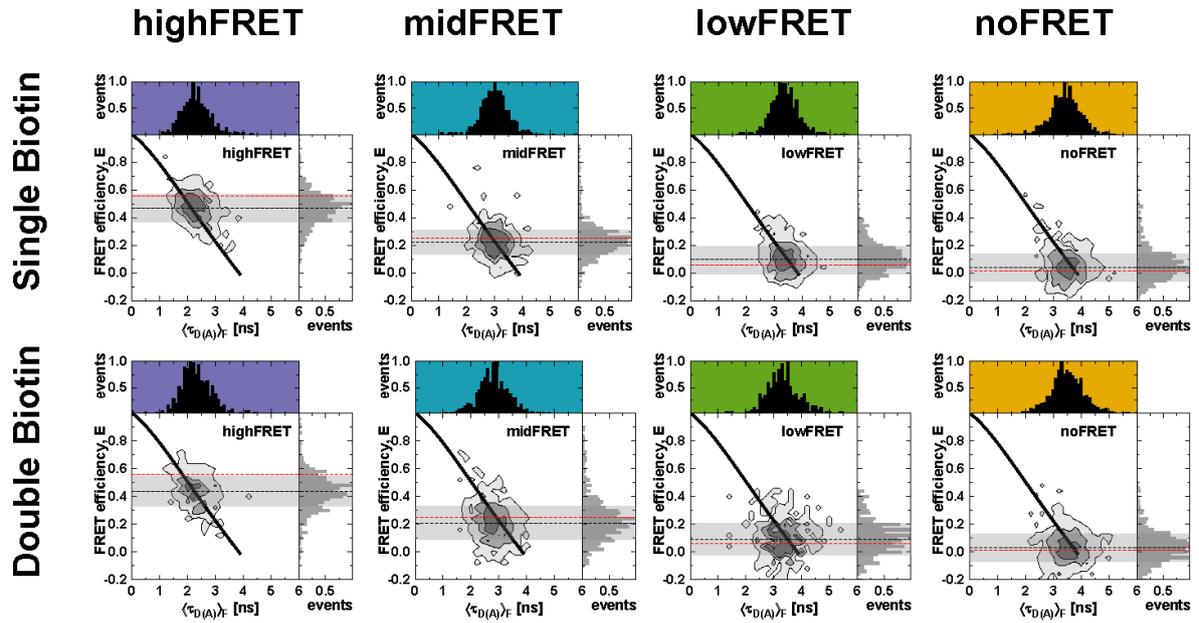

# Surface immobilization: NHS- PEG- Biotin

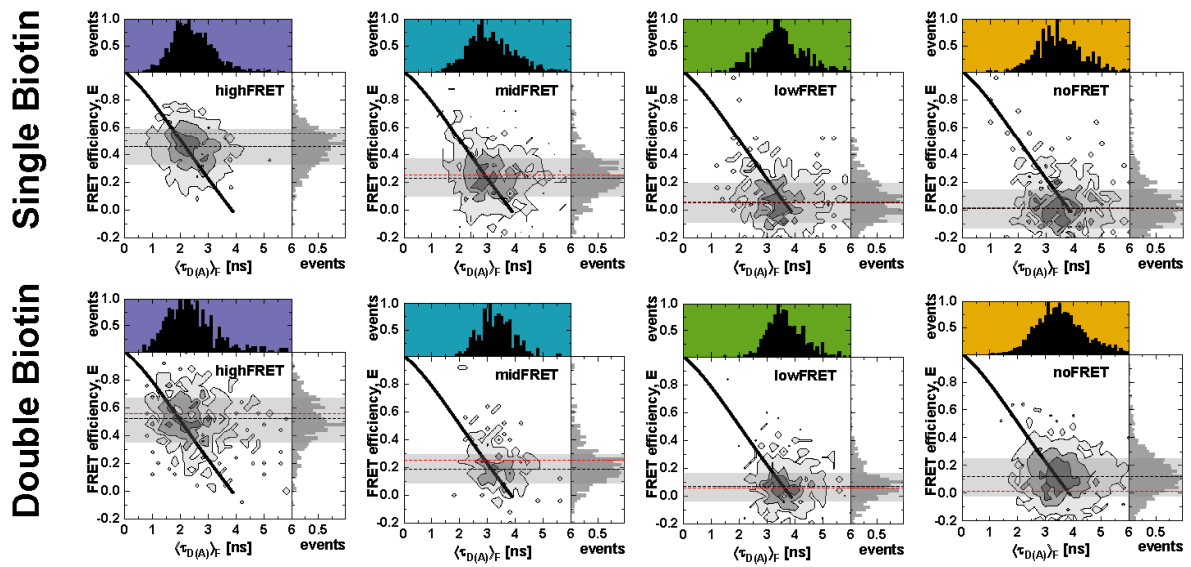

**Supplementary Fig. 19: FRET nanoscopy of dsDNA with two-dimensional frequency histograms of E and donor lifetime.** Two-dimensional frequency histograms of the ROI-integrated intensity-based FRET efficiency, E, and the fluorescence-weighted average donor fluorescence lifetime, $\langle \tau_{D(A)} \rangle_F$. The ROI populations lie on a static FRET line (solid line, given by eq.10 in the Methods section of the main text using the parameters in Supplementary Tab. 20). Single biotin and corresponding double biotin samples for each immobilization are colored equivalently (see Fig. 4 for a detailed description). FRET related values are reported in Supplementary Tab. 10.

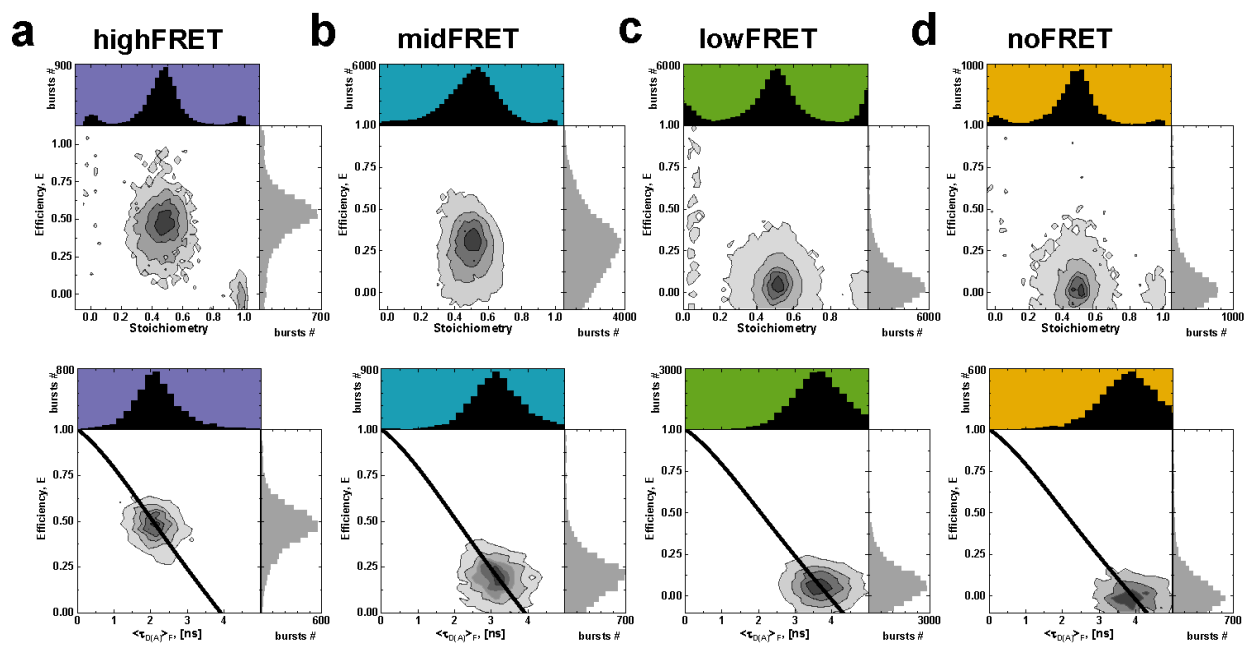

**Supplementary Fig. 20: Confocal sm measurements of dsDNA with two-dimensional frequency histograms of E and donor lifetime.** For each sample dsD(HF) **(a)**, dsD(MF) **(b)**, dsD(LF) **(c)**, dsD(NF) **(d)**. Two-dimensional frequency histograms of single-molecule bursts for intensity-based FRET efficiency, E, and stoichiometry are plotted (first row). Double labeled species were selected via stoichiometry cut between $S$ = 0.3 and $S$ = 0.7, leading to corresponding FRET efficiency versus fluorescence weighted lifetime plots (second row). For each sample, the FRET species lie on the static FRET line (black, given by eq.10 in the Methods section of the main text using the parameters in Supplementary Tab. 20). Sub-ensemble lifetime fits (see *'Confocal single-molecule spectroscopy with multiparameter fluorescence detection'*) of the selected populations were performed and compiled in Supplementary Tab. 9.

# BSA-Biotin- Immobilization

## Single biotin anchor          ## Double biotin anchor

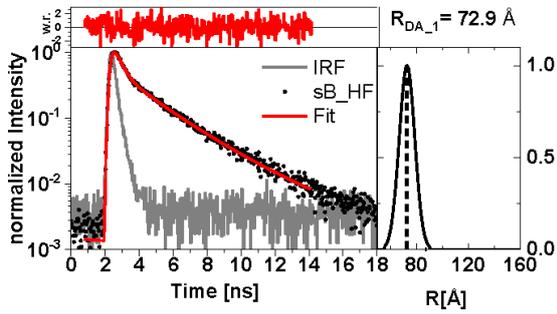
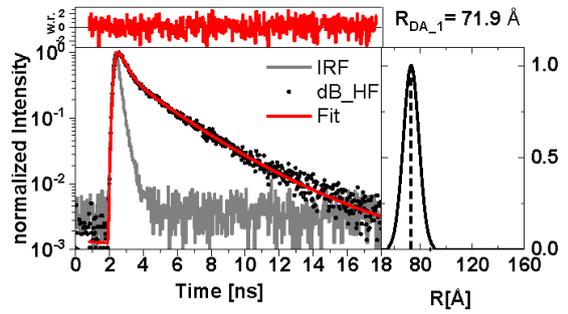

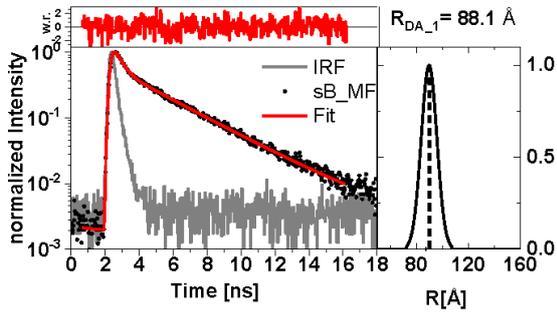
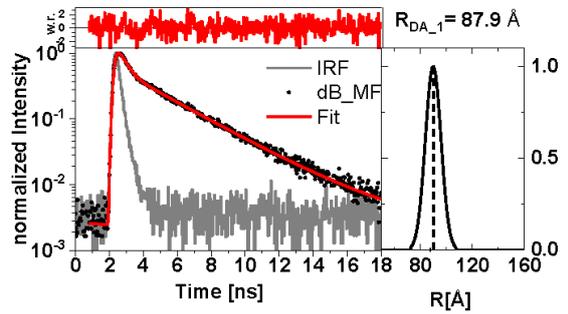

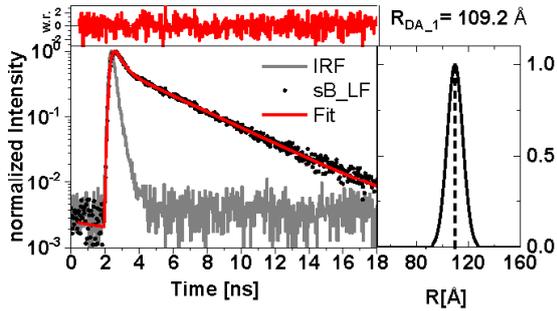
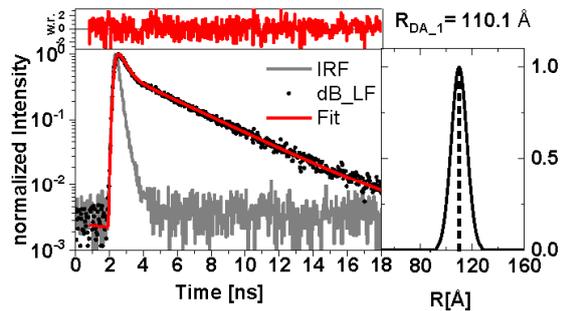

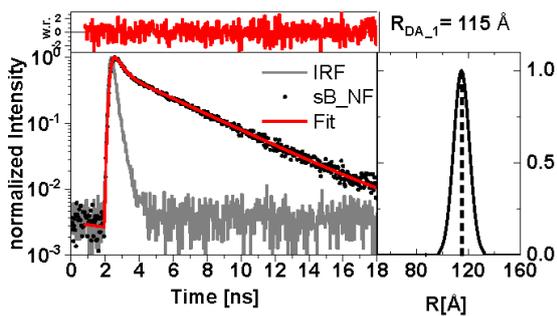
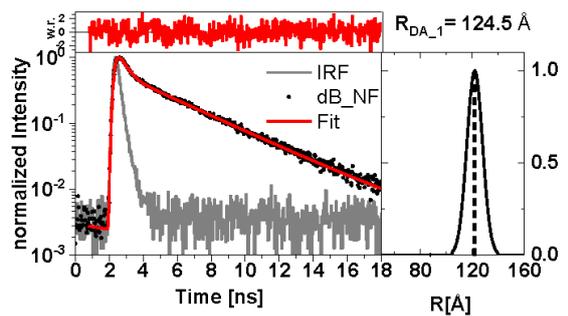

# NHS-PEG-Biotin- Immobilization

**Single biotin anchor**                    **Double biotin anchor**

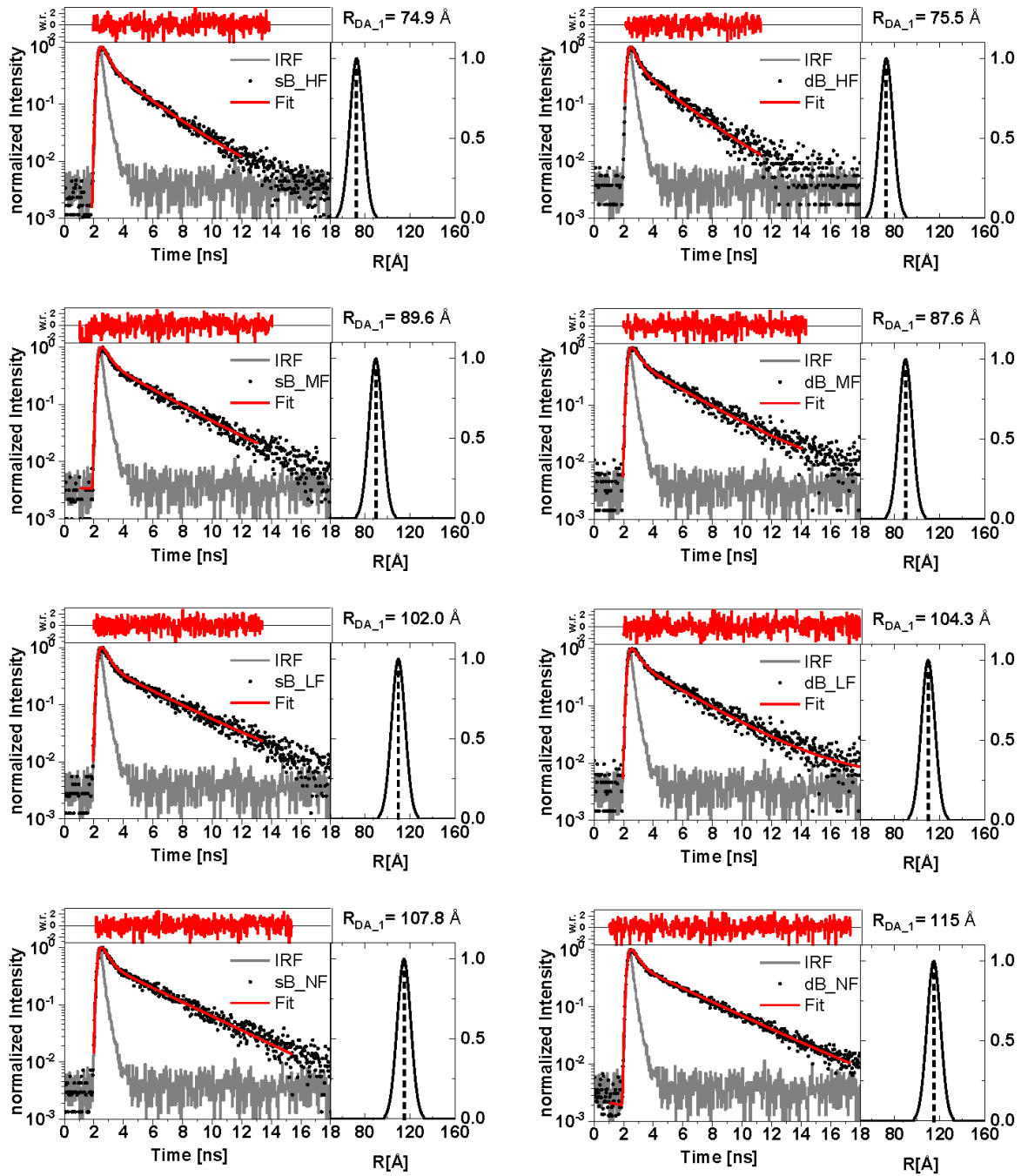

***Supplementary Fig. 21: Sub-ensemble fluorescence decay analysis of dsDNA rulers.*** Summary of sub-ensemble lifetime fit with Gaussian distributed distances model (see section '*Spectroscopy and image analysis, Sub-ensemble fluorescence decay analysis', eq. S19*). Bottom left: Sub-ensemble decay (scatter) convoluted with instrument response function (IRF, grey). Top left: weighted residuals. Bottom right: Gaussian model centered $R_{DA}$ with $\sigma$ = 6 nm. All fit results are compiled in Supplementary Tab. 9.

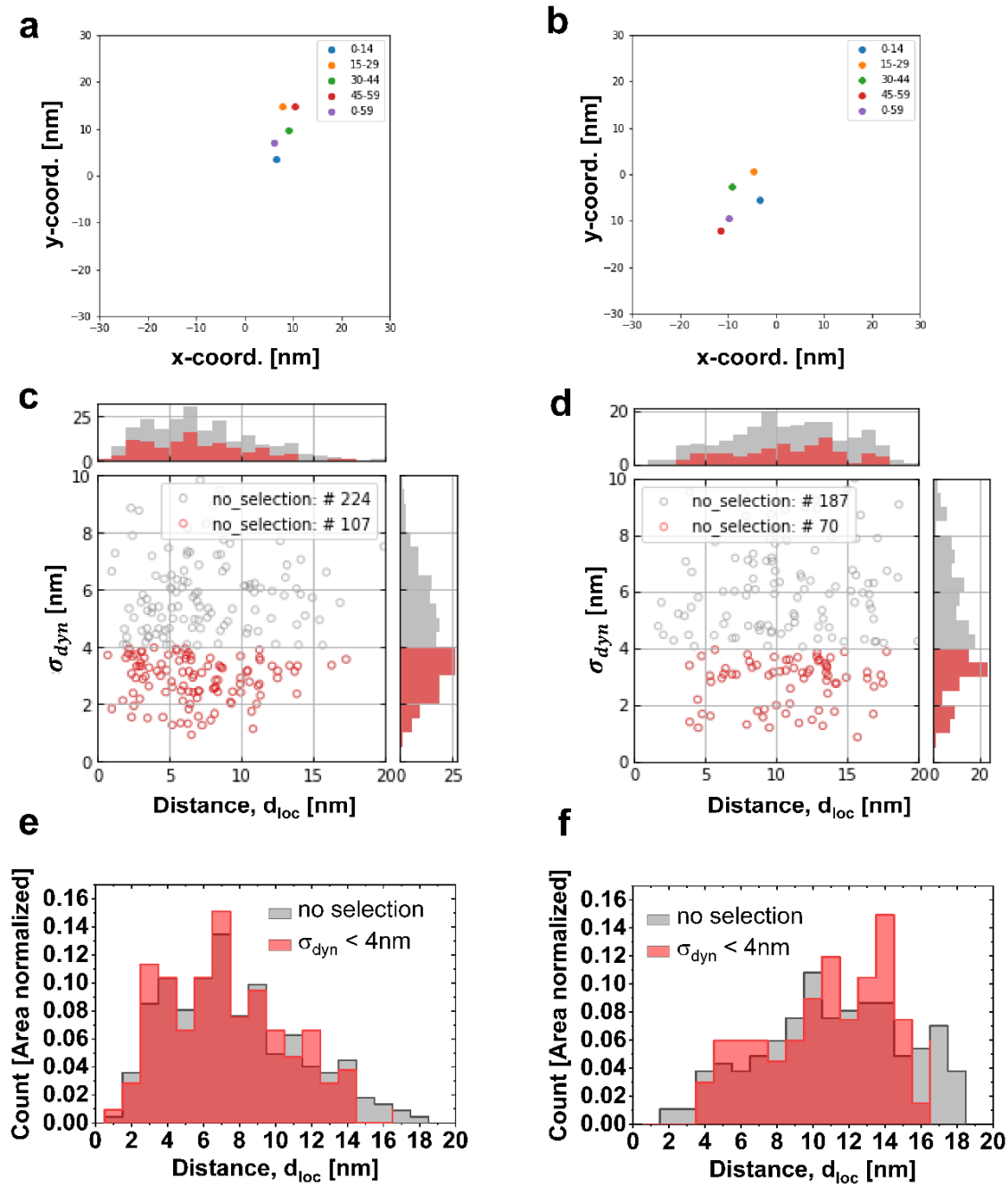

**Supplementary Fig. 22: The position of dsDNA rulers remains constant during the acquisition time. a-b)** Exemplary distance vector plots (three selected spots) of single (a) and double biotin (b) labeled dsD(NF) sample immobilized by BSA-biotin. Small displacement of colored spots indicates fast dynamic of dsDNA on the surface. **c-d)** Visualization of dynamic sigma selection for dsD(NF)T sample with single biotin (c) and double biotin (d) immobilization. The red colored selection corresponds to a dynamic sigma smaller than 4 nm. **e-f)** The corresponding distance histograms of the different selections are colored as in c-d. The procedures are described in Supplementary Note 5.

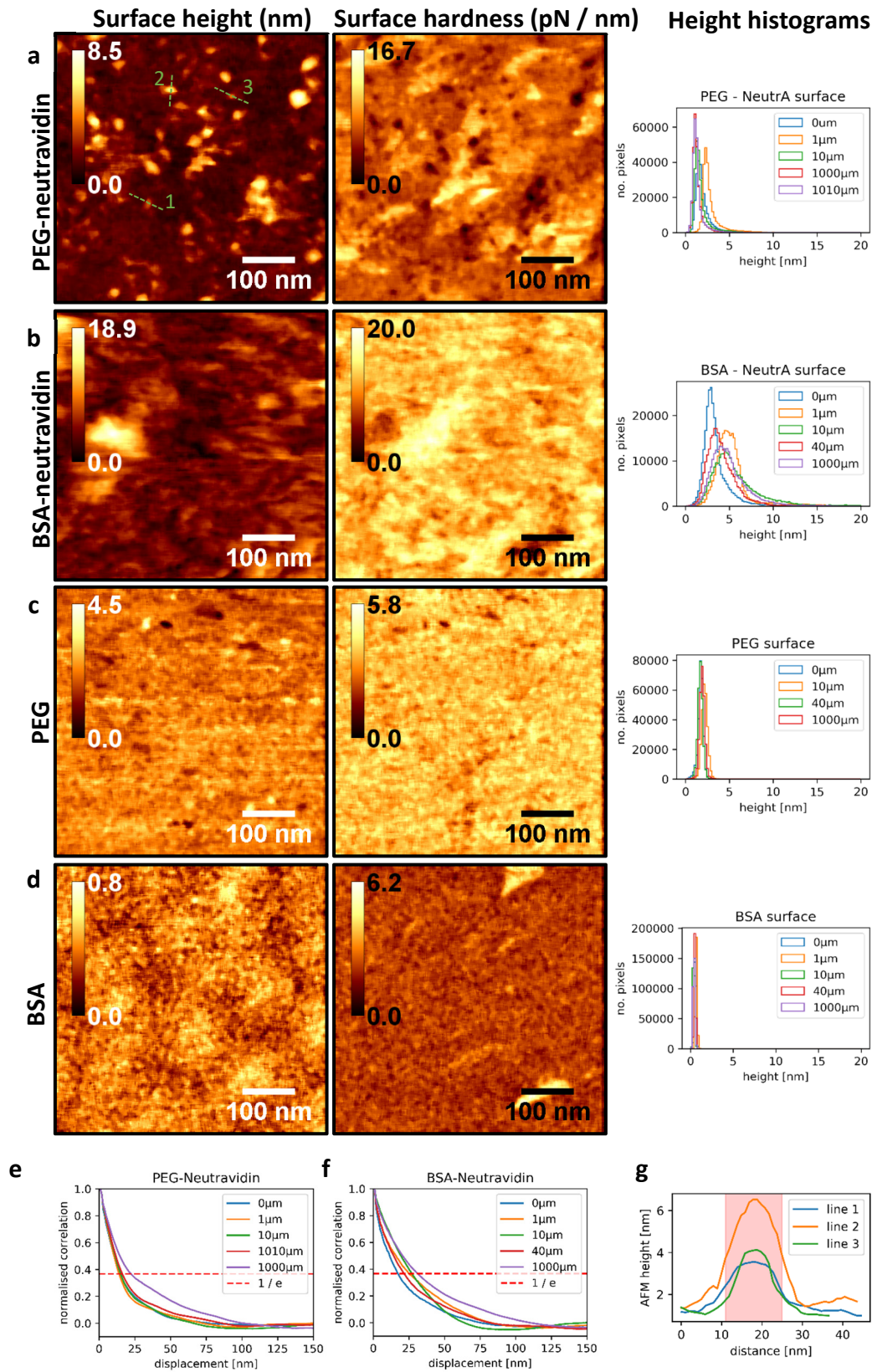

**Supplementary Fig. 23: AFM imaging of functionalized surfaces.** Biotin-functionalized surfaces using PEG and BSA after (**a-b**) and before (**c-d**) addition of neutravidin. The AFM was operated in QI mode [29] (see '*Atomic force microscopy*'). Height and hardness information was filtered using a median filter of 10x10 pixels and afterwards the zero-level was set to the lowest value in the image. For each preparation, the surface was recorded on multiple locations so that the obtained images and analysis parameters in panels a, e and f were labelled according to their relative positions. **a**) PEG-neutravidin surfaces show elevated features that are identified as individual neutravidin molecules because the height matches the known height from neutravidin proteins (4 nm) and the size is resolution-limited by the dimension of the AFM tip. Green lines correspond to lines profiles shown in **(e)**. Furthermore, elevated features (white) correlate with softer surfaces (black) indicating that these proteins are soft with respect to their surroundings. Additional hard features are visible in the hardness, but not in the height, indicating that they are either shallow or covered. **b**) BSA-neutravidin surfaces show higher features, typically 10 nm and up to 30 nm. This potentially indicates crosslinking of BSA and neutravidin, as each BSA protein has up to 12 biotins and neutravidin has four bindings sites. In addition, BSA might form aggregates. By comparing the surface roughness of (**c**) and (**d**), it is clear that our measurement setup is capable of resolving these height profiles and that additional heterogeneity is the consequence of surface roughness. **e-f)** Normalized image autocorrelation functions of the height profiles (Supplementary eqs. 50a-d) for PEG-Neutravidin (e) and BSA-Neutravidin (f) surfaces. Autocorrelations in 2D were radially integrated to obtain a 1D profile. The first value in the autocorrelation function is skipped as it represents a constant offset, normalization is done on the second value. Fig. 5 in the main text shows the average of the 5 displayed BSA and PEG curves. The cross section with 1/e indicates the correlation length. **g)** Line profiles of the smallest features identified throughout are used to determine the xy-resolution of the AFM images. Red box indicates the approximate half-width of the features to be 14nm. Note that the feature size is not significantly affected by the median filter.

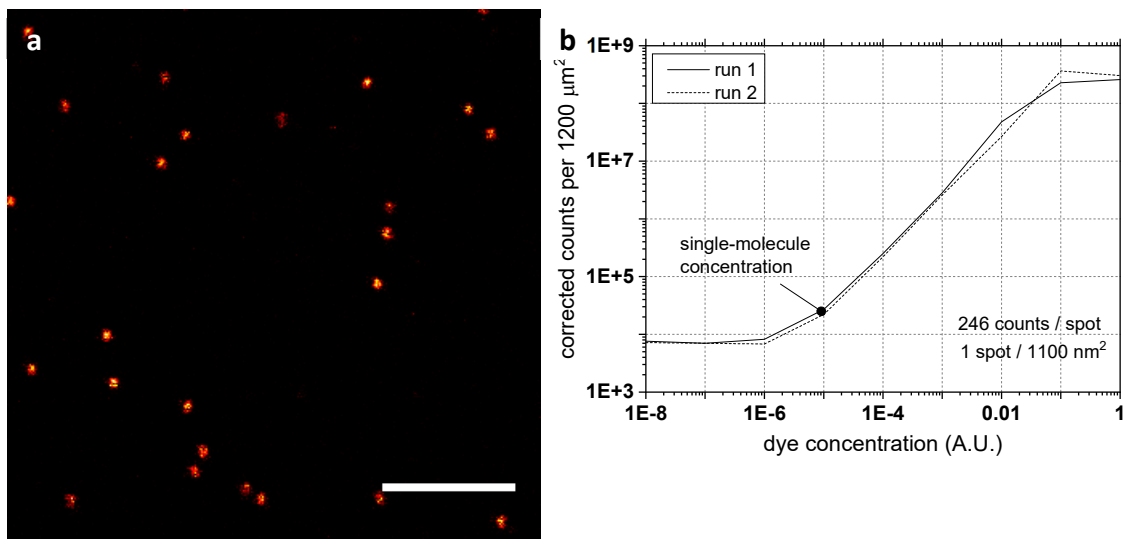

***Supplementary Fig. 24: Estimating the neutravidin density on the surface.*** **a**) Atto647N-biotin in low concentration is added to a PEG-neutravidin functionalized surface to achieve single-molecule concentration. The average molecular brightness is determined from the number of emitters and the background corrected intensity according to Supplementary Note 8. Scale bar is 5 μm. **b**) The solution is consequently exposed to higher concentration of Atto647N-biotin until surface saturation is reached (relative concentrations 0.1 and 1 A.U.). At higher dye concentrations the excitation power is reduced to avoid detector saturation. The counts are corrected for the lower excitation power assuming a linear dependence between brightness and excitation power. Two repetitions yield a saturation level of 300 and 250 Mcounts, of which the average has been taken. At saturation level individual spots can no longer be counted. The maximum spot density is calculated from the saturation brightness divided by the average brightness per emitter. Finally, the density is obtained by dividing with the surface area.

# Immobilization: BSA-Biotin

### Single Biotin                    ### Double Biotin

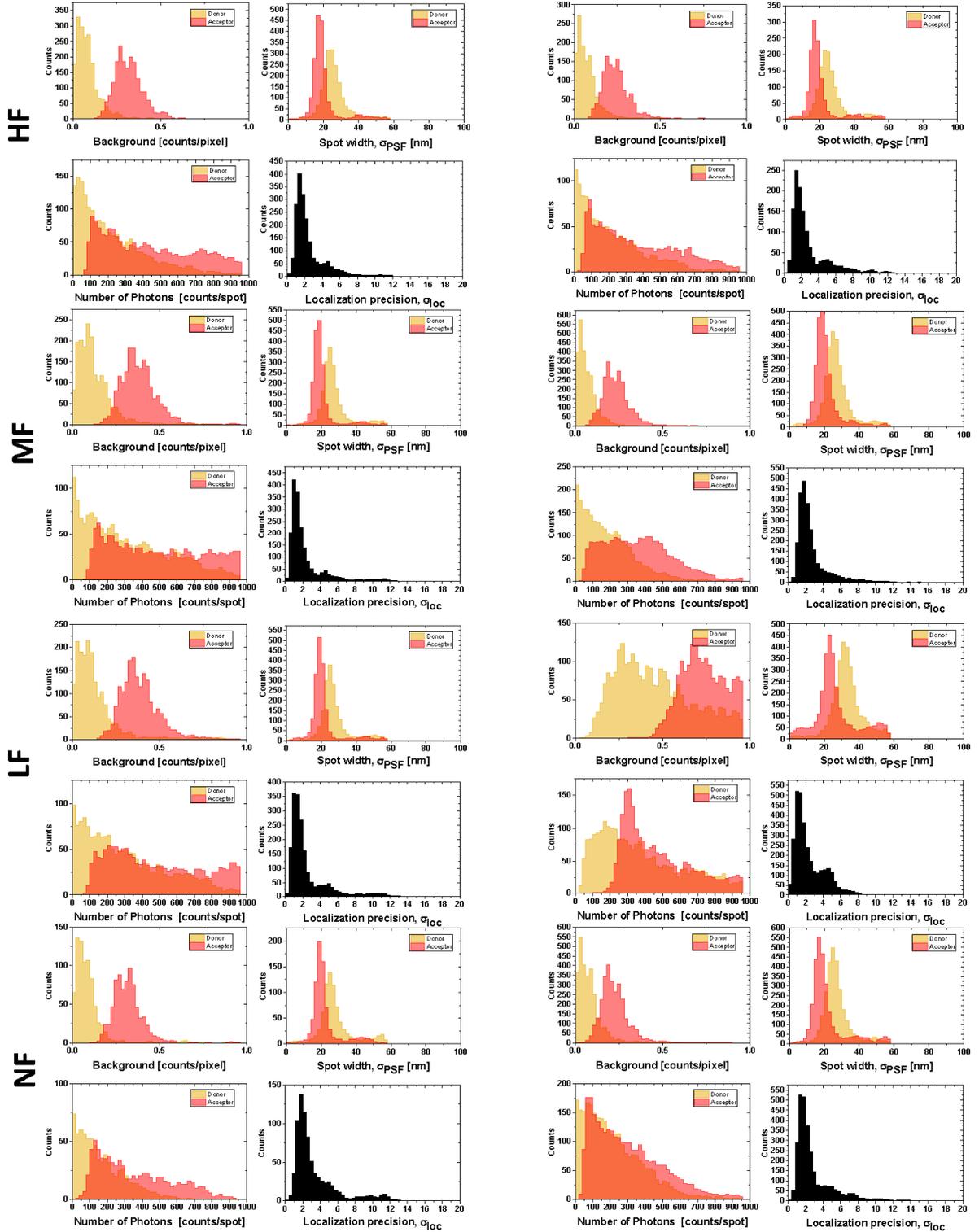

# Immobilization: PEG-Biotin

### Single Biotin          Double Biotin

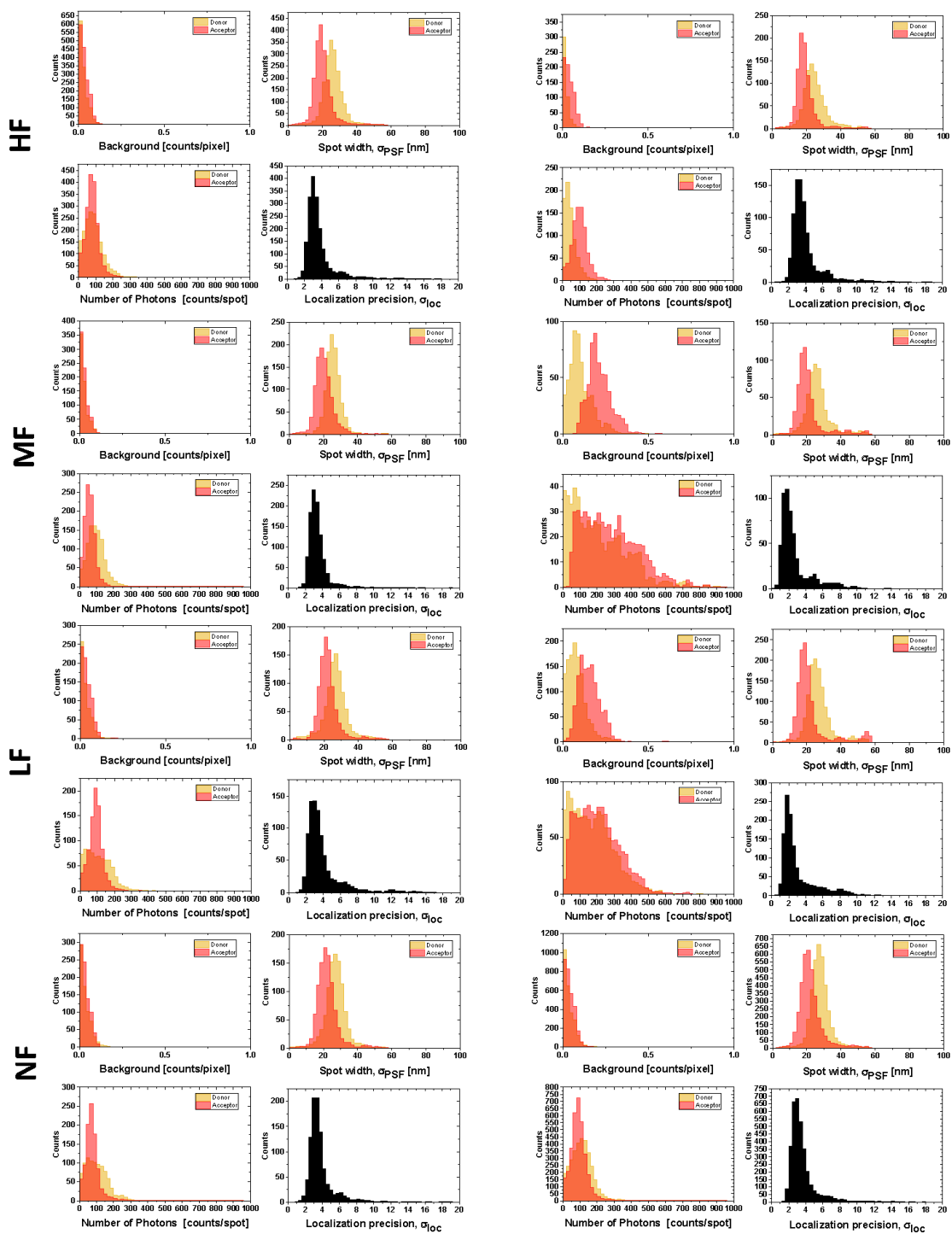

**Supplementary Fig. 25: Localization precision of dsDNA rulers.** Spot-wise localization characteristics for all dsDNA samples for BSA-Biotin and NHS-PEG-Biotin immobilization. The localization precision (bottom right) is calculated from equation S29 and depends on the number of photons per spot (bottom left), the background counts (top left) and the fitted spot width $\sigma_{\mathrm{PSF}}$ (top right). All data on microscope resolution and predicted precision are compiled in Supplementary Tab. 5.

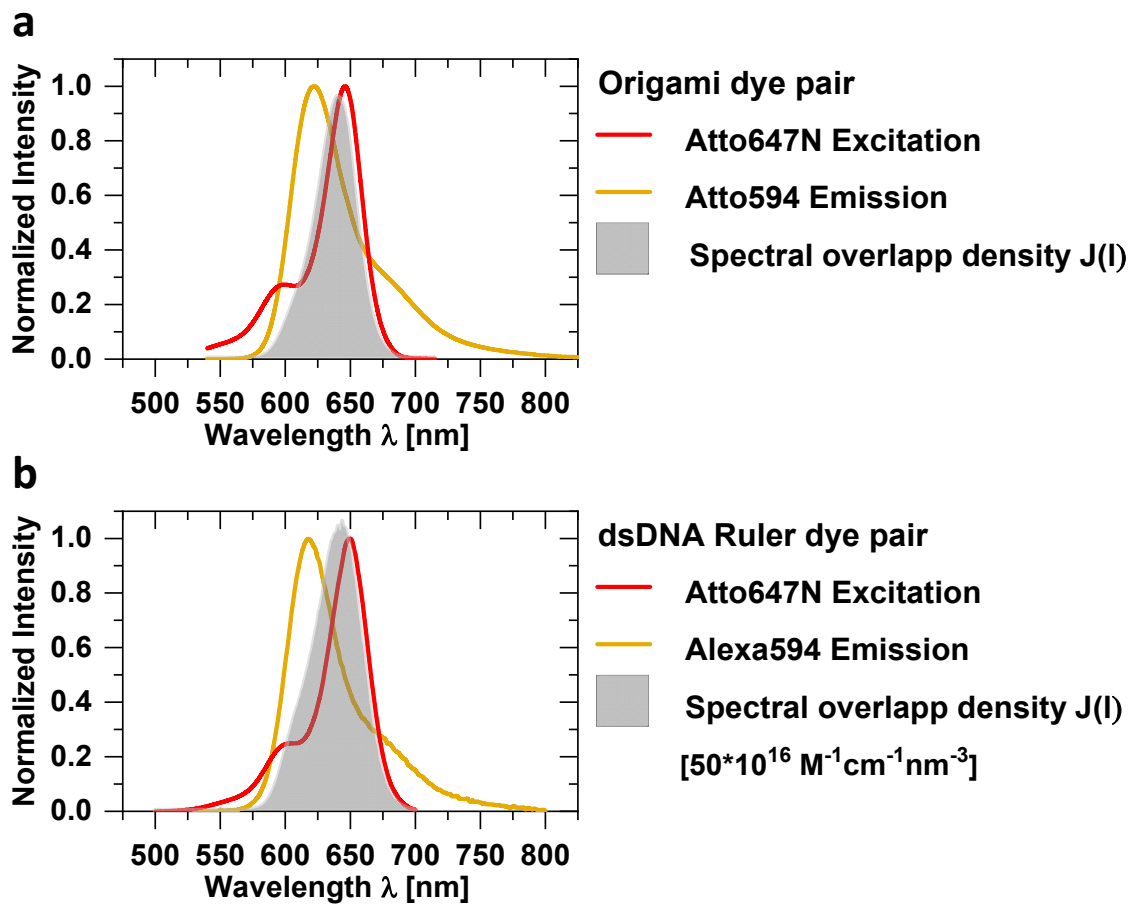

**Supplementary Fig. 26: Spectral overlap integrals.** Spectral overlap (grey area) of donor emission (orange) and acceptor excitation spectra (red) for **a)** origamis and **b)** dsDNA rulers. The calculated Förster radii $R_0$ are reported in Supplementary Tab 4.

**hGBP1 protein figures**

a

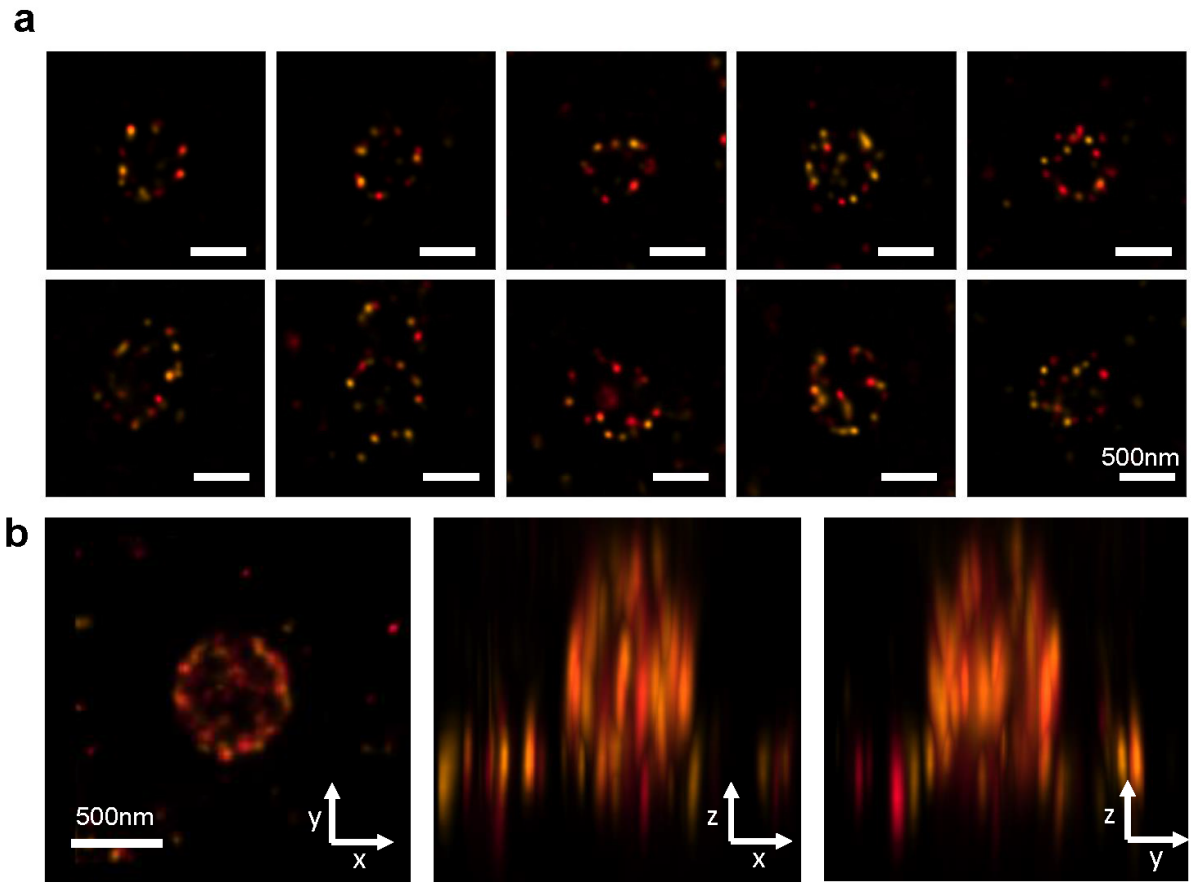

b

**Supplementary Fig. 27: 3D STED images of hGBP1 fibers.** a) Selection of deconvolved STED images. hGBP1 (18-577) was diluted 1:100 with unlabeled wild type hGBP1. The hGBP1 molecule is labeled with Alexa594 (donor) and Atto647N (acceptor). First row shows hGBP1 forming isolated ring-like structures, while the association of multiple ring structures is observed in second row. b) 3D illustration of hGBP1 ring structure in xy, xz and yz plane shows the 3D ring structure formed in oligomeric state. The procedures for the colocalization analysis of hGBP1 are described in Supplementary Note 10.

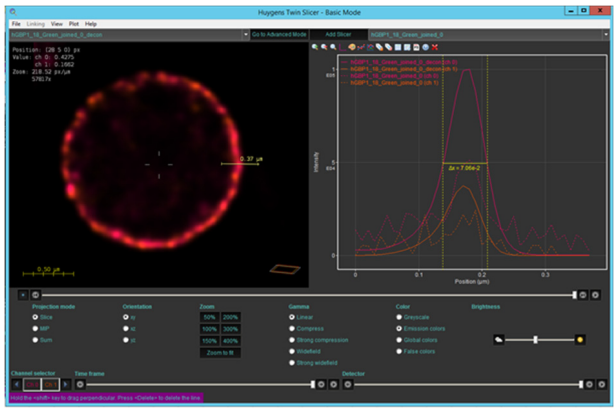 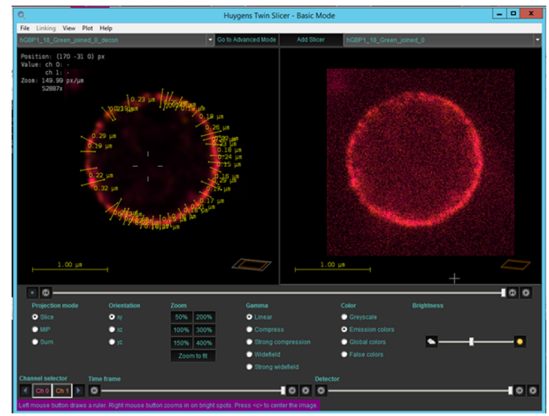

***Supplementary Fig. 28: Determination of hGBP1 fiber diameter.*** Determination of fiber diameter using Huygens deconvolution software (see section 'Assessment of hGBP1 fiber diameter'). The measured lineprofile is drawn perpendicular to the ring profile.

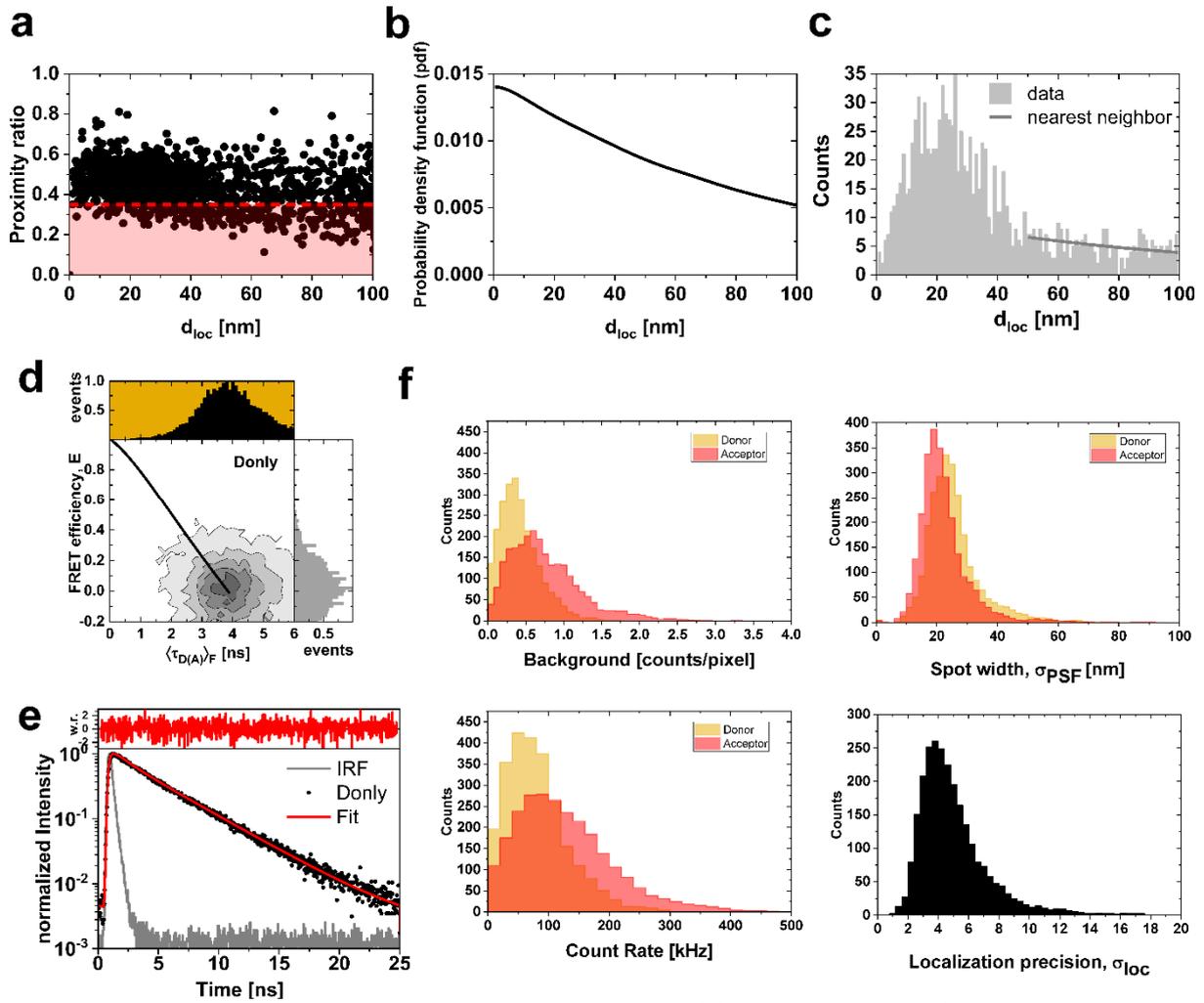

**Supplementary Fig. 29: Colocalization analysis of hGBP1 by cSTED. a)** Proximity ratio versus localization distance. Only those spots are selected with proximity ratio P > 0.35. **b)** Probability density function (pdf) versus localization distance. Simulation of randomly distributed Donor and acceptor describing the shape of the baseline of randomly distributed spots on the ring (repeated 1000 times). **c)** hGBP1 interdye distance histogram used to determine the fraction of baseline (black line). **d)** Two-dimensional frequency histogram of the ROI-integrated intensity-based FRET efficiency, E, and the fluorescence-weighted average donor fluorescence lifetime, $\langle \tau_{D(A)} \rangle_F$ for the hGBP1 18-577 donor only (Alexa594) sample. The population lies on static FRET line (solid line, given by eq.10 in the Methods section of the main text using the parameters in Supplementary Tab. 20). **e)** The corresponding sub-ensemble lifetime decay. Measured data (black scatter) is fitted with a two lifetime model yielding a donor only lifetime of 3.9 ns determined by fitting model eq. S19. **f)** For given ROI selection the histograms of donor and acceptor background photons, the Gaussian fitted sigma, count rate and predicted precision is shown. See Supplementary Note 10 for details.